\documentclass[twocolumn]{aastex701}

\usepackage{amsmath}
\usepackage{CJK}

\begin{document}
\begin{CJK*}{UTF8}{gbsn}

\title{Less massive black holes in faint hosts: SED fitting of Little Red Dots within the Black Hole Star scenario}

\author[0009-0004-6405-3440]{Haitong Liu (刘海桐)}
\affiliation{Department of Astronomy, University of Science and Technology of China, Hefei 230026, China}
\affiliation{School of Astronomy and Space Science, University of Science and Technology of China, Hefei 230026, China}
\email{liuhaitong@mail.ustc.edu.cn}

\author[0000-0003-4200-4432]{Lulu Fan (范璐璐)}
\affiliation{Department of Astronomy, University of Science and Technology of China, Hefei 230026, China}
\affiliation{School of Astronomy and Space Science, University of Science and Technology of China, Hefei 230026, China}
\affiliation{College of Physics, Guizhou University, Guiyang 550025, China}
\affiliation{Deep Space Exploration Laboratory, Hefei 230088, China}
\email[show]{llfan@ustc.edu.cn}

\author[0000-0002-2547-0434]{Yunkun Han (韩云坤)}
\affiliation{Yunnan Observatories, Chinese Academy of Sciences, 396 Yangfangwang, Guandu District, Kunming 650216, P. R. China}
\affiliation{International Centre of Supernovae (ICESUN), Yunnan Key Laboratory of Supernova Research, Yunnan Observatories, Chinese Academy of Sciences (CAS), Kunming 650216, China}
\email[show]{hanyk@ynao.ac.cn}

\author[0000-0002-8079-6525]{Yibin Luo (罗毅彬)}
\affiliation{School of Information, Hunan University of Humanities, Science and Technology, Loudi 417000, China}
\email{yibinluo@mail.ustc.edu.cn}

\begin{abstract}

The James Webb Space Telescope (JWST) has revealed a population of compact sources known as ``little red dots'' (LRDs), characterized by a distinctive V-shaped spectral energy distribution (SED) from UV to optical wavelengths. The physical nature of these sources remains debated. The ``black hole star'' (BH*) scenario, wherein a super-Eddington accreting black hole is embedded in a dense gaseous envelope, has been proposed to explain them. We assemble a sample of 22 LRDs at $4.0 \lesssim z \lesssim 9.5$ from the ASTRODEEP-JWST database. We construct a composite ``host galaxy + BH*'' model using the \texttt{Cloudy} photoionization code, and perform SED fitting with \textsc{BayeSED3}. Our modeling reproduces the observed V-shaped features: the UV continuum blueward of the Balmer break is well reproduced by stellar emission within this decomposition, while the optical regime is dominated by the BH* component. Our fits reveal that LRDs are black holes hosted by faint, low-mass galaxies with median stellar and black hole masses of $M_* = 10^{8.7}\,M_\odot$ and $M_{\rm BH} = 10^{6.4}\,M_\odot$, respectively. These black holes are moderately overmassive relative to local scaling relations but less extreme than broad-line estimates. The median bolometric luminosity $L_{\rm BH^*} \sim 10^{44.9}\,\rm erg\,s^{-1}$ mitigates the reported number density tension from 1--2 orders of magnitude to a factor of a few. The host galaxies lie near the high-redshift star-forming main sequence, and stellar extinction correlates positively with star formation rate (SFR). The BH* scenario provides a self-consistent paradigm for LRDs, offering insights into the rapid growth of supermassive black hole seeds and their co-evolution with host galaxies during the cosmic dawn.

\end{abstract}


\section{INTRODUCTION} 

The James Webb Space Telescope (JWST) is revolutionizing our exploration of the high-redshift universe. Among its most remarkable discoveries are  the Little Red Dots (LRDs), a population of compact sources that has garnered significant attention due to their enigmatic nature \citep{2024ApJ...963..129M}. These objects are characterized by a compact morphology and a distinctive spectral energy distribution (SED), featuring a steep red spectral slope in the rest-frame optical and a fainter blue continuum in the rest-frame ultraviolet \citep{Kocevski_2023,Hviding,2025ApJ...991...37A}. Consequently, LRDs represent a unique high redshift population whose fundamental physical nature remains highly uncertain.

Regarding the blue UV component, its origin remains debated. Proposed scenarios include unobscured star formation in the host galaxy, scattered light from an AGN accretion disk, and direct leakage through a non-unity covering envelope \citep{perezgonzalez2024naturelittlereddots,Taylor_2025,2026arXiv260403370J}. Evidence for a stellar origin comes from narrow emission lines and UV line ratios photoionized by massive stars \citep{2025ApJ...991...37A, asada2026originsuvcontinuumbalmer}, and from the correlation of the narrow-line Balmer decrement with the UV slope, which indicates galaxy-scale dust reddening \citep{2026arXiv260913444D}. Evidence for a compact central contribution includes the point-source fraction in stacked rest-UV morphologies \citep{2026arXiv260324700C}, UV continua that are redder and more compact than those of matched star-forming galaxies \citep{2026arXiv260603522A}, and broad Ly$\alpha$ emission in the far-UV of individual sources \citep{2026arXiv260403370J,2026arXiv260814534K}. For the red optical component, although obscured accretion disk models can fit the rest-frame optical and near-infrared (NIR) data, stellar-dominated models often provide a superior fit \citep{perezgonzalez2024naturelittlereddots}. Alternatively,  \citet{Li_2025} proposed that AGNs embedded in a dusty medium with a gray extinction curve could reproduce the observed SEDs without invoking stellar emission or scattered AGN light. Initially, driven by the red rest-frame optical continua and the detection of broad Balmer lines in a significant fraction ($>80\%$) of LRDs, it was widely hypothesized that these sources are heavily dust-reddened AGNs \citep{Greene_2024,Hviding}. 

Nevertheless, recent studies indicate that these traditional models are likely inadequate or, at the very least, incomplete. First, if LRDs were heavily dust-obscured AGNs, the dust would inevitably produce strong near-to-mid-infrared (NIR-MIR) re-emission, which is notably absent in JWST/MIRI observations \citep{2025ApJ...991...37A,Setton_2025,Leung_2025}. Second, LRDs display pronounced Balmer breaks near the rest-frame 3635 \AA, which, in certain instances, cannot be adequately reproduced by the continua of evolved stellar populations \citep{naidu2025blackholestarreveals,Taylor_2025}. Third, the supermassive black hole (SMBH) masses estimated from the widths of the H$\alpha$ and H$\beta$ emission lines, under the assumption of a dust-reddened AGN, are anomalously overmassive relative to their host galaxies \citep{Kokorev_2023,Greene_2024}.

A new physical paradigm has recently emerged to address these discrepancies. \citet{Inayoshi_2025} demonstrated that the Balmer break and absorption features observed in LRDs can originate from dense, neutral gas clumps surrounding the accretion disk, rather than stellar populations. This provides a compelling explanation for the spectral characteristics of LRDs and potentially resolves the discrepancies in inferred supermassive black hole masses and number densities. Building upon this, \cite{naidu2025blackholestarreveals} introduces the “black hole star” (BH*) model, motivated by a specific source exhibiting extreme properties, including the largest hydrogen Balmer breaks reported at any redshift, broad multi-peaked H$\beta$ emission, and Balmer line absorption across multiple transitions. The BH* model posits that a SMBH is enveloped by extremely dense, turbulent, and dust-free gaseous atmosphere. This gas naturally produces the strong Balmer absorption and smooth Balmer break observed in LRDs. Crucially, this framework successfully reproduces key observational features, such as the Balmer break strength, H$\beta$ and H$\gamma$ line profiles, and UV-faintness, without invoking significant dust attenuation. Subsequent work has further shown that the Balmer line profiles are shaped by radiative transfer in the envelope, and that the envelope column density, and hence its covering factor, varies systematically across the population \citep{2026arXiv260317667M}. When combined with a star-forming host galaxy, this composite model accurately matches the unique properties of LRDs, including their V-shaped SEDs, Balmer breaks, and X-ray and infrared (IR) weakness. Furthermore, this model suggests that SMBH masses in LRDs may be overestimated by up to two orders of magnitude in previous studies, and that their red optical color is driven by gas opacity rather than dust extinction.

An increasing body of observational and theoretical evidence supports the scenario that LRDs are powered by accreting black holes enshrouded by dense, optically thick envelopes. JWST/NIRSpec spectroscopy reveals that these sources are AGN embedded in such gas envelopes, which produce a characteristic red optical continuum with properties tightly coupled to their strong emission lines \citep{degraaff2025littlereddotshost}. Interestingly, three local analogs of LRDs have recently been discovered in the nearby universe, exhibiting similar properties, including absorption features indicative of cool gas envelopes surrounding local AGNs \citep{lin2025discoverylittlereddots}. Theoretically, the unique properties of LRDs can be explained by a dense, extended envelope surrounding a SMBH. This envelope confines the outflow from super-Eddington accretion and reprocesses its energy into blackbody radiation, naturally producing the observed red optical bump. This mechanism also suppresses radiative feedback, thereby facilitating rapid black hole growth in the early universe  \citep{kido}. Furthermore, recent studies indicate that the UV continuum and H$\alpha$ emission lines in LRDs are powered by young stellar populations rather than the central AGN, providing direct observational validation for models in which the black hole is obscured by a dense, opaque gas envelope \citep{asada2026originsuvcontinuumbalmer}.

A quantitative multi-band SED fitting for a sample of LRDs within the BH* framework will be valuable for decomposing the physical properties of the host galaxies and their central SMBHs. We develop a composite ``host galaxy + BH*'' model, wherein the BH* component is constructed using the \texttt{Cloudy} photoionization code, following the prescriptions outlined by \citet{naidu2025blackholestarreveals}. We perform  SED fitting on a sample of 22 LRDs utilizing the \textsc{BayeSED3} code. Our dataset comprises photometric catalogs from the publicly available ASTRODEEP-JWST database \citep{Merlin2024}, supplemented by NIRSpec prism spectroscopy from DAWN JWST Archive (DJA; \citealt{2025A&A...699A.358V}). The remainder of this paper is organized as follows. In Section \ref{sec:style} , we present our data and sample selection criteria. Section \ref{sec:SED fitting} details the Bayesian SED fitting methodology and the specific model configurations employed. In Section \ref{sec:results} , we present the fitting results of our SED fitting and the derived black hole masses. Section \ref{sec:DISCUSSION} discusses the physical nature of LRDs, the origin of the rest-UV emission, star formation and dust properties, redshift evolution of $M_{\rm BH}/M_{\star}$, LRD number density, the evolution of LRDs, and comparison with alternative LRD continuum scenarios.  Finally, we summarize our key findings in Section \ref{sec:summary}. Throughout this work, we assume a flat $\Lambda$CDM cosmology, with $H_0 = 70\ \mathrm{km\ s^{-1}\ Mpc^{-1}}$, $\Omega_M = 0.3$, and $\Omega_\Lambda = 0.7$ \citep{Komatsu_2011}.

\section{Data} 
\label{sec:style}

\subsection{Photometry Catalog} 
\label{subsec:Photometry Catalog}

We utilize the ASTRODEEP-JWST database, a comprehensive public release\footnote{\url{http://www.astrodeep.eu/astrodeep-jwst-catalogs/}}  \citep{Merlin2024} that facilitates detailed investigations of galaxy populations in the high-redshift universe. The photometric catalogs encompass 16 bands spanning 0.44--4.44 $\mu$m, comprising 8 JWST NIRCam filters (F090W, F115W, F150W, F200W, F277W, F356W, F410M, and F444W) and 8 HST filters (4 ACS bands: F435W, F606W, F775W, F814W; and 4 WFC3 bands: F105W, F125W, F140W, F160W). These data cover approximately 0.2 square degrees across six public extragalactic deep fields, totaling $\sim$500,000 sources. The six fields and their corresponding JWST programs are: the A2744 field, including JWST observations from GLASS-JWST (ERS 1324, \citealt{2022ApJ...935..110T}), UNCOVER (GO 2561, \citealt{2024ApJ...974...92B}), DDT 2756, and GO 3990 \citep{Morishita_2025}; the EGS field, observed by CEERS (ERS 1345, \citealt{Finkelstein_2025}); the GOODS-South and GOODS-North footprints, covered by the JADES-GS and JADES-GN fields (GTO 1180 and GTO 1210, \citealt{eisenstein2023overviewjwstadvanceddeep}), respectively, including FRESCO data (GO 1895, \citealt{10.1093/mnras/stad2411}); the first-epoch imaging of the NGDEEP field \citep{Bagley_2024}; and the UDS and COSMOS fields in CANDELS, observed by PRIMER (GO-1837; \citealt{2011ApJS..197...35G}; \citealt{2011ApJS..197...36K}). 

Source detection was performed on weighted F356W+F444W stacks using \textsc{SExtractor} with signal-to-noise ratio (S/N) threshold of $\sim$2. Photometry was measured using \texttt{a-phot} through point-spread function (PSF)-matched images and empirical ``\"uber'' PSFs constructed from stacked stars across all fields. Total fluxes were derived by scaling aperture colors to Kron elliptical apertures. The catalogs reach 5$\sigma$ depths of $\sim$28--30.5 mag depending on band and field, with NGDEEP being the deepest (50\% completeness at $\sim$30.5 mag) and PRIMER-UDS the shallowest ($\sim$29 mag). Astrometric calibration aligned all HST images to the JWST NIRCam F150W reference frame, while fluxes were homogenized to $\mu$Jy with a zero-point of 23.9. Root-mean-square (RMS) maps were validated via artificial source injection and rescaled to ensure accurate error estimates. Galactic extinction corrections were applied using values from the NASA/IPAC Extragalactic Database (NED). Among the several catalogs available for each field, we adopt the ``optap'' catalog, which assigns colors computed in the ``optimal'' aperture for each source and serves as the recommended reference.

\subsection{NIRSpec spectra} 
\label{subsec:NIRSpec spectra}

Our spectroscopic data are drawn from version 4 of the NIRSpec Merged Table in DAWN JWST Archive (DJA; \citealt{2025A&A...699A.358V})\footnote{\url{https://dawn-cph.github.io/dja/index.html}}. These data were processed using the \texttt{msaexp} and \texttt{grizli} pipelines, following the procedures described in \citet{refId0} and \citet{Heintz}. Version 4 of the spectral extractions significantly extends the wavelength coverage to regions that may suffer contamination by overlapping spectral orders. The catalog contains 80,367 entries, each corresponding to a unique NIRSpec spectrum. For every spectrum, the table provides source ID, sky coordinates, grating/filter setup, exposure time, redshift estimates, emission-line fluxes and photometric matches. The results are organized into eight grating-filter combinations, ranging from PRISM-CLEAR (34,949 spectra) to the high-resolution G395H-F290LP mode (13,606 spectra), with corresponding FITS files containing concatenated 1D spectra and all derived measurements. 

\subsection{Sample Selection} 
\label{subsec:Sample Selection}

Our sample selection follows the methodology of \citet{kocevski2025risefaintredagn}. LRDs are initially identified by fitting the continuum slope on both sides of the Balmer break, utilizing redshift-shifted JWST and HST bandpasses. The selection criteria for photometric candidates are as follows: (1) a red optical spectral slope of $\beta_{\rm opt} > 0$; (2) a blue UV spectral slope of $\beta_{\rm UV} < -0.37$; (3) an S/N $\geq 12$ in the F444W band; and (4) a UV spectral slope lower limit of $\beta_{\rm UV} > -2.8$ to exclude brown dwarfs, which produce V-shaped spectra resembling LRDs but with excessively blue UV colors. Furthermore, candidates must appear compact in F444W imaging, with {\sc SExtractor}-derived half-light radius smaller than 1.5 times that of the stellar locus. To exclude sources where red optical slopes might be driven by strong emission lines rather than the underlying stellar continuum, we impose $\beta_{\rm F277W-F356W} > -1$ and $\beta_{\rm F277W-F410M} > -1$. These criteria yield a parent sample of 504 LRD candidates.

For spectroscopic analysis, we restrict our sample to spectra with a reliability flag of \texttt{grade=3}, indicating robust redshifts determined from one or more emission or absorption features. Because subsequent SED fitting requires broad wavelength coverage at moderate spectral resolution, we specifically focus on the NIRSpec/PRISM spectra which have low resolution ($R \sim 100$) but cover a broad wavelength range ($\lambda \sim 0.6$--$5.3\,\mu\mathrm{m}$), yielding 115 \texttt{grade=3} PRISM spectra. We apply sequential quality cuts for reliable SED fitting: (i) average S/N $> 2.5$; (ii) broad wavelength coverage spanning the rest-frame UV to near-infrared. We then exclude 91 LRDs, comprising 66 with insufficient S/N, 9 with limited wavelength coverage, 11 with spectral morphologies inconsistent with LRDs (likely due to photometric contaminants or coordinate matching errors), and 5 with poor SED fitting quality ($\chi^2 > 10$). We arrive at a sample of 24 bona fide LRDs. We note that three objects in our spectroscopic sample, ABELL2744-10798,
ABELL2744-12048, and ABELL2744-12781, are not independent sources. The three
spectra have nearly identical spectroscopic redshifts ($z=7.038$, $7.038$,
and $7.040$, respectively), and \citet{2024Natur.628...57F} identified them as the three lensed images of a single LRD, Abell2744-QSO1. To avoid multiple counting of the same object, we retain only the spectrum of the highest-S/N image (ABELL2744-12048) for SED fitting and treat the three spectra as one unique source in all subsequent analyses. Our final sample therefore comprises 22 unique LRDs.

Six unique sources in our final sample lie in the ABELL2744 field, observed by the GLASS-JWST (ERS 1324), UNCOVER (GO 2561), DDT 2756, and GO 3990 programs. Because this is a strong-lensing cluster field, the observed fluxes of these background sources are magnified by the cluster potential, and lensing
magnification corrections are mandatory before their spectra can be used for
SED fitting. For each source we obtained the magnification $\mu$ from the public
UNCOVER strong-lensing data products\footnote{\url{https://jwst-uncover.github.io/}}. Sources observed as NIRSpec/MSA targets in the UNCOVER program adopt $\mu$ from the DR4 spectroscopic magnification catalog \citep{2025ApJ...982...51P}, while the remaining sources adopt $\mu$ from the SUPER photometric magnification catalog \citep{2024ApJ...976..101S}. Both are computed with the UNCOVER strong-lensing model of \citet{2023MNRAS.523.4568F}. Note that $\mu$ is achromatic, it multiplies all wavelengths by the same factor, so the spectral shapes and all relative quantities inferred from them (e.g., continuum slopes, Balmer break strengths, and $M_{\rm BH}/M_{\star}$) are unaffected by lensing, only absolute fluxes, luminosities, and masses are scaled by $1/\mu$. We therefore divided the flux, flux uncertainty, and sky columns of the 1D PRISM spectra of these sources by the corresponding $\mu$, and all SED fitting presented in this work is performed on the delensed spectra. The physical properties of this final sample are summarized in Table \ref{tab:sample}.

\section{SED fitting} \label{sec:SED fitting}

We perform SED fitting on the NIRSpec/PRISM data of our final sample using the \textsc{BayeSED3} code.  Our composite model incorporates two primary components: a stellar component and a BH* component. The instrumental resolution curve for NIRSpec/PRISM, obtained from the official JWST documentation\footnote{\url{https://jwst-docs.stsci.edu/jwst-near-infrared-spectrograph/nirspec-instrumentation/nirspec-dispersers-and-filters\#gsc.tab=0}}, is employed to convolve the model spectra prior to comparing them with the observed data. The prior ranges for all model parameters are summarized in Table \ref{tab:Priors}.

\subsection{BayeSED3} 
\label{subsec:BayeSED3}

\textsc{BayeSED3} \citep{Han_2012,Han_2014,Han_2019,Han_2023} is a Bayesian SED synthesis and analysis tool specifically designed for galaxies and AGNs. The code performs full Bayesian parameter estimation by deriving posterior probability distributions and conducts model comparison via Bayesian evidence calculations. It integrates physical SED models with machine learning emulation and nested sampling inference to derive physical properties from both photometric and spectroscopic observations. Specifically, \textsc{BayeSED3} employs the \texttt{MultiNest} nested sampling algorithm to perform Bayesian inference on the physical properties of diverse astronomical sources across. This approach simultaneously constrains model parameters with full posterior distributions, evaluates competing models through Bayesian evidence, and quantifies uncertainties and parameter degeneracies arising from both broad-band photometry and high-resolution spectroscopy.

\begin{deluxetable*}{lccccccc}
\tabletypesize{\scriptsize}
\tablewidth{0pt} 
\tablecaption{Summary of our Final Sample \label{tab:sample}}
\tablehead{
\colhead{ID} & \colhead{RA} & \colhead{DEC} & \colhead{$z_{\text{spec}}$} & \colhead{$m_{\text{f444w}}$} & \colhead{$\mu$} \\
\colhead{} & 
\colhead{(deg)} & 
\colhead{(deg)} & 
\colhead{} & 
\colhead{(AB mag)} 
}
\startdata
ABELL2744-4992 & 3.619200 & -30.423266 & 5.837 & $25.02 \pm 0.01$ & $1.61^{+0.00}_{-0.01}$ \\
ABELL2744-8864 & 3.556702 & -30.408194 & 4.075 & $23.43 \pm 0.00$ & $1.54^{+0.00}_{-0.01}$ \\
ABELL2744-9196 & 3.550837 & -30.406600 & 5.057 & $24.40 \pm 0.00$ & $1.00^{+0.00}_{-0.00}$ \\
ABELL2744-12048 & 3.583537 & -30.396675 & 7.038 & $25.19 \pm 0.01$ & $6.74^{+0.57}_{-0.34}$ \\
ABELL2744-12351 & 3.545785 & -30.395727 & 5.206 & $25.50 \pm 0.01$ & $1.84^{+0.00}_{-0.03}$ \\
ABELL2744-25363 & 3.530005 & -30.358015 & 4.960 & $24.89 \pm 0.00$ & $2.25^{+0.09}_{-0.10}$ \\
CEERS-21058 & 214.899693 & 52.812842 & 5.500 & $25.39 \pm 0.01$ \\
CEERS-29193 & 214.880159 & 52.812562 & 7.075 & $25.00 \pm 0.01$ \\
CEERS-31837 & 214.924124 & 52.849048 & 4.223 & $24.44 \pm 0.01$ \\
CEERS-47313 & 214.892475 & 52.856887 & 6.738 & $25.67 \pm 0.03$ \\
CEERS-47680 & 214.795358 & 52.788850 & 5.268 & $23.25 \pm 0.00$ \\
CEERS-55630 & 214.892238 & 52.877407 & 6.684 & $24.06 \pm 0.00$ \\
CEERS-64408 & 214.972438 & 52.962194 & 5.077 & $25.81 \pm 0.02$ \\
CEERS-72934 & 214.876144 & 52.880825 & 8.353 & $25.81 \pm 0.01$ \\
JADES-GN-10395 & 189.197412 & 62.177232 & 4.133 & $25.16 \pm 0.01$ \\
JADES-GN-41323 & 189.151960 & 62.259635 & 6.760 & $25.26 \pm 0.02$ \\
JADES-GS-6608 & 53.107225 & -27.890602 & 5.575 & $26.51 \pm 0.02$ \\
PRIMER-COSMOS-1223 & 150.132300 & 2.169688 & 5.650 & $25.99 \pm 0.02$ \\
PRIMER-COSMOS-53040 & 150.124021 & 2.306835 & 5.575 & $26.00 \pm 0.01$ \\
PRIMER-COSMOS-53670 & 150.136251 & 2.308031 & 9.450 & $26.35 \pm 0.01$ \\
PRIMER-UDS-47866 & 34.264582 & -5.232545 & 5.676 & $24.91 \pm 0.02$ \\
PRIMER-UDS-129331 & 34.261914 & -5.105169 & 5.325 & $25.48 \pm 0.03$ \\
\enddata
\tablecomments{The last three columns list the lensing magnification $\mu$ and the 16th and 84th percentiles of the magnification posterior. These quantities are provided only for the six sources in the ABELL2744 field, which is a strong-lensing cluster field and thus requires lensing magnification corrections.}
\end{deluxetable*}

\begin{deluxetable*}{lcccccccc}
\tabletypesize{\scriptsize}
\tablewidth{0pt} 
\tablecaption{BayeSED Priors for Stellar--BH* Model \label{tab:Priors}}
\tablehead{
\colhead{Component} & \colhead{Parameter} & \colhead{Description} & \colhead{Range} 
} 
\startdata 
Stellar & $\log(\mathrm{age/yr})$ & Age of stellar population & (8, 10.13) \\
{} & $\log(\tau/\text{yr})$ & Delayed decay time & (6, 10) \\
{} & $\log(Z/{Z_{\odot}})$ & Metallicity in $Z_{\odot}$ & (-2.3, 0.7) \\
{} & $A_{V}$ & Visual extinction in magnitude & (0.0, 7.0) \\
{} & $v_{\rm d}^{\rm eml}\,[{\rm km/s}]$ & Velocity dispersion of emission lines & (100, 500) \\
\hline
BHstar & log(T$_{BB}$/K) & AGN BB effective temperature & (4.0, 6.0) \\
{} & a(ox) & Optical to X-ray index & (-3.0, -1.0) \\
{} & a(uv) & UV slope & (-2.5, -0.1) \\
{} & a(x) & X-ray slope & (-2.5, -0.1) \\
{} & log(U) & Ionization parameter & (-2.5, -0.5) \\
{} & log(n$_{H}$/cm$^3$) & Gas density & (9.0, 12.0) \\
{} & log(Z$_{gas}$/Z$_{sun}$) & Gas metallicity & (-2.0, 0) \\
{} & log(N$_{H}$/cm$^2$) & Cloud column density & (21.0, 26.0) \\
{} & log(v$_{turb}$) & Turbulent velocity & (1.0, 3.0) \\
{} & $A_{V}$ & Visual extinction in magnitude & (0.0, 7.0) \\
\enddata
\tablecomments{This table lists the BayeSED priors for our Stellar--BH* model.}
\end{deluxetable*}

\begin{deluxetable*}{lccccccccccccc}
\tabletypesize{\scriptsize}
\tablewidth{0pt} 
\tablecaption{Summary of Posterior Inferences from SED Modeling of the LRD Sample \label{tab:Posterior}}
\tablehead{
\colhead{ID} & 
\colhead{$\log(\mathrm{age_{\star}})$} & 
\colhead{$\log(M_{\star})$} & 
\colhead{$\log(L_{\star})$} & 
\colhead{$A_{V,\star}$} & 
\colhead{$\log(\mathrm{SFR})$} & 
\colhead{$\log(T_{\rm BB})$} & 
\colhead{$\log(U)$} & 
\colhead{$\log(n_{\rm H})$} & 
\colhead{$\log(N_{\rm H})$} & 
\colhead{$\log(v_{\rm turb})$} & 
\colhead{$A_{V,BH*}$} & 
\colhead{$\log(L_{BH*})$} & 
\colhead{$\log(M_{\rm BH})$} \\
\colhead{} & 
\colhead{(yr)} & 
\colhead{($M_{\odot}$)} & 
\colhead{($\text{erg}\, \text{s}^{-1})$} & 
\colhead{(mag)} & 
\colhead{($M_\odot\,\mathrm{yr}^{-1})$} & 
\colhead{(K)} & 
\colhead{} & 
\colhead{(cm$^{-3}$)} & 
\colhead{(cm$^{-2}$)} & 
\colhead{(\text{km}\, $\text{s}^{-1})$} & 
\colhead{(mag)} & 
\colhead{(\text{erg}\, $\text{s}^{-1})$} & 
\colhead{($M_{\odot}$)}
}
\startdata
ABELL2744-4992 & $8.17^{+0.49}_{-0.00}$ & $8.62^{+0.09}_{-0.00}$ & $44.37^{+0.00}_{-0.27}$ & $0.94^{+0.01}_{-0.30}$ & $0.54^{+0.00}_{-0.36}$ & $4.97^{+0.15}_{-0.00}$ & $-0.88^{+0.01}_{-0.14}$ & $9.08^{+0.00}_{-0.03}$ & $24.99^{+0.08}_{-0.00}$ & $1.71^{+0.00}_{-0.05}$ & $2.91^{+0.01}_{-0.49}$ & $44.94^{+0.00}_{-0.05}$ & $6.37^{+0.00}_{-0.05}$ \\
ABELL2744-8864 & $9.17^{+0.00}_{-0.01}$ & $9.84^{+0.03}_{-0.02}$ & $44.02^{+0.03}_{-0.02}$ & $0.07^{+0.01}_{-0.02}$ & $-0.65^{+0.03}_{-0.02}$ & $5.99^{+0.01}_{-0.01}$ & $-0.57^{+0.05}_{-0.12}$ & $11.3^{+0.04}_{-0.06}$ & $25.99^{+0.01}_{-0.00}$ & $1.94^{+0.02}_{-0.03}$ & $0.25^{+0.5}_{-0.5}$ & $44.00^{+0.03}_{-0.05}$ & $5.43^{+0.03}_{-0.05}$ \\
ABELL2744-9196 & $8.74^{+0.03}_{-0.03}$ & $9.25^{+0.04}_{-0.04}$ & $44.62^{+0.04}_{-0.03}$ & $0.89^{+0.05}_{-0.04}$ & $0.63^{+0.04}_{-0.03}$ & $4.92^{+0.08}_{-0.08}$ & $-0.54^{+0.03}_{-0.03}$ & $9.56^{+0.10}_{-0.07}$ & $25.96^{+0.03}_{-0.06}$ & $1.67^{+0.07}_{-0.09}$ & $1.16^{+0.17}_{-0.13}$ & $44.51^{+0.04}_{-0.04}$ & $5.94^{+0.04}_{-0.04}$ \\
ABELL2744-12048 & $8.00^{+0.00}_{-0.00}$ & $8.03^{+0.01}_{-0.00}$ & $43.87^{+0.00}_{-0.00}$ & $0.87^{+0.00}_{-0.01}$ & $0.12^{+0.00}_{-0.00}$ & $4.99^{+0.00}_{-0.01}$ & $-1.89^{+0.00}_{-0.00}$ & $9.59^{+0.00}_{-0.00}$ & $25.23^{+0.00}_{-0.00}$ & $1.03^{+0.00}_{-0.00}$ & $1.82^{+0.01}_{-0.00}$ & $44.94^{+0.00}_{-0.00}$ & $6.37^{+0.00}_{-0.00}$ \\
ABELL2744-12351 & $8.22^{+0.45}_{-0.06}$ & $7.87^{+0.27}_{-0.20}$ & $43.47^{+0.20}_{-0.16}$ & $0.69^{+0.24}_{-0.20}$ & $-0.38^{+0.20}_{-0.18}$ & $5.19^{+0.16}_{-0.14}$ & $-0.80^{+0.17}_{-0.19}$ & $10.37^{+0.17}_{-0.17}$ & $25.36^{+0.15}_{-0.14}$ & $1.43^{+0.29}_{-0.36}$ & $2.45^{+0.20}_{-0.22}$ & $44.29^{+0.06}_{-0.05}$ & $5.72^{+0.06}_{-0.05}$ \\
ABELL2744-25363 & $8.30^{+0.00}_{-0.00}$ & $7.86^{+0.00}_{-0.00}$ & $43.45^{+0.00}_{-0.00}$ & $0.11^{+0.00}_{-0.00}$ & $-0.36^{+0.00}_{-0.00}$ & $5.87^{+0.00}_{-0.08}$ & $-1.70^{+0.01}_{-0.00}$ & $9.24^{+0.03}_{-0.07}$ & $24.03^{+0.09}_{-0.01}$ & $1.07^{+0.01}_{-0.00}$ & $2.78^{+0.02}_{-0.08}$ & $44.37^{+0.05}_{-0.00}$ & $5.80^{+0.00}_{-0.00}$ \\
CEERS-21058 & $8.22^{+0.09}_{-0.10}$ & $9.15^{+0.12}_{-0.12}$ & $44.02^{+0.12}_{-0.13}$ & $0.79^{+0.17}_{-0.16}$ & $-0.70^{+0.96}_{-6.02}$ & $5.42^{+0.31}_{-0.30}$ & $-1.35^{+0.39}_{-0.44}$ & $11.74^{+0.17}_{-0.26}$ & $25.20^{+0.44}_{-0.35}$ & $1.98^{+0.19}_{-0.19}$ & $2.35^{+0.28}_{-0.39}$ & $44.75^{+0.10}_{-0.10}$ & $6.18^{+0.10}_{-0.10}$ \\
CEERS-29193 & $8.10^{+0.00}_{-0.06}$ & $8.70^{+0.24}_{-0.00}$ & $44.00^{+0.27}_{-0.00}$ & $0.05^{+0.25}_{-0.00}$ & $0.49^{+0.43}_{-0.00}$ & $4.99^{+0.00}_{-0.28}$ & $-1.65^{+0.10}_{-0.02}$ & $9.25^{+1.61}_{-0.00}$ & $21.42^{+1.50}_{-0.00}$ & $1.45^{+0.30}_{-0.00}$ & $0.24^{+0.07}_{-0.01}$ & $45.51^{+0.57}_{-0.00}$ & $6.94^{+0.57}_{-0.00}$ \\
CEERS-31837 & $8.63^{+0.13}_{-0.52}$ & $8.52^{+0.27}_{-0.22}$ & $44.03^{+0.24}_{-0.20}$ & $1.15^{+0.35}_{-0.28}$ & $0.15^{+0.29}_{-0.28}$ & $5.55^{+0.15}_{-0.11}$ & $-1.18^{+0.25}_{-0.17}$ & $10.37^{+0.10}_{-0.11}$ & $25.73^{+0.12}_{-0.13}$ & $1.15^{+0.31}_{-0.12}$ & $1.58^{+0.16}_{-0.14}$ & $44.26^{+0.05}_{-0.05}$ & $5.69^{+0.05}_{-0.05}$ \\
CEERS-47313 & $8.01^{+0.00}_{-0.01}$ & $9.46^{+0.01}_{-0.01}$ & $45.29^{+0.01}_{-0.02}$ & $1.43^{+0.03}_{-0.02}$ & $1.54^{+0.01}_{-0.02}$ & $5.07^{+0.39}_{-0.33}$ & $-1.88^{+0.48}_{-0.31}$ & $9.66^{+0.48}_{-0.52}$ & $24.86^{+0.79}_{-0.74}$ & $1.52^{+0.18}_{-0.25}$ & $6.49^{+0.31}_{-0.49}$ & $46.79^{+0.35}_{-0.32}$ & $8.22^{+0.35}_{-0.32}$ \\
CEERS-47680 & $8.01^{+0.01}_{-0.01}$ & $9.85^{+0.05}_{-0.02}$ & $45.69^{+0.05}_{-0.02}$ & $2.12^{+0.07}_{-0.01}$ & $1.93^{+0.04}_{-0.01}$ & $5.00^{+0.20}_{-0.05}$ & $-0.78^{+0.08}_{-0.17}$ & $9.79^{+0.06}_{-0.23}$ & $25.22^{+0.09}_{-0.22}$ & $1.17^{+0.31}_{-0.09}$ & $1.48^{+0.22}_{-0.25}$ & $44.88^{+0.05}_{-0.10}$ & $6.31^{+0.05}_{-0.10}$ \\
CEERS-55630 & $8.33^{+0.06}_{-0.01}$ & $9.78^{+0.00}_{-0.03}$ & $45.35^{+0.00}_{-0.08}$ & $1.57^{+0.00}_{-0.07}$ & $1.54^{+0.00}_{-0.08}$ & $5.24^{+0.00}_{-0.08}$ & $-1.08^{+0.00}_{-0.05}$ & $9.01^{+0.21}_{-0.00}$ & $25.05^{+0.01}_{-0.19}$ & $1.73^{+0.00}_{-0.09}$ & $1.74^{+0.00}_{-0.17}$ & $45.34^{+0.01}_{-0.00}$ & $6.77^{+0.01}_{-0.00}$ \\
CEERS-64408 & $8.03^{+0.03}_{-0.02}$ & $8.69^{+0.11}_{-0.07}$ & $44.53^{+0.08}_{-0.06}$ & $1.32^{+0.14}_{-0.10}$ & $0.75^{+0.10}_{-0.07}$ & $5.61^{+0.23}_{-0.29}$ & $-0.75^{+0.15}_{-0.17}$ & $9.61^{+0.18}_{-0.35}$ & $22.17^{+0.23}_{-0.24}$ & $1.07^{+0.28}_{-0.05}$ & $3.98^{+0.23}_{-0.25}$ & $45.35^{+0.27}_{-0.18}$ & $6.78^{+0.27}_{-0.18}$ \\
CEERS-72934 & $8.01^{+0.01}_{-0.00}$ & $9.32^{+0.03}_{-0.04}$ & $44.42^{+0.02}_{-0.04}$ & $0.42^{+0.02}_{-0.05}$ & $1.26^{+0.08}_{-0.08}$ & $5.15^{+0.09}_{-0.10}$ & $-1.46^{+0.00}_{-0.16}$ & $9.18^{+0.30}_{-0.03}$ & $25.95^{+0.01}_{-0.00}$ & $1.54^{+0.18}_{-0.00}$ & $2.96^{+0.02}_{-0.31}$ & $45.51^{+0.00}_{-0.12}$ & $6.94^{+0.00}_{-0.12}$ \\
JADES-GN-10395 & $8.01^{+0.01}_{-0.00}$ & $8.32^{+0.04}_{-0.03}$ & $44.15^{+0.03}_{-0.04}$ & $0.51^{+0.04}_{-0.03}$ & $0.39^{+0.04}_{-0.03}$ & $5.44^{+0.07}_{-0.07}$ & $-2.06^{+0.05}_{-0.07}$ & $9.08^{+0.09}_{-0.05}$ & $25.97^{+0.02}_{-0.03}$ & $1.10^{+0.05}_{-0.04}$ & $0.82^{+0.10}_{-0.09}$ & $44.08^{+0.02}_{-0.01}$ & $5.51^{+0.02}_{-0.01}$ \\
JADES-GN-41323 & $8.05^{+0.00}_{-0.01}$ & $8.95^{+0.06}_{-0.00}$ & $44.76^{+0.05}_{-0.00}$ & $0.69^{+0.05}_{-0.01}$ & $0.99^{+0.06}_{-0.30}$ & $5.98^{+0.06}_{-0.06}$ & $-0.81^{+0.01}_{-0.00}$ & $9.74^{+0.07}_{-0.00}$ & $23.52^{+0.03}_{-0.00}$ & $1.00^{+0.06}_{-0.05}$ & $4.49^{+0.31}_{-0.00}$ & $46.11^{+0.14}_{-0.00}$ & $7.54^{+0.14}_{-0.00}$ \\
JADES-GS-6608 & $8.03^{+0.03}_{-0.02}$ & $8.59^{+0.04}_{-0.04}$ & $44.05^{+0.08}_{-0.14}$ & $0.19^{+0.08}_{-0.12}$ & $0.60^{+0.07}_{-0.09}$ & $4.41^{+0.33}_{-0.24}$ & $-1.17^{+0.45}_{-0.63}$ & $11.65^{+0.25}_{-0.40}$ & $23.56^{+0.63}_{-0.80}$ & $2.23^{+0.56}_{-0.54}$ & $3.65^{+0.56}_{-0.74}$ & $45.01^{+0.19}_{-0.17}$ & $6.44^{+0.19}_{-0.17}$ \\
PRIMER-COSMOS-1223 & $8.02^{+0.01}_{-0.00}$ & $7.97^{+0.05}_{-0.02}$ & $43.57^{+0.01}_{-0.01}$ & $0.04^{+0.09}_{-0.00}$ & $0.04^{+0.03}_{-0.05}$ & $4.38^{+0.11}_{-0.09}$ & $-2.37^{+0.07}_{-0.02}$ & $9.41^{+0.10}_{-0.36}$ & $22.71^{+0.02}_{-0.28}$ & $1.17^{+0.18}_{-0.13}$ & $0.03^{+0.00}_{-0.03}$ & $46.17^{+0.27}_{-0.12}$ & $7.60^{+0.27}_{-0.12}$ \\
PRIMER-COSMOS-53040 & $8.26^{+0.18}_{-0.17}$ & $7.85^{+0.16}_{-0.14}$ & $43.45^{+0.11}_{-0.07}$ & $0.09^{+0.11}_{-0.06}$ & $-0.38^{+0.16}_{-0.08}$ & $4.32^{+0.35}_{-0.19}$ & $-1.16^{+0.43}_{-0.58}$ & $10.58^{+0.90}_{-1.14}$ & $22.15^{+0.71}_{-0.74}$ & $1.64^{+0.55}_{-0.41}$ & $1.28^{+0.54}_{-0.48}$ & $44.50^{+0.20}_{-0.15}$ & $5.93^{+0.20}_{-0.15}$ \\
PRIMER-COSMOS-53670 & $8.10^{+0.20}_{-0.08}$ & $9.46^{+0.09}_{-0.09}$ & $44.43^{+0.15}_{-0.17}$ & $0.75^{+0.17}_{-0.25}$ & $0.78^{+0.56}_{-0.21}$ & $5.15^{+0.28}_{-0.28}$ & $-0.79^{+0.21}_{-0.41}$ & $11.64^{+0.23}_{-0.34}$ & $25.81^{+0.14}_{-0.31}$ & $2.13^{+0.24}_{-0.32}$ & $1.86^{+0.26}_{-0.24}$ & $45.27^{+0.12}_{-0.09}$ & $6.70^{+0.12}_{-0.09}$ \\
PRIMER-UDS-47866 & $8.03^{+0.01}_{-0.02}$ & $9.19^{+0.03}_{-0.03}$ & $45.00^{+0.03}_{-0.03}$ & $1.21^{+0.04}_{-0.05}$ & $1.25^{+0.03}_{-0.03}$ & $5.75^{+0.16}_{-0.16}$ & $-1.32^{+0.12}_{-0.15}$ & $9.59^{+0.13}_{-0.26}$ & $23.98^{+0.11}_{-0.09}$ & $1.11^{+0.34}_{-0.08}$ & $5.13^{+0.34}_{-0.35}$ & $45.83^{+0.13}_{-0.12}$ & $7.26^{+0.13}_{-0.12}$ \\
PRIMER-UDS-129331 & $8.27^{+0.29}_{-0.15}$ & $8.20^{+0.22}_{-0.19}$ & $43.71^{+0.13}_{-0.16}$ & $0.40^{+0.17}_{-0.18}$ & $-0.04^{+0.16}_{-0.16}$ & $5.12^{+0.43}_{-0.42}$ & $-1.19^{+0.41}_{-0.52}$ & $10.91^{+0.56}_{-0.51}$ & $24.50^{+0.41}_{-0.43}$ & $1.93^{+0.24}_{-0.35}$ & $1.57^{+0.37}_{-0.39}$ & $44.50^{+0.08}_{-0.09}$ & $5.93^{+0.08}_{-0.09}$ \\
\enddata
\tablecomments{Values are medians with 16th and 84th percentile uncertainties for LRD sample. Note that $\log(M_{\rm BH})$ is computed from Equation (1) in Section 4.2. $(L_\star$ and $L_{\rm BH^\star}$ denote the bolometric luminosities contributed by the stellar and BH* components, respectively.}
\end{deluxetable*}

\subsection{Stellar Component
\label{subsec:Stellar Component}}

For the stellar component of our composite model, we adopt the \cite{BC2003} simple stellar population (SSP). The stellar emission is further complemented by nebular continuum and line emission. We assume a \citet{Chabrier_2003} initial mass function (IMF), which provides a more realistic description of the low-mass stellar population compared to single-power-law alternatives. We adopt an exponentially declining star formation history (SFH) where the parameter $\tau$ characterizes the e-folding time and is allowed to vary between $10^{6}$ and $10^{10}$ yr. The age of the stellar population ranges from $10^{8}$ to $1.35\times10^{10}$ yr, spanning from young, actively star-forming systems to mature galaxies. The stellar metallicity $Z$ is constrained to the range of $10^{-2.3}$ to $10^{0.7} Z_\odot$. We adopt the \citet{Calzetti_2000} dust attenuation law due to its wide applicability in studies of high-redshift galaxies \citep{2023ApJ...949L..18P,2024AJ....168..113C, perezgonzalez2024naturelittlereddots}. The visual extinction $A_V$ varies from 0.0 to 7.0 mag. The velocity dispersion of emission lines $v_{\rm d}^{\rm emi}$ is fixed between 100 and 500 km s$^{-1}$. This range captures the typical kinematic broadening of nebular emission lines in high-redshift galaxies, reflecting both gravitational potential and turbulent motions. The spectral resolution of the stellar template is set to R=2000.

\subsection{BH* Component
\label{subsec:Black Hole Star Component}}

Building directly on the \texttt{Cloudy}\footnote{\url{https://nublado.org/}} \citep{2017RMxAA..53..385F} modeling strategy described in \citet{naidu2025blackholestarreveals}, we construct a one-dimensional, spherical grid in \texttt{Cloudy} that embeds a multi-power-law ``big-bump'' AGN continuum inside a dense, turbulent, metal-poor shell. The blackbody temperature of the incident spectrum is allowed to vary between $10^{4}$ and $10^{6}$ {K}, enabling the peak of the accretion-disk emission can shift from the extreme-UV to the soft X-ray range. This covers both sub-Eddington and super-Eddington regimes expected for $10^{6}$--$10^{8}\ M_{\odot}$ black holes at $z\sim8$. The optical-to-X-ray index $\alpha_{\mathrm{ox}}$ is stepped from $-3.0$ to $-1.0$, bracketing the X-ray-weak population common at high redshift while still reaching the canonical value for luminous quasars. The UV slope $\alpha_{\mathrm{UV}}$ and the X-ray slope $\alpha_{\mathrm{x}}$ are both sampled from $-2.5$ to $-0.1$, providing the model with sufficient flexibility to match either an intrinsically blue continuum or the extremely red $f_{\nu} \propto \lambda^{-0.8}$ shape observed in sources like MoM-BH*-1 once hydrogen opacity is included.

The gas shell is positioned at $\sim 40\ \mathrm{AU}$ from the black hole, with the gas density $n_{\mathrm{H}}$ and cloud column density $N_{\mathrm{H}}$ scanned over the ranges of $10^{9}-10^{12} \mathrm{cm}^{-3}$ and  $10^{21}-10^{26} \mathrm{cm}^{-2}$, respectively. The lower density bound probes normal Seyfert-like environments, while the upper bound reaches the extreme conditions ($10^{12}\ \mathrm{cm}^{-3}$ and $10^{26}\ \mathrm{cm}^{-2}$) required to populate the $n = 2$ level of hydrogen and produce a Balmer jump exceeding a factor of 20, as seen in the data. A turbulent velocity between 10 and 1000 km s$^{-1}$ is included. The observed 500 km s$^{-1}$ width of the central H$\beta$ absorption trough pins the preferred value near $\log v_{\mathrm{turb}} \approx 2.7$. The irradiation of the cloud is modulated by an ionization parameter $\log U$, which is stepped from $-2.5$ to $-0.5$ to ensure that the predicted H$\beta$ equivalent width falls within the $30$--$45\ \mathrm{\AA}$ range demanded by the NIRSpec spectra. The gas metallicity is varied from $10^{-2} Z_{\odot}$ to $Z_{\odot}$. The low end tests primordial compositions expected in a $z = 7.8$ dwarf, while the high end checks whether metal-line cooling modifies the continuum; we find that it has a negligible impact at these wavelengths. Following each \texttt{Cloudy} run, an external SMC-type dust screen is applied with $0\leq A_{V}\leq 7\ \mathrm{mag}$. We also include a fractional flux-calibration systematic error $f_{\rm sys}$, added in quadrature to the statistical uncertainties, ranging from $0.01$ to $0.1$.

\section{RESULTS} \label{sec:results}

\subsection{Fitting Result} 
\label{subsec:Fitting Result}

\setkeys{Gin}{draft=false}

\begin{figure*}
        \centering
    \includegraphics[width=0.8\linewidth]{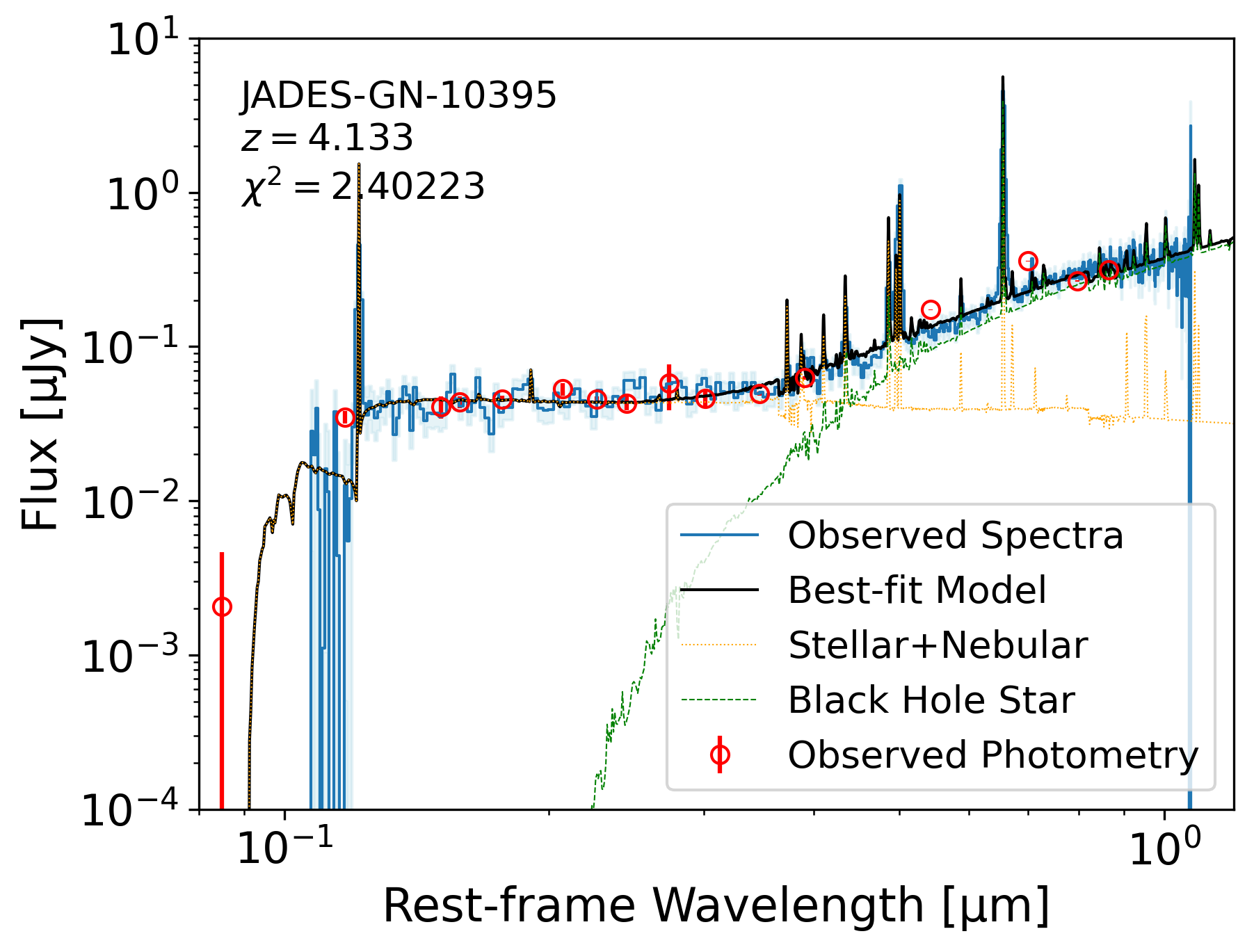}
    \caption{Best-fit SED for a representative LRD candidate, JADES-GN-10395 ($z = 4.133$). Red points denote the observed photometry, while the blue solid line and shaded region represent the observed spectrum and its $1\sigma$ uncertainty. The green and yellow lines represent the modeled emission from BH* and stellar components, respectively. The black solid line indicates the best-fit composite model.}
    \label{fig:example}
\end{figure*}

In this section, we present the comprehensive SED fitting results for our full LRD sample. Figure~\ref{fig:example} illustrates the fitting outcome for a representative LRD candidate, JADES-GN-10395 at $z=4.133$. The spectroscopic data (blue solid line with shaded uncertainties) are well reproduced by our best-fit model (black solid line), yielding excellent agreement across the entire rest-frame wavelength range of $\sim 0.1$--$1.0\,\mu\text{m}$. This decomposition reveals distinct emission contributions: the Stellar+Nebular component (yellow line) dominates the UV continuum blueward of the Balmer break, while the BH* component (green line) primarily accounts for the optical emission redward of $\lambda_{\mathrm{rest}}> 0.36\,\mu\text{m}$. This bimodal decomposition successfully captures the characteristic spectral break and the rising continuum in the near-infrared, with the total model accurately tracing both the prominent emission line features and the underlying continuum shape.

\begin{figure*}[htbp]
    \centering
    \begin{minipage}[t]{0.49\linewidth}  
        \centering
        \includegraphics[width=\linewidth]{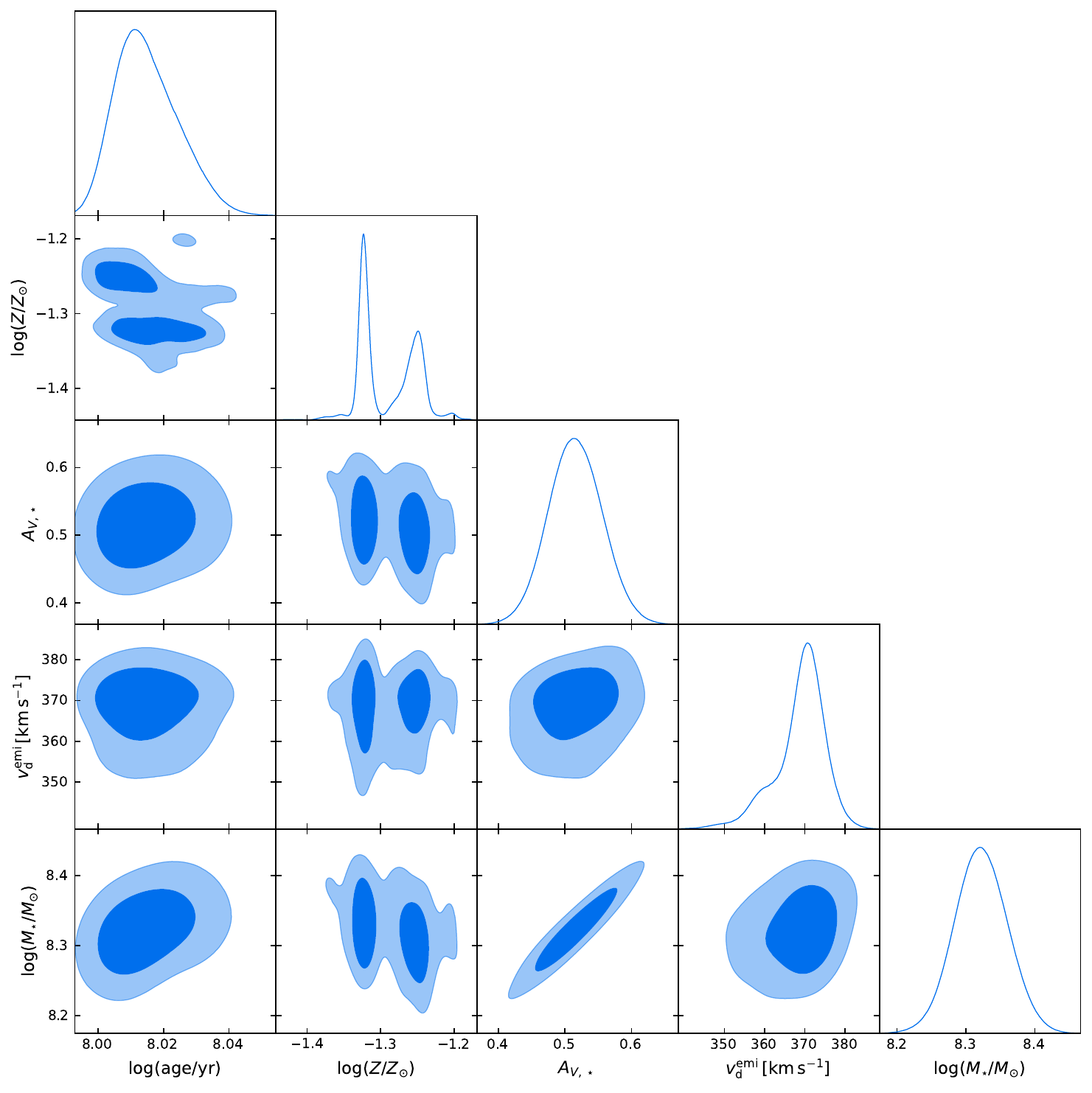}
        \label{fig:left}
    \end{minipage}
    \hfill 
    \begin{minipage}[t]{0.49\linewidth}
        \centering
        \includegraphics[width=\linewidth]{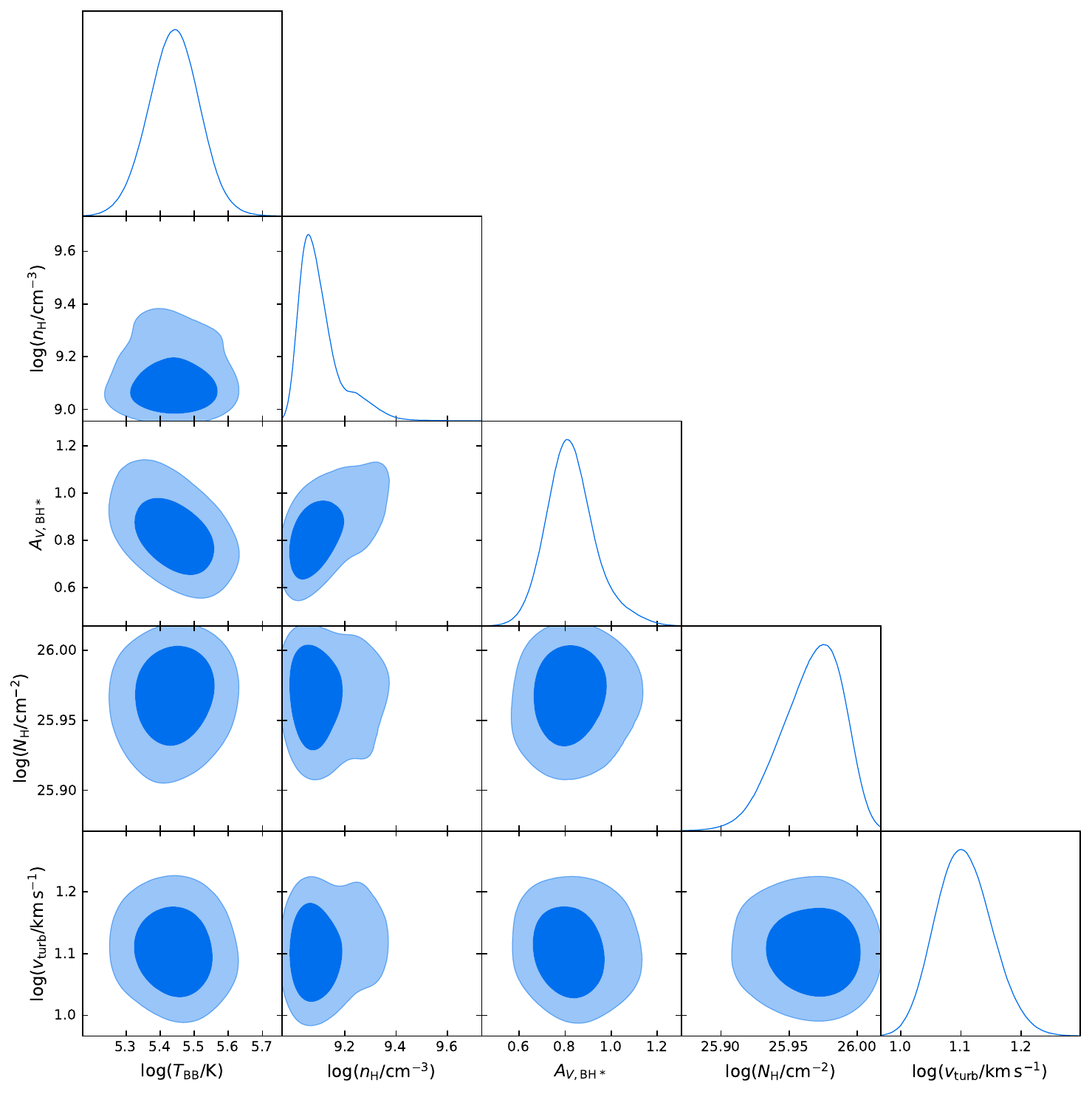}
        \label{fig:right}
    \end{minipage}
    \caption{Example posterior probability distributions derived from the SED fitting. Left panels: Posterior distributions for the stellar component parameters. The top panels in each column displays the 1D marginalized PDF for the corresponding parameters, while the lower panels show the 2D marginalized PDFs for each parameter pair. Right panels: Corresponding posterior distributions for the BH* component parameters.}
    \label{fig:both}
\end{figure*}

The robustness of our fitting procedure is further demonstrated by the posterior probability distributions shown in Figure~\ref{fig:both}. The corner plots display well-constrained 1D marginalized probability density functions (PDFs) and 2D confidence contours for key physical parameters of both the stellar (left panels) and BH* (right panels) components. The unimodal, symmetric distributions in the diagonal panels indicate well-behaved MCMC convergence. The off-diagonal contours show that the envelope parameters ($T_{\rm BB}$, $n_{\rm H}$, $N_{\rm H}$, and $A_{V,{\rm BH*}}$) are mutually covariant, as expected physically: combinations of cooler $T_{\rm BB}$, higher $n_{\rm H}$ and $A_{V,{\rm BH*}}$, and lower $N_{\rm H}$ can produce similar red-optical continua. Whether such covariance prevents accurate parameter recovery cannot be judged from a single posterior, we therefore verify it directly with mock recovery tests in Section \ref{sec:recovery}.

The completeness and consistency of our fitting approach are systematically validated across the entire sample of 22 LRDs, as presented in Appendix~\ref{appendix:full sample} (Figures~\ref{fig:12} and \ref{fig:24}). These figures demonstrate that our composite model successfully reproduces the observed SEDs across a broad redshift range ($4.0 \lesssim z \lesssim 9.5$) and diverse spectral morphologies. The close match between the observed spectra (blue) and best-fit models (black) across all 22 sources confirms that our two-component model provides an adequate and physically motivated description of the LRD population. Our analysis reveals distinct wavelength-dependent contributions from the two physical components, consistent with recent findings in the literature \citep{degraaff2025littlereddotshost,sun2026littlereddot,asada2026originsuvcontinuumbalmer}. For the most of our LRDs, the UV continuum blueward of the Balmer break ($\lambda_{\rm rest} \lesssim 0.36\,\mu\text{m}$) is predominantly stellar in origin, reflecting active star formation in these compact systems, whereas the black hole accretion disk component dominates the optical portion of the spectrum, producing the characteristic red continuum slope that defines the LRD phenotype. PRIMER-COSMOS-1223 is the only source in our sample where the BH* component is non-negligible in the rest-UV, foreshadowing the heterogeneity of UV origins discussed in Section \ref{subsec:The origin of the rest-UV emission}.

The median posterior values for all fitted parameters, summarized in Table~\ref{tab:Posterior}, reveal a coherent physical picture of compact star-forming galaxies hosting actively accreting central black holes. The consistency of the fitting residuals---showing no systematic deviations as a function of wavelength or redshift, further validates that our model adequately captures the essential astrophysical processes governing LRD spectral properties. Collectively, these results demonstrate that our SED fitting procedure is statistically robust and physically sound, providing reliable measurements of stellar masses, black hole masses, and dust properties for the full LRD sample.

\subsection{Parameter Recovery Test}
\label{sec:recovery}

The reliability of our parameter inference depends on whether the composite host $+$ BH* model can recover accurate parameter estimates from data with the S/N of our sample. This is not trivial: as discussed in Section \ref{subsec:Fitting Result}, the envelope parameters (the blackbody temperature $T_{\rm BB}$, the gas density $n_{\rm H}$, the column density $N_{\rm H}$, and the dust attenuation $A_{V,{\rm BH*}}$) are mutually covariant, since combinations of cooler $T_{\rm BB}$, higher $n_{\rm H}$, higher $A_{V,{\rm BH*}}$, and lower $N_{\rm H}$ can produce similar red-optical continua. To verify the recovery and to calibrate the uncertainties quoted in Table~\ref{tab:Posterior}, we perform a mock recovery test in which synthetic spectra with known input parameters are generated and refitted with the same pipeline.

For each of the 22 sources in our sample, mock spectra are constructed
in three steps. (i) The noiseless spectrum is the best-fit model
spectrum of the source produced by BayeSED3 itself, already convolved
to the instrumental resolution, attenuated, and resampled onto the
observed wavelength grid. Because the mock and the fitter share the
identical forward model, the test isolates the statistical recovery of
the parameters, free of any model mismatch. (ii) Gaussian noise is added with the standard deviation set by the actual error spectrum of the source's NIRSpec/PRISM observation, median-filtered with a 51-pixel kernel to remove pixel-level spikes while preserving its full wavelength dependence. The mock spectra therefore reproduce the S/N distribution of the real data. (iii) This procedure is repeated with 20 independent random realizations per source.

Each mock spectrum is fitted with the identical priors and
configuration as the real data. For every parameter we compute the bias $b=\langle\hat{\theta}\rangle-\theta_{\rm true}$ across the 20 realizations. Table~\ref{tab:recovery} summarizes the recovery error of the eight key parameters across the 22 sources. The recovery error is defined as the offset between the mean of the posterior medians recovered from the 20 mock noise realizations and the maximum-likelihood (MAL) estimate of the real-data fit, namely the precise point in parameter space at which the mock spectrum is generated. It therefore measures how far an individual fit can be displaced from the true solution by the observational noise. Three points emerge. First, the marginal recovery is accurate. The median errors are $0.131$~dex in $\log(M_\star/M_\odot)$, $0.070$~mag in $A_{V,\star}$, $0.238$~dex in $\log(T_{\rm BB})$, $0.165$~dex in $\log(n_{\rm H})$, and $0.203$~mag in $A_{V,{\rm BH*}}$, with the continuum slopes $\alpha_{\rm ox}$, $\alpha_{\rm uv}$, and $\alpha_{\rm x}$ recovered to $0.17$--$0.29$~dex. No parameter has a median recovery error exceeding $0.29$. These values are far smaller than the adopted prior ranges, so the constraints are driven by the data rather than the priors. Second, the covariance noted in Section \ref{subsec:Fitting Result} does not prevent accurate recovery. The typical displacement of $T_{\rm BB}$ is only $\sim0.24$~dex, an order of magnitude smaller than its prior range of two dex. The tight posterior on $T_{\rm BB}$ in Figure~\ref{fig:both} is therefore genuine, not an artifact of the prior. Third, a small number of sources show large errors in individual parameters ($A_{V,{\rm BH*}}$ up to $3.01$~mag; $\log n_{\rm H}$ up to $1.56$~dex), preferentially the lowest-S/N spectra. Our main conclusions are unaffected: the median recovery errors are well below the one-to-two-order-of-magnitude offsets that define our results.

\begin{table}
\centering
\caption{Recovery error of the key model parameters across the 22
sources of our sample.\label{tab:recovery}}
\begin{tabular}{lcccc}
\hline\hline
Parameter & Unit & Max & Min & Median \\
\hline
$\log(M_\star/M_\odot)$         & dex & 0.568 & 0.007 & 0.131 \\
$A_{V,\star}$                   & mag & 0.460 & 0.007 & 0.070 \\
$\log(T_{\rm BB}/{\rm K})$      & dex & 1.000 & 0.004 & 0.238 \\
$\alpha_{\rm ox}$               & dex & 1.137 & 0.020 & 0.166 \\
$\alpha_{\rm uv}$               & dex & 0.801 & 0.019 & 0.179 \\
$\alpha_{\rm x}$                & dex & 0.780 & 0.027 & 0.285 \\
$\log(n_{\rm H}/{\rm cm^{-3}})$ & dex & 1.559 & 0.014 & 0.165 \\
$A_{V,\rm BH^*}$                & mag & 3.013 & 0.006 & 0.203 \\
\hline
\end{tabular}
\begin{minipage}{0.85\linewidth}
\small \vspace{2pt}
Note: The recovery error is the offset between the mean of the posterior
medians recovered from $N=20$ mock noise realizations of a given source
and the maximum-likelihood (MAL) estimate of the real-data fit. The three columns give the maximum, minimum, and median of this error across the 22
sources, each at its native S/N.
\end{minipage}
\end{table}

We note that the intrinsic covariance between the envelope parameters, which arises because combinations of cooler $T_{\rm BB}$, higher $n_{\rm H}$, higher $A_{V,\rm BH^*}$ and lower $N_{\rm H}$ producing similar red-optical continua, is fully expressed in the posterior, rather than being averaged over or ignored. The uncertainties quoted in Table \ref{tab:Posterior} are the
16th--84th percentiles of the marginalized one-dimensional posteriors, namely the parameter ranges obtained after integrating over the covariant
directions. They are therefore conservative by construction. The
degeneracy inflates these intervals but does not displace their centers:
as the parameter recovery test demonstrates, the ensemble of recovered
posterior medians remains centered on the input values even though
individual realizations scatter along the degenerate direction. The
nested-sampling exploration of the covariant parameter space is
efficient by construction, so the quoted widths are not artifacts of
under-sampling.

\subsection{Black hole mass estimation} 
\label{subsec:Black hole mass estimation}

In Table~\ref{tab:Posterior}, we present the bolometric luminosity of the BH* component, $\log(L_{\rm BH*})$, for each LRD by integrating the BH* model spectrum over the wavelength range $0.01$--$100\,\mu$m, which encompasses the bulk of the AGN radiative output from the far-UV to the far-infrared. This bolometric luminosity subsequently enables black hole mass estimation. The physical connection between bolometric luminosity and black hole mass is expressed through the following relation:
\begin{equation}
\begin{split}
M_{\rm BH}  
\simeq 0.81 \times 10^5\ M_\odot \lambda_{\rm Edd}^{-1} \left(\frac{L_{\rm BH*}}{10^{43}\ {\rm erg\ s}^{-1}}\right)
\end{split}
\end{equation}
Here, $\lambda_{\rm Edd}$ represents the Eddington ratio.

In the context of the BH* model, \citet{naidu2025blackholestarreveals} adopted an Eddington ratio of $\lambda_{\rm Edd} \approx 5-10$. Cosmological radiation-hydrodynamic simulations presented by \citet{chon2026rapidemergenceovermassiveblack} suggest that LRDs can be identified with heavy black hole seeds undergoing a brief phase of super-Eddington accretion at $\lambda_{\rm Edd} \approx 3-30$, with a characteristic value around $\lambda_{\rm Edd} \approx 10$. Conversely, several studies estimate the Eddington ratio for LRDs to be $\lesssim 1$ \citep{Umeda_2026}. Following the analysis of \citet{marszewski2026littlereddotsfire}, we adopt a fiducial Eddington ratio of $\lambda_{\rm Edd} = 3$ for our LRDs. 
In this scenario, LRDs are characterized as super-Eddington accretors with black hole masses $M_{\rm BH} \gtrsim 2\times 10^5 M_\odot$ hosted in galaxies with stellar masses $M_\star \gtrsim 2\times 10^7 M_\odot$. 
While the exact value of $\lambda_{\rm Edd}$ remains uncertain, a value of 3 represents a physically plausible super-Eddington accretion rate that is consistent with GRMHD simulations \citep{Jiang_2014,Zhang_2025}. These simulations demonstrate that black holes can sustain accretion rates several to tens of times the Eddington limit while maintaining luminosities near $L_{\rm Edd}$. 

\begin{figure}
    \centering
    \includegraphics[width=0.9\linewidth]{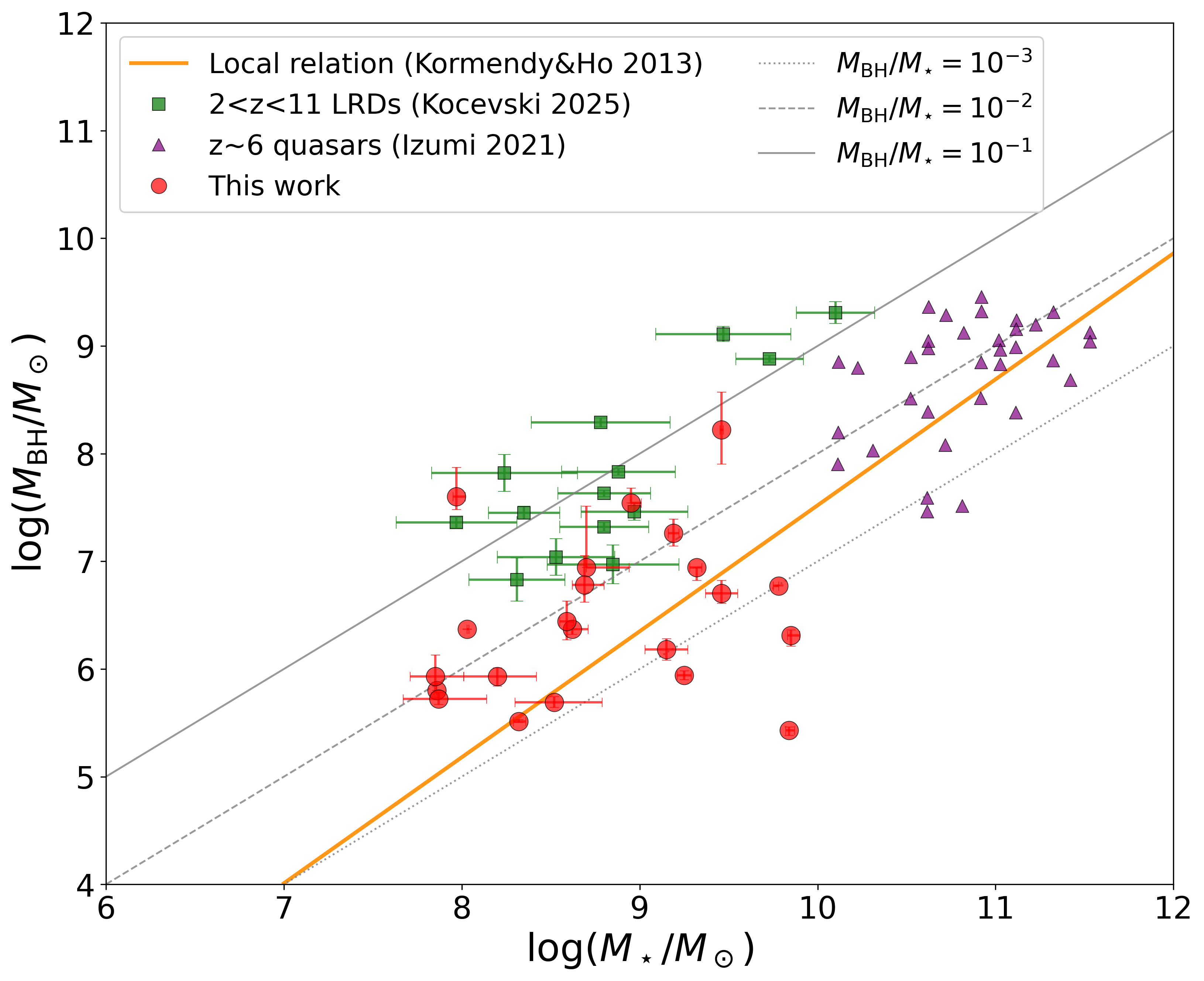}
    \caption{Relation between the stellar mass ($M_{\star}$) and the black hole mass $M_{\rm BH}$ for our LRD sample (red circles) within the BH* framework. For comparison,the local $M_{\rm BH}-M_{\star}$ relation (orange line; \citealt{2013ARA&A..51..511K}) and measurements of $z \sim 6$ (purple triangles; \citealt{Izumi_2021}) are overplotted. The LRD population exhibits moderately overmassive black holes relative to the local relation. Black hole masses inferred from the H$\alpha$-based single-epoch method \citep{kocevski2025risefaintredagn} are also plotted (green squares). The dotted, dashed and solid gray lines indicate $M_{\rm BH}/M_{\star} = 10^{-3},\ 10^{-2},\ \text{and}\ 10^{-1}$, respectively.}
    \label{fig:bhmass}
\end{figure}

Leveraging our black hole mass measurements, we re-examine the stellar mass and black hole mass relation based on our SED fitting results. Figure \ref{fig:bhmass} presents the $M_{\rm BH}-M_{\star}$ relation for our LRD sample (red circles) derived from the BH* model. For comparison, we also plot the local $M_{\rm BH}-M_{\star}$ relation from \cite{2013ARA&A..51..511K} (orange line) and the measurements of $z \sim 6$ quasars \citep{Izumi_2021} (purple triangles). Our results indicate that LRDs generally exhibit systematically higher black hole masses relative to their stellar masses compared to the local relation and $z \sim 6$ quasars, though the deviation remains moderate. Comparing our findings with LRD $M_{\rm BH}/M_{\star}$ measurements from \cite{kocevski2025risefaintredagn}—obtained through empirical correlations between $M_{\rm BH}$ and the broad emission lines, we observe that $M_{\rm BH}/M_{\star}$ measurements based on BH* model are approximately one to two orders of magnitude lower than those inferred from broad-line relations at fixed stellar mass, acknowledging that local scaling relations may overestimate the black hole mass by one to two orders of magnitude due to resonant scattering in the dense gas envelope.

\section{DISCUSSION} 
\label{sec:DISCUSSION}

\subsection{LRDs as BH*s in faint host galaxies} 
\label{subsec:LRDs as BH*s in faint host galaxies}

Assuming LRDs follow the host galaxy+BH* paradigm, our SED modeling results summarized in Table \ref{tab:Posterior} reveal a system architecture consisting of four key components: (i) A central SMBH with $M_{\rm BH}\approx10^{5.4-8.2}\,M_\odot$ (median $M_{\rm BH}=10^{6.4}\,M_\odot$) that provides the primary energy source through super-Eddington accretion, yielding a bolometric luminosity $L_{\rm BH*}\approx10^{44.0-46.8}\,\rm erg\,s^{-1}$ (median $L_{\rm BH*}=10^{44.9}\,\rm erg\,s^{-1}$). The AGN continuum is characterized by a big bump temperature $T_{\rm BB}\approx10^{4.3-6.0}\,\rm K$ (median $T_{\rm BB}=10^{5.2}\,\rm K$) and a high ionization parameter $\log(U)\approx-2.37$ to $-0.54$ (median $\log(U)=-1.18$), indicating an intense ionizing radiation field that powers the broad emission lines. (ii) A partially ionized, dense inner region with an electron density $n_{\rm H} \approx 10^{9.0-11.7}~{\rm cm}^{-3}$ (median $n_{\rm H}=10^{9.6}~{\rm cm}^{-3}$) and column density $N_{\rm H} \approx 10^{21.4-26.0}~{\rm cm}^{-2}$ (median $N_{\rm H}=10^{25.0}~{\rm cm}^{-2}$) embedded within the gaseous envelope. This region is optically thick to Balmer transitions, exhibiting collisional excitation and resonant scattering, while turbulent velocities of $v_{\rm turb} \approx 10^{1.0-2.2}~\text{km}\, ~\text{s}^{-1}$ (median $v_{\rm turb}=10^{1.5}~\text{km}\, ~\text{s}^{-1}$) broaden the line profiles via electron scattering. (iii) An optically thick, convective gaseous envelope with visual extinction $A_{V,\rm BH*}\approx0-6.5$ mag (median $A_{V,\rm BH*}=1.8$ mag). The envelope thermalizes the accretion disk radiation into a near-blackbody continuum, producing the characteristic red optical continuum and Balmer break through a large population of $n=2$ hydrogen atoms. (iv) A faint, low-mass, star-forming host galaxy with stellar mass $M_* \approx 10^{7.8-9.9}M_\odot$ (median $M_*=10^{8.7}M_\odot$), a young age $t_* \approx 10^{8.0-9.2}~{\rm yr}$ (median $t_*=10^{8.1}~{\rm yr}$), a 100-Myr averaged star formation rate $\mathrm{SFR} \approx 10^{-0.7-1.9}~M_\odot\,~\mathrm{yr}^{-1}$ (median $\mathrm{SFR}=10^{0.62}~M_\odot\,~\mathrm{yr}^{-1}$), and modest extinction $A_{\rm V,*} \approx 0$--$2.1$~mag (median $A_{\rm V,*}=0.72$~mag). The host contributes to UV continuum and narrow [O~III] emission but remains subdominant at optical wavelengths due to the overwhelming AGN luminosity, with the host-to-AGN flux ratio varying across the population.

The decoupled extinction components---gas-dominated $A_{\rm V,BH*}$ for the envelope versus modest $A_{\rm V,*}$ for the host, along with the high-density, high-column parameters of the inner region, provide quantitative evidence that LRDs are not dust-reddened AGNs. Instead, they represent a distinct population of BH* with unique spectral signatures arising from their dense, hydrostatic gaseous envelopes. The characteristic V-shaped SED arises from the combination of a red optical continuum reprocessed by the optically thick photosphere and a blue UV continuum dominated by host galaxy starlight or photoionized nebular emission. 

\subsection{The origin of the rest-UV emission: possible multi-component contributions}
\label{subsec:The origin of the rest-UV emission}

In our decomposition, the rest-UV continuum is attributed to the stellar component in all but one source (PRIMER-COSMOS-1223), consistent with evidence that the narrow emission lines and the UV continuum originate from the dusty host galaxy. The narrow-line Balmer decrement correlates with both $\beta_{\rm UV}$ and the break strength \citep{2026arXiv260913444D}, and dust-attenuated stellar templates reproduce the UV continuum for most sources ($\beta_{\rm UV}\lesssim-1.5$), while the optical continuum is unaffected by the same dust screen \citep{2026arXiv260913444D}, a geometry naturally produced by our dusty host plus dust-poor envelope picture. Our modest $A_{V,\star}$ (median $0.72$~mag) and the main-sequence position of the hosts further support this interpretation.

However, a compact component may contribute significantly to the rest-UV of at least a subset of LRDs. Stacking 181 LRDs, \citet{2026arXiv260324700C} found a point-source fraction rising from $\sim25\%$ at $1500$~\AA\ to $\sim50\%$ at $3000$~\AA\ , with $\sim45\%$ of sources compact across the full UV--optical range. \citet{2026arXiv260603522A} showed that LRD UV continua are redder and $\sim4\times$ more compact than those of matched star-forming galaxies, with break strengths correlating with UV slope and size. \citet{2026arXiv260403370J} detected broad Ly$\alpha$ in the far-UV of one source, requiring holes in the envelope, although \citet{2026arXiv260814534K} constrain such broad emission to $\sim10\%$ in stacks. Notably, \citet{2026arXiv260913444D} found that stellar templates alone fail for the reddest UV slopes ($\beta_{\rm UV}\gtrsim-1.5$), and we caution that a point-like component need not be AGN-powered: compact circumnuclear star formation or dense nebular continuum would be morphologically indistinguishable.

Physically, \citet{2026arXiv260317667M} showed that the absorber velocity correlates with the UV--optical color, implying that the envelope column density, and hence the covering factor, varies systematically across the population. This correlation is primarily driven by the red end of the color ratio and is therefore not in tension with a stellar-dominated UV. But sightlines through low-column-density gas would leak central UV photons. Our model does not include this effect. We adopt a static, spherical CLOUDY grid with full covering. The turbulent velocity $v_{\rm turb}$ reproduces the absorption width and the grid contains no bulk flows.

The impact of such leakage on our results is bounded and predominantly one-sided. Because our model has no third blue component, any non-stellar UV flux is by construction absorbed into the stellar component, biasing $M_\star$ and SFR high by $\lesssim+0.13$--$0.30$~dex for a contamination fraction of 25--50\%. In contrast, $M_{\rm BH}$ is robust, as $\log(L_{\rm BH*})$ is integrated over $0.01$--$100~\mu$m with the optical BH*-dominated in every fit. $M_{\rm BH}/M_\star$ would shift upward by $\lesssim0.3$~dex, making the hosts even fainter relative to their black holes, without affecting our conclusion that these ratios lie one to two orders of magnitude below broad-line estimates. The SFR--$A_{V,\star}$ correlation is qualitatively unchanged.

Our own sample illustrates this heterogeneity: PRIMER-COSMOS-1223, the least massive host ($\log(M_\star/M_\odot)=7.97$) with the most luminous BH* ($\log(L_{\rm BH*})=46.17$) and essentially no dust screen ($A_{V,{\rm BH*}}\simeq0.03$~mag), resembles the break-compact population of \citet{2026arXiv260324700C}. Our two-component fits thus capture the population-averaged tendency, while individual sources vary with the covering factor. Spatially resolved tests -- UV point-source decomposition, variability, UV line diagnostics, polarimetry, and medium-resolution spectroscopy of the absorber kinematics, will be needed to go beyond the one-dimensional PRISM spectra used here.

\subsection{Star Formation and Dust Properties} 
\label{subsec:Star Formation and Dust Properties}

\begin{figure*}[htbp]
    \centering
    \begin{minipage}[t]{0.49\linewidth}  
        \centering
        \includegraphics[width=\linewidth]{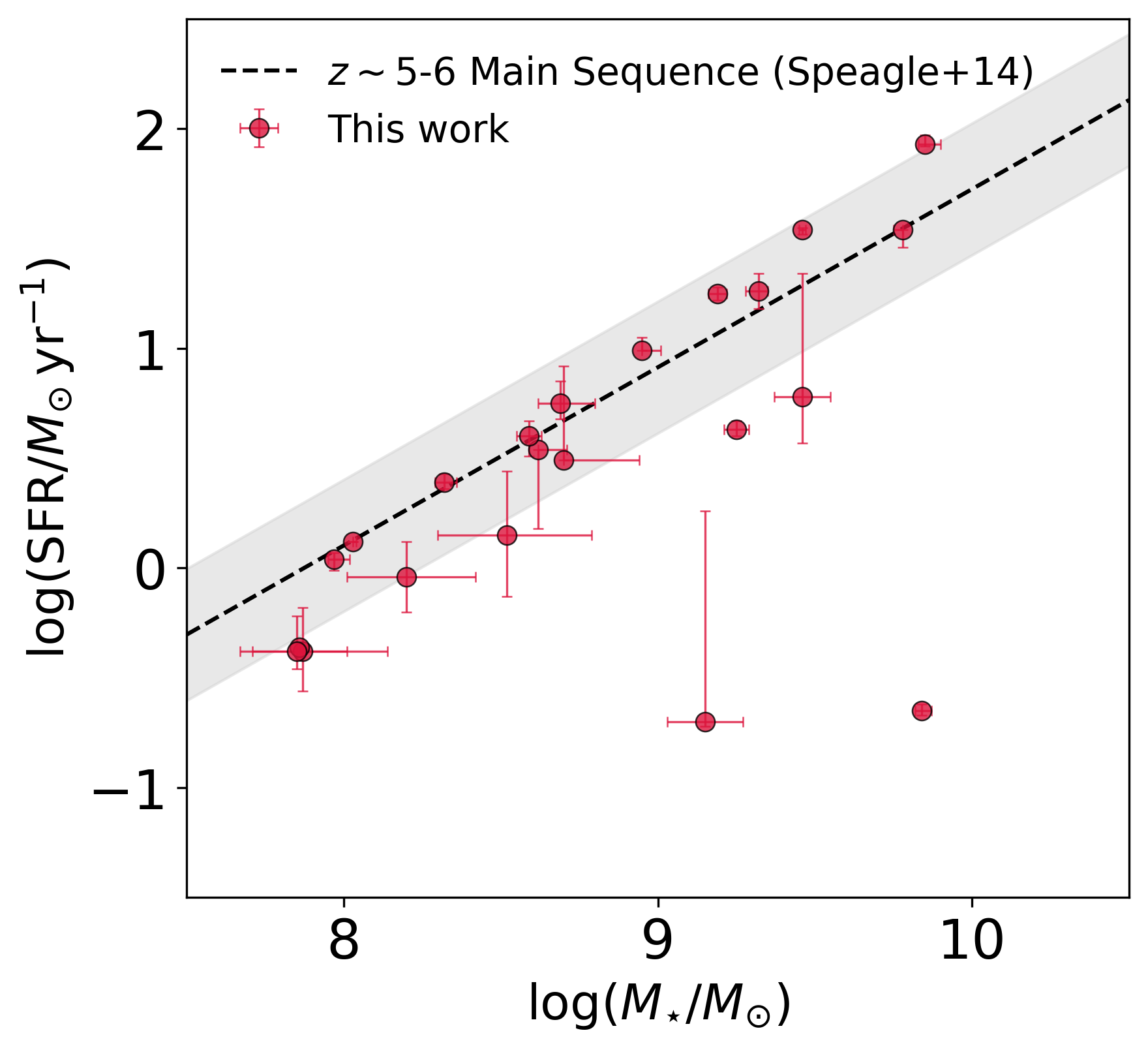}
        \label{fig:sfr_mstar}
    \end{minipage}
    \hfill 
    \begin{minipage}[t]{0.49\linewidth}
        \centering
        \includegraphics[width=\linewidth]{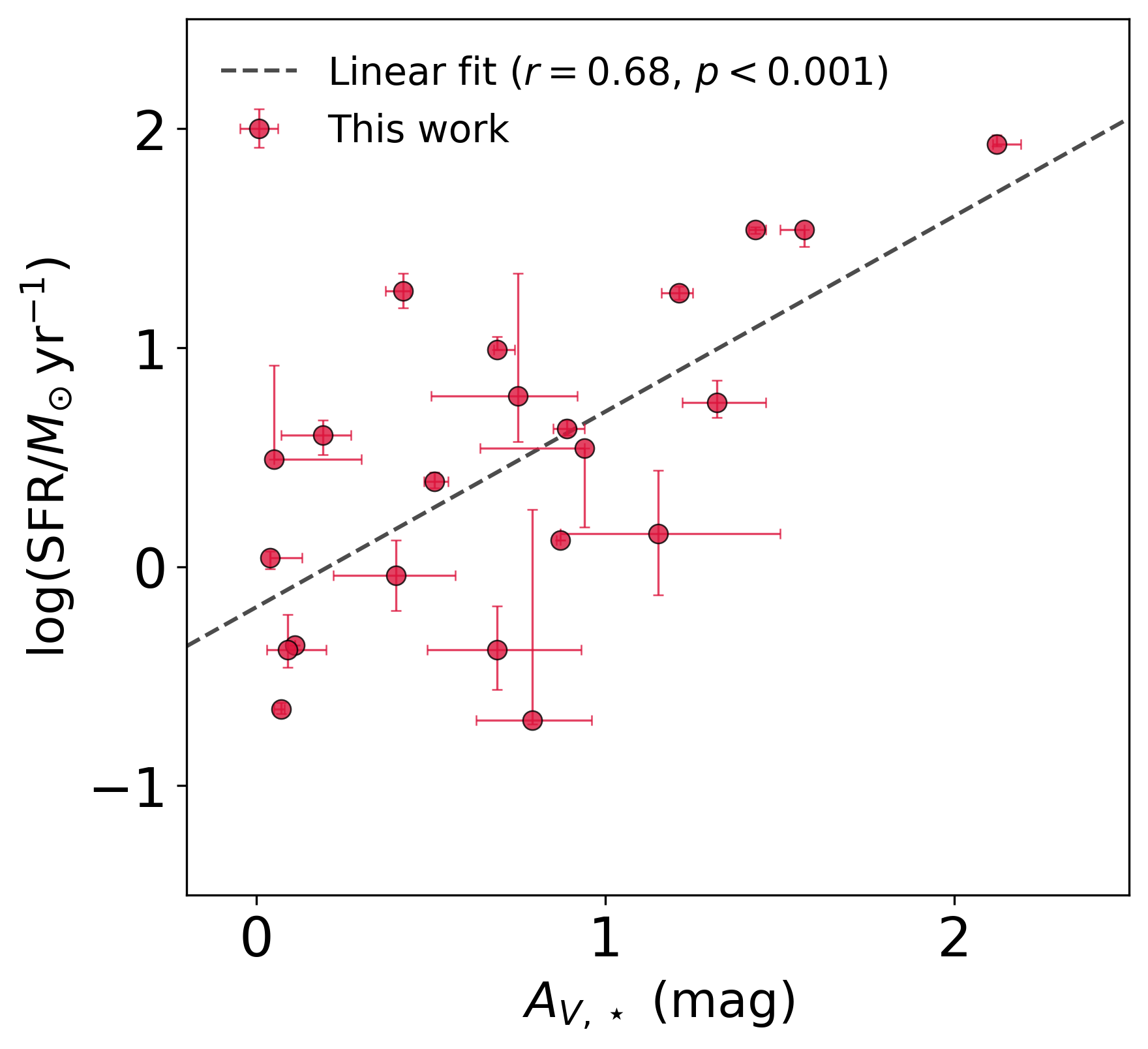}
        \label{fig:sfr_avstar}
    \end{minipage}
    \caption{Left: SFR versus $M_\star$ for the 22 LRD host galaxies. Red circles with error bars denote the posterior medians and 16th--84th percentile uncertainties derived from the SED fits. The dashed line and shaded region represent the $z\sim5$--$6$ star-forming main sequence from \citet{Speagle2014} and its typical $0.3$\,dex scatter. The majority of LRD hosts lie within or near the main sequence. Right: SFR versus $A_{V,\star}$ for the LRD host galaxies. A significant positive correlation is found between SFR and $A_{V,\star}$ (Pearson $r=0.68$, $p=0.001$; Spearman $r_s=0.68$, $p=0.001$).}
    \label{fig:sfr}
\end{figure*}

The left panel of Figure~\ref{fig:sfr} illustrates the SFR versus $M_\star$ relation for the 22 LRD host galaxies derived from our SED fits. The sample spans $\log(M_\star/M_\odot)=7.85$--$9.85$ (median: $8.70$) , and $\log({\rm SFR}/M_\odot\,~{\rm yr^{-1}})=-0.70$--$1.93$ (median: $0.62$). The majority of the LRD hosts reside within or near the $0.3$\,dex scatter of the $z\sim5$--$6$ star-forming main sequence \citep{Speagle2014}, indicating that their star formation activity is broadly consistent with that of normal star-forming galaxies at comparable redshifts. This supports a picture of coeval growth in which the black hole and stellar population accumulate mass simultaneously \citep{2014ApJ...782....9H,2014ARA&A..52..373L}.

The right panel of Figure~\ref{fig:sfr} presents the relationship between SFR and the stellar extinction $A_{V,\star}$. The stellar extinction spans $A_{V,\star}\approx0.04$--$2.12$\,mag with a median of $0.72$\,mag. A Pearson correlation analysis yields a coefficient of $r=0.68$ ($p=0.001$), indicating a statistically significant positive linear correlation between these two quantities. This trend is further corroborated by a Spearman rank correlation test, which yields $r_s=0.68$ ($p=0.001$), confirming that the correlation is robust and monotonic even without assuming a strictly linear relationship. The agreement between the Pearson and Spearman coefficients indicates that this trend is not driven by outliers, but rather reflects a genuine physical property across the population \citep{2015ApJ...806..259R, 2023ApJ...951...29L}. This positive correlation aligns with the well-established paradigm that dustier galaxies host more vigorous star formation, a trend observed in both local starbursts \citep{1996ApJ...457..645W, 2001ApJ...558...72S, 2002A&A...393...33B} and high-redshift star-forming galaxies \citep{2009ApJ...698L.116P, 2013MNRAS.429.1113H, 2014MNRAS.437.1268H}. Furthermore, \citet{2024A&A...690A.290C} recently reported a positive correlation between $A_V$ and the star formation rate surface density ($\Sigma_{\rm SFR}$) for galaxies at $z\sim4$--$10$, with no evidence for a decline in $A_V$ even in extremely star-forming systems, reinforcing the physical connection between high dust content and intense star formation activity across cosmic time.

\subsection{Redshift Evolution of $M_{\rm BH}/M_{\star}$ } 
\label{subsec:Redshift evolution}

\begin{figure*}
    \centering
    \includegraphics[width=0.90\linewidth]{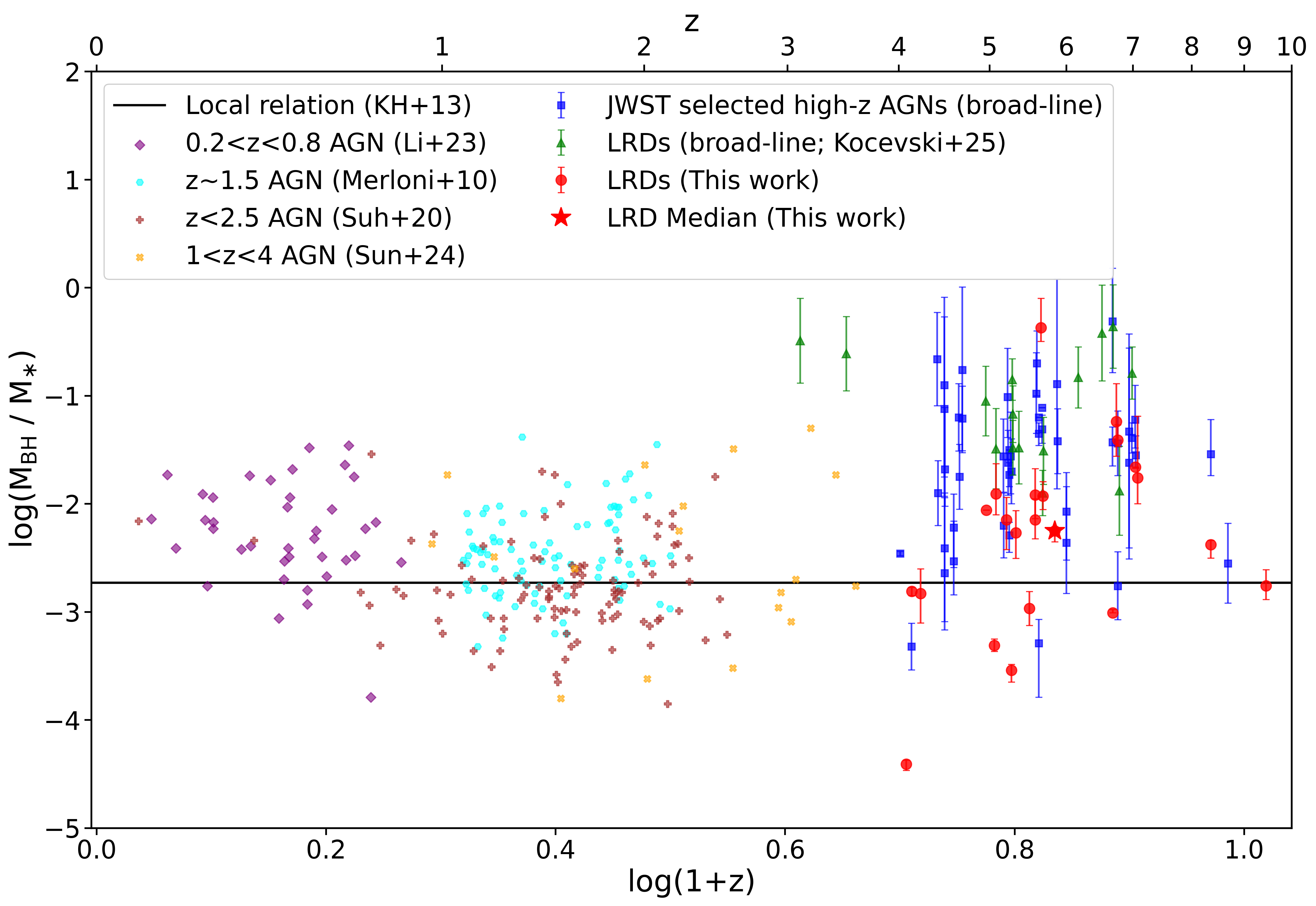}
    \caption{Redshift evolution of the black hole-to-stellar mass ratio ($M_{\mathrm{BH}}/M_{\star}$).  The black solid line denotes the benchmark local relation ($\log_{10}(M_{\rm BH}/M_{*}) = -2.73$), corresponding to the typical  $M_{\mathrm{BH}}/M_{\mathrm{bulge}}$ relation for black holes in the mass range $10^{6}-10^{10}\,M_\odot$ \citep{2013ARA&A..51..511K}. Colored symbols represent AGN samples at increasing redshifts: $0.2<z<0.8$ (\citealt{Li_2023}, purple diamonds), $z\sim1.5$ (cyan circles), $z<2.5$ (\citealt{Suh_2020}, brown crosses), $1<z<4$ (\citealt{Sun_2025}, orange crosses), and JWST-selected high-$z$ AGN at $z\sim4$--$8$ (\citealt{Larson_2023,Kocevski_2023,Harikane_2023,Wang_2024,Kokorev_2024,Maiolino}, blue squares). Green triangles denote single-epoch LRDs \citep{kocevski2025risefaintredagn}, while red circles show our LRD sample. The red star marks the median of our sample. The literature comparison points are broad-line black hole mass estimates unless otherwise noted. The green-to-red LRD offset primarily reflects the change from broad-line single-epoch masses to BH*-based SED masses, whereas comparisons with the blue JWST AGN points may also include selection effects.}
    \label{fig:bhmass-z}
\end{figure*}

Using these estimates of stellar masses and black hole masses, we compute a black hole-to-stellar mass ratio ($M_{\rm BH}/M_{\star}$) for our sample. We present these results in Figure \ref{fig:bhmass-z}, where our LRDs are plotted alongside recently identified JWST selected normal AGNs and LRDs \citep{Larson_2023,Kocevski_2023,Harikane_2023,Wang_2024,Kokorev_2024,Maiolino},  AGNs at $0.2<z<0.8$ \citep{Li_2023},  AGNs at $z<2.5$ \citep{Suh_2020},  AGNs at $1<z<4$ \citep{Sun_2025} and local relation \citep{2013ARA&A..51..511K}. The $M_{\rm BH}/M_{\star}$ measurements for LRD from \cite{kocevski2025risefaintredagn}, derived using empirical relations between $M_{\rm BH}$ and the broad emission lines, are also included. For $\log(M_{\rm BH}/M_{\rm \star})$ of the 22 LRDs, the 16th, 50th, and 84th percentiles are -2.99, -2.20, and -1.70, respectively. The red star marked the median $\log(M_{\rm BH}/M_{\rm \star})$ of the 22 LRDs. Figure \ref{fig:bhmass-z} illustrates the redshift evolution of the black hole-to-host galaxy mass ratio ($M_{\rm BH}/M_\star$) from the local universe to $z\sim10$. At redshifts $z \lesssim 4$, various AGN samples are broadly consistent with the local scaling relation (horizontal line at $\log(M_{\rm BH}/M_\star) = -2.73$). A significant departure from this relation emerges at higher redshifts ($z > 4$), where JWST-selected broad-line AGNs (blue squares) exhibit systematically elevated $M_{\rm BH}/M_\star$ values. In contrast, our LRD sample does not follow this extreme trend. While the BH* model infers moderately overmassive black holes for our LRDs relative to their hosts, this offset is significantly less extreme than estimates derived from broad-line empirical relations, and the inferred ratios are substantially closer to the local scaling relation.

This discrepancy primarily reflects differences in mass estimation methodology. The single-epoch method assumes that broad emission lines originate from the traditional broad-line region (BLR), which tends to significantly overestimate black hole masses when applied to LRDs (\citealt{Kokorev_2023}; \citealt{Greene_2024}). By contrast, our BH* model yields more conservative estimates that are less subject to this systematic bias. We note that the current sample of LRDs with detected broad lines may be biased toward overmassive black holes, because only the most luminous broad-line systems are detectable at the survey depth. The key comparison for LRDs is therefore between the green broad-line points and the red BH*-based points, because this isolates the change in black hole mass inference at fixed LRD selection. Offsets relative to the blue JWST-selected AGN points should not be interpreted solely as redshift evolution, because those samples are selected differently and are biased toward detectable broad-line systems.

From an evolutionary perspective, the relatively modest $M_{\rm BH}/M_\star$ ratios observed in our LRDs suggest that they may represent an early evolutionary stage of JWST-selected broad-line AGNs. As young systems in the initial phase of super-Eddington accretion with short accretion histories, the central black holes in LRDs have not yet had sufficient time to accumulate mass. Consequently, at comparable redshifts, LRDs exhibit lower $M_{\rm BH}/M_\star$ than more evolved AGNs, whose black holes have undergone longer durations of sustained growth. Over time, continued accretion may drive these LRDs toward the higher $M_{\rm BH}/M_\star$ ratios characteristic of the JWST broad-line AGN population.

\subsection{The LRD number density} 
\label{subsec:The LRD number density}

Previous studies found that LRDs at $z \simeq 4$--$7$ have an inferred comoving number density of $n \simeq 10^{-5}$--$10^{-4}$ cMpc$^{-3}$ \citep{Kocevski_2023,2025ApJ...991...37A}.  This abundance is orders of magnitude higher than that of luminous quasars known at similar cosmic epochs \citep{McGreer_2018,Matsuoka_2018,Niida_2020,Matsuoka_2023}. According to our SED modeling, LRDs harbor relatively low-mass black holes and reside in faint, low-mass host galaxies. This implies that a large population of such systems can naturally arise in the early universe without requiring rare, massive galaxies or extremely massive black holes. However, to rigorously assess their cosmic abundance, it is essential to benchmark their number density against the appropriate luminosity function rather than that of bright quasars. We emphasize that we do not perform a formal effective-volume calculation of the number density: with only 22 unique objects, selected through a heterogeneous spectroscopic selection whose completeness is not characterized, any such estimate would be dominated by small-number statistics and selection biases. Instead, the comparison below is an order-of-magnitude consistency check at matched luminosity, not a density measurement.

The bolometric luminosity of the BH* component $L_{\rm BH*}$ ranges from $10^{44.0}$ to $10^{46.8}\,\rm erg\,~s^{-1}$ with a median of $10^{44.9}\,\rm erg\,~s^{-1}$. According to the quasar luminosity function presented by \citet{2025ApJ...991...37A}, at $L_{\rm bol}\sim10^{45}\,\rm erg\,~s^{-1}$, the representative LRD number density reaches $\phi\sim2$--$3\times10^{-5}\,\rm dex^{-1}\,~Mpc^{-3}$, while the quasar number density is $\phi\sim5$--$7\times10^{-6}\,\rm dex^{-1}\,~Mpc^{-3}$. Although a discrepancy of a factor of $\sim$3--5 remains, this tension is substantially mitigated compared with the previously reported difference of 1--2 orders of magnitude, which arose from erroneously benchmarking LRDs against the bright end of the quasar luminosity function. 

The residual excess can be understood in the context of observational selection effects. Traditional quasar surveys are heavily optimized for point-like, UV-bright sources with minimal contamination from host galaxy emission. In contrast, LRDs are composite systems comprising a low-mass SMBH accreting at super-Eddington rates, enshrouded in a dense gaseous cocoon, and hosted by a faint, main-sequence star-forming galaxy. Their compact rest-optical morphologies and red UV-to-optical colors are driven by the BH* component, while their rest-UV emission is dominated by the host galaxy. Consequently, LRDs fail to satisfy standard quasar selection criteria, and have been systematically missed by previous AGN censuses \citep{kocevski2025risefaintredagn,2024ApJ...963..129M}. The BH* framework, which self-consistently accounts for this composite nature, provides a natural explanation for why LRDs appear overabundant relative to the known quasar luminosity function: they represent a distinct, low-luminosity AGN phase that has been largely omitted from traditional quasar surveys.

\subsection{The evolution of LRDs} 
\label{subsec:The evolution of LRDs}

LRDs represent a distinctive population of high-redshift galaxies that likely trace an early, rapid-growth phase of SMBH evolution. Initially, LRDs appear as super-Eddington accreting SMBHs enshrouded in dense, dust-poor gas with densities of $n_H \sim 10^9\text{--}10^{12}{\rm cm}^{-3}$. This dense envelope suppresses the escape of X-ray photons from the hot corona (\citealt{2025MNRAS.538.1921M}; \citealt{2024ApJ...974L..26Y}), obscures mid-infrared emission from hot dust (\citealt{2024ApJ...975L...4C}; \citealt{2025ApJ...991L..10S}), and weakens radio emission \citep{2025A&A...693L...2P}, making them appear fundamentally different from typical AGNs. \cite{fu2025discoverylittlereddots} reported two unusual LRDs at redshift z = 2.868 and 2.925 that were in a transitional phase towards typical AGNs. They retained LRD-like properties including V-shaped SEDs, compact optical morphologies, and broad emission lines, while simultaneously exhibiting intense X-ray, radio, and mid-infrared radiation characteristic of normal AGNs.

The population of JWST-discovered AGN may represent an evolutionary sequence from LRDs to Little Blue Dots (LBDs, \citealt{brazzini2026littlebluereddots}; \citealt{madau2026littlereddotsobscured}), both governed by a common central engine---a SMBH embedded within a dense gaseous envelope. In the early LRD phase, as described by our BH* interpretation, the black hole is enshrouded by a massive, optically thick gas cocoon that absorbs most of the ionizing UV radiation, thermalizing it to produce the characteristic v-shaped, red optical continuum. This same dense envelope imprints the prominent Balmer absorption features and heavily extinguishes the X-ray emission, rendering LRDs extremely X-ray weak (\citealt{2025MNRAS.538.1921M}; \citealt{brazzini2026littlebluereddots}). As the system evolves, the gas envelope gradually disperses, perhaps through radiation pressure-driven outflows or accretion-driven contraction \citep{2026ApJ..1000...90I}, allowing progressively more direct sightlines to the central accretion disk. This transition produces the LBDs phase: the optical continuum becomes dominated by the intrinsic blue power-law emission from the accretion disk \citep{brazzini2026littlebluereddots}, while the weakening gas cocoon eliminates the Balmer absorption features \citep{2026MNRAS.545f2117D}. 

Ultimately, these objects will evolve into normal quasars as the dense envelope completely clears and the AGN becomes fully visible across all wavelengths. This evolutionary pathway provides a crucial link between some LRDs and the established AGN population, indicating that LRDs do not persist as a distinct population but rather represent the initial growth phase of SMBHs that subsequently join the canonical AGN population.

\subsection{Comparison with alternative LRD continuum scenarios}
\label{sec:model_compare}

A complete exploration of all proposed LRD models is beyond the scope of this work. Instead, we benchmark the “host galaxy + BH*” interpretation against alternative scenarios using the observational constraints that are already available for our sample and for the broader LRD population. We emphasize that this comparison is not a claim that the BH* model is uniquely favored for every LRD. Rather, it identifies which physical ingredients are required by the data and where the model is likely to be incomplete.

The classical heavily dust-reddened broad-line AGN scenario can reproduce the red optical slope of individual LRDs, but it is challenged by the weak NIR--MIR re-emission, X-ray weakness, and the requirement of extremely high Balmer-break strengths in some sources \citep{2025ApJ...991...37A, Setton_2025, Leung_2025}. It also preserves the broad-line mass calibration that has been argued to overestimate black hole masses in LRD-like dense envelopes \citep{Kokorev_2023,Greene_2024, naidu2025blackholestarreveals}. Pure stellar or stellar + nebular models can explain the rest-UV continuum and narrow emission in many objects, but they do not naturally produce the combination of a strong red optical bump, high-order Balmer absorption, and weak X-ray/MIR emission when these features are required simultaneously \citep{naidu2025blackholestarreveals, Taylor_2025, degraaff2025littlereddotshost}. Gray-extinction AGN models provide useful flexibility for the SED shape \citep{Li_2025}, but by themselves they do not specify the origin of the velocity-dependent Balmer absorption or the gas-envelope line--continuum coupling emphasized by the BH* picture.

The main alternative we cannot exclude with the present PRISM data is a minority direct or scattered AGN component in the rest-UV. Such a component would be most relevant for objects with weaker Balmer absorption, lower absorber velocities, or UV-bright point-like morphologies, and may connect individual LRDs to transitional LRD/LBD systems \citep{brazzini2026littlebluereddots, madau2026littlereddotsobscured, 2026MNRAS.545f2117D}. A multi-component UV model with an explicit $f_{\rm UV,AGN}$ parameter should therefore be explored in future work. For the present sample we treat the stellar-dominated UV solution as the default decomposition, not as evidence that the UV emission has a single origin.

\section{SUMMARY} 
\label{sec:summary}

In this study, we investigate the physical nature of LRDs using a sample of 22 sources drawn from the ASTRODEEP-JWST database. We construct a composite “host galaxy + BH*” model and performed Bayesian SED fitting to NIRSpec spectra from the DAWN JWST Archive of our sample. We summarize our main results as follows. 

\begin{itemize}
    \item \textbf{V-shaped SED origins.} The distinctive V-shaped SEDs of LRDs arise from two distinct physical components: the UV continuum blueward of the Balmer break is well reproduced by stellar emission within our two-component decomposition, while the optical spectrum is dominated by emission from the BH* component. A compact non-stellar contribution to the rest-UV cannot be excluded with the current data, but its effect is bounded: $M_\star$ and SFR could be overestimated by $\lesssim0.3$~dex, while $M_{\rm BH}$ and our main conclusions are unaffected.

    \item \textbf{Physical nature of LRDs.} LRDs comprise four key components: (i) a central SMBH with a median sample mass of $M_{\rm BH}\sim10^{6.4}\,M_\odot$ accreting at super-Eddington rates, characterized by an AGN big bump temperature of $T_{\rm BB}\approx10^{5.2}\,$K; (ii) a partially ionized, dense inner region with a high median electron density of $n_H\approx10^{9.6}\,\rm cm^{-3}$; (iii) an optically thick, convective gaseous envelope that thermalizes the accretion disk radiation into a near-blackbody continuum; and (iv) a faint, low-mass star-forming host galaxy with a median stellar mass of $M_\star \approx 10^{8.7}M_\odot$, a young stellar population, and modest extinction. The decoupled extinction properties between the gas-dominated envelope and the host galaxy provide quantitative evidence that LRDs are BH*s hosted by faint galaxies rather than heavily dust-reddened AGNs.

    \item \textbf{$M_{\rm BH}$--$M_\star$ relation.} Our derived scaling relation indicates that the inferred black hole masses for our LRDs are moderately overmassive relative to their hosts. However, this offset is significantly less extreme than estimates derived from broad-line empirical relations, and is substantially closer to the local $M_{\rm BH}$-$M_\star$ relation. Specifically, our BH*-based $M_{\rm BH}/M_\star$ measurements are approximately one to two orders of magnitude lower than those inferred from broad-line empirical correlations at fixed stellar mass, suggesting that the latter may overestimate black hole masses for this population.

    \item \textbf{Host galaxy properties.} The LRD host galaxies are faint, low-mass systems with a median stellar mass of $M_\star\approx10^{8.7}\,M_\odot$ and a median SFR of ${\rm SFR}\approx10^{0.6}\,M_\odot\,~{\rm yr^{-1}}$. The majority of hosts lie on or near the $z\sim5$--$6$ star-forming main sequence, indicating that their star formation is not strongly suppressed by AGN feedback. The stellar extinction spans $A_{V,\star}\approx0.04$--$2.12$\,mag with a median of $0.72$\,mag, and shows a significant positive correlation with SFR.

    \item \textbf{Number density.} Our BH* model yields a median bolometric luminosity $L_{\rm BH*}\sim10^{44.9}\,\rm erg\,~s^{-1}$. At this luminosity, the LRD number density is a factor of a few above the low-luminosity end of the quasar luminosity function, substantially alleviating the previously reported tension of 1--2 orders of magnitude. 

    \item \textbf{Evolutionary scenario.} LRDs likely represent an early, rapid-growth phase of SMBH evolution, wherein super-Eddington accretion occurs within dense, dust-poor gaseous envelopes. As these systems evolve, the envelope gradually disperses, transitioning LRDs through a LBD phase before ultimately emerging as normal quasars. This evolutionary sequence provides a crucial link between LRDs and the established AGN population, indicating that LRDs do not persist as a distinct class but rather represent the initial growth phase of SMBHs.
\end{itemize}

Despite the success of our model in reproducing the overall SED shapes of LRDs, certain spectral features in individual sources are not perfectly fitted. This limitation arises because our analysis relies on low-resolution PRISM spectra, which restricts our ability to resolve individual emission lines and tightly constrain the kinematics and ionization conditions of the gas. Future investigations will benefit from high-resolution NIRSpec spectroscopy of LRDs, enabling detailed analysis of emission line profiles and origins. Furthermore, we caution that this framework may not capture the full diversity of the LRD population. Two further idealizations deserve note. First, our CLOUDY grid is static, spherical, and fully covering. It reproduces the width of the Balmer absorption through $v_{\rm turb}$ but contains no bulk flows, and the envelope covering factor may in reality vary across the population \citep{2026arXiv260317667M}. Second, the decomposition attributes the rest-UV to starlight alone, whereas stacking analyses reveal a compact UV component in at least a subset of LRDs \citep{2026arXiv260324700C,2026arXiv260603522A,2026arXiv260403370J}. We have quantified the resulting systematic uncertainties in Section~\ref{subsec:The origin of the rest-UV emission}. Resolving them will require spatially resolved and medium-resolution observations. LRDs likely represent a heterogeneous group of objects at various evolutionary stages, ranging from heavily gas-enshrouded systems to more unobscured AGN. While our current model is best suited for sources dominated by dense gaseous cocoons, future studies utilizing larger statistical samples will be essential to clarify the physical diversity and evolutionary pathways of this enigmatic population.

\begin{acknowledgments}

We are grateful to the anonymous referee for their valuable comments, which have greatly helped us improve our paper. This work is supported by National Key Research and Development Program of China (2025YFF0511003). The authors gratefully acknowledge the support provided by the National Natural Science Foundation of China (NSFC, Grant Nos. 12673026, 12673028, 12173037, 12233008), the CAS Project for Young Scientists in Basic Research (No. YSBR-092), Guizhou Provincial Major Scientific and Technological Programs XKBF (2025)010 and XKBF (2025)011 and Cyrus Chun Ying Tang Foundations. 

This work is based on observations made with the NASA/ESA/CSA James Webb Space Telescope. The data were obtained from the Mikulski Archive for Space Telescopes at the Space for Space Telescopes at the Space Telescope Science Institute, which is operated by the Association of Universities for Research in
Astronomy, Inc., under NASA contract NAS 5-03127 for JWST.  We acknowledge the use of the ASTRODEEP-JWST photometric catalogs from \cite{Merlin2024}, publicly available at \url{http://www.astrodeep.eu/astrodeep-jwst-catalogs/}. These observations are associated with programs: GLASS-JWST (ERS 1324, P.I. Treu), UNCOVER (GO 2561, P.I. Labbe), DDT 2756 (P.I. Chen), and GO 3990 (P.I. Morishita); CEERS (ERS 1345, P.I. Finkelstein) on the EGS field; the JADES-GS and JADES-GN fields (GTO 1180 and GTO 1210, P.I. Eisenstein) including FRESCO data (GO 1895, P.I. Oesch); the first-epoch imaging of the NGDEEP field (Co-PIs Finkelstein, Papovich, Pirzkal); and the PRIMER (GO-1837, P.I. Dunlop) observations of the UDS and COSMOS fields in CANDELS.
\end{acknowledgments}

\facilities{JWST, HST}

\software{\textsc{BayeSED3} \citep{Han_2012,Han_2014,Han_2019,Han_2023},  
          \texttt{Cloudy} \citep{2017RMxAA..53..385F}
          }

\appendix
\section{Best-fit SED for our sample} \label{appendix:full sample}

\renewcommand{\thefigure}{A\arabic{figure}}
\setcounter{figure}{0}

Here we include some supplementary figures that show the best-fit SED for our full LRD sample.

\begin{figure}[htbp]
    \centering
    \begin{minipage}[t]{0.32\linewidth}
        \centering
        \includegraphics[width=\linewidth]{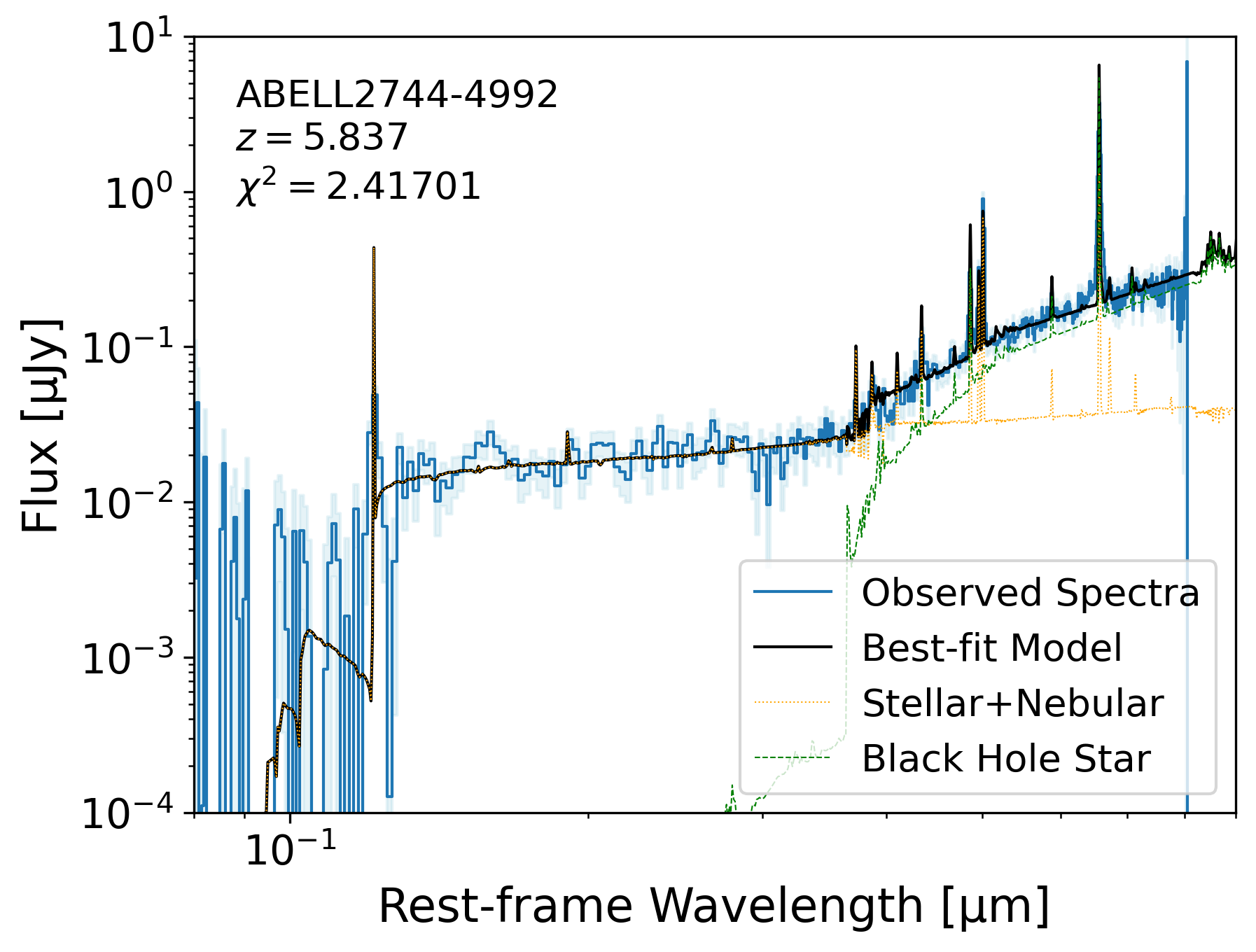}
    \end{minipage}
    \hfill
    \begin{minipage}[t]{0.32\linewidth}
        \centering
        \includegraphics[width=\linewidth]{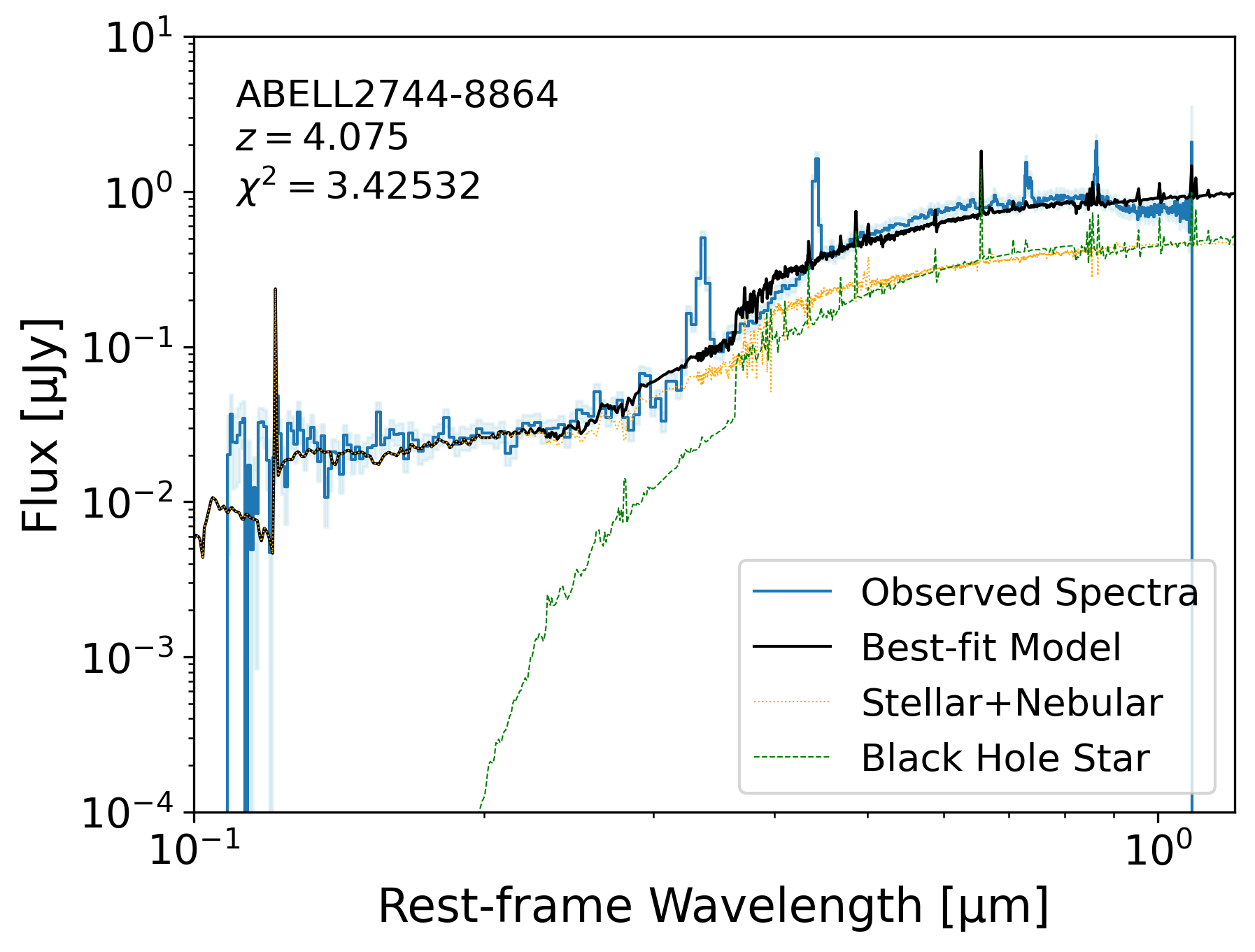}
    \end{minipage}
    \hfill
    \begin{minipage}[t]{0.32\linewidth}
        \centering
        \includegraphics[width=\linewidth]{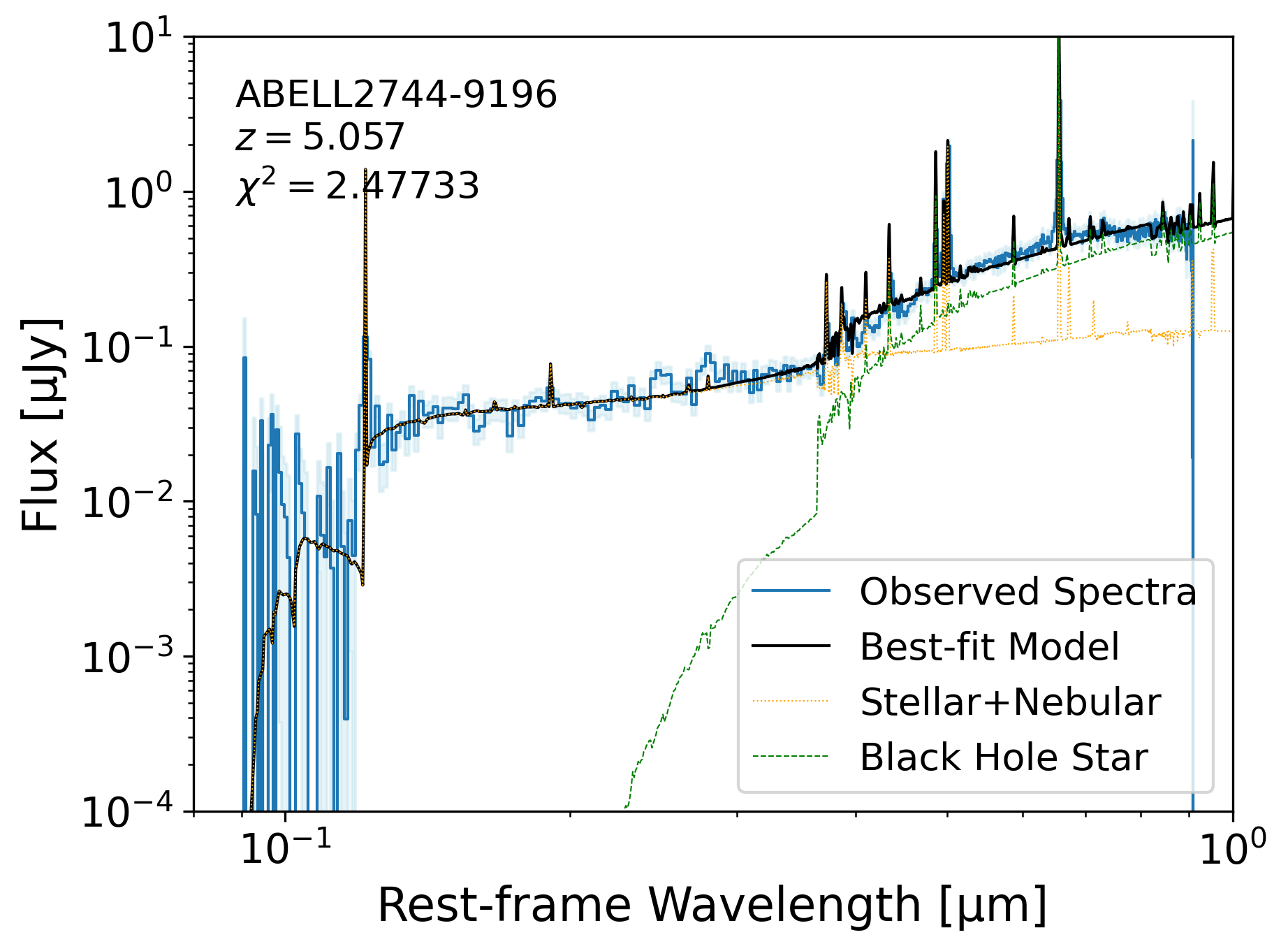}
    \end{minipage}

    \vspace{0.5cm}  

    \begin{minipage}[t]{0.32\linewidth}
        \centering
        \includegraphics[width=\linewidth]{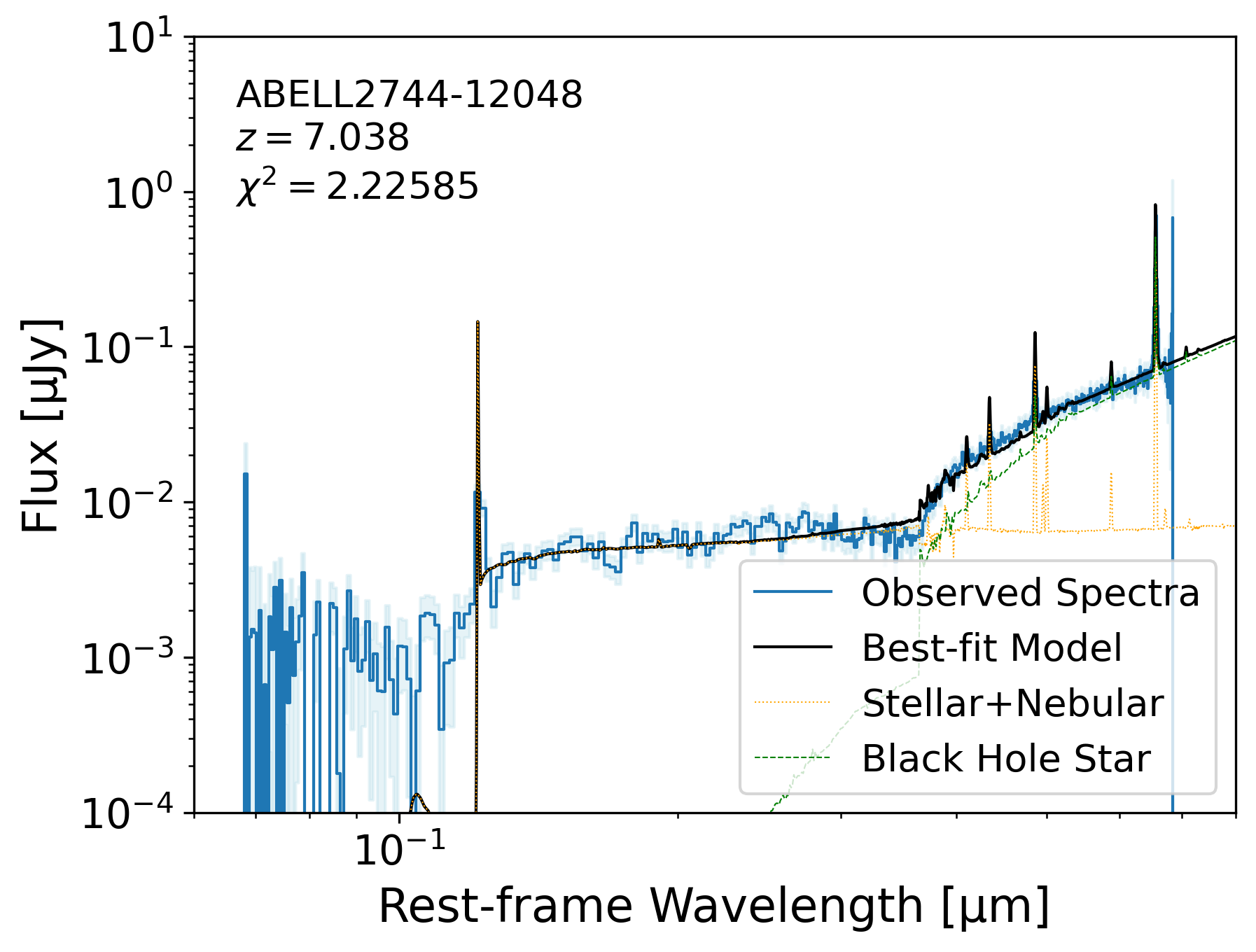}
    \end{minipage}
    \hfill
    \begin{minipage}[t]{0.32\linewidth}
        \centering
        \includegraphics[width=\linewidth]{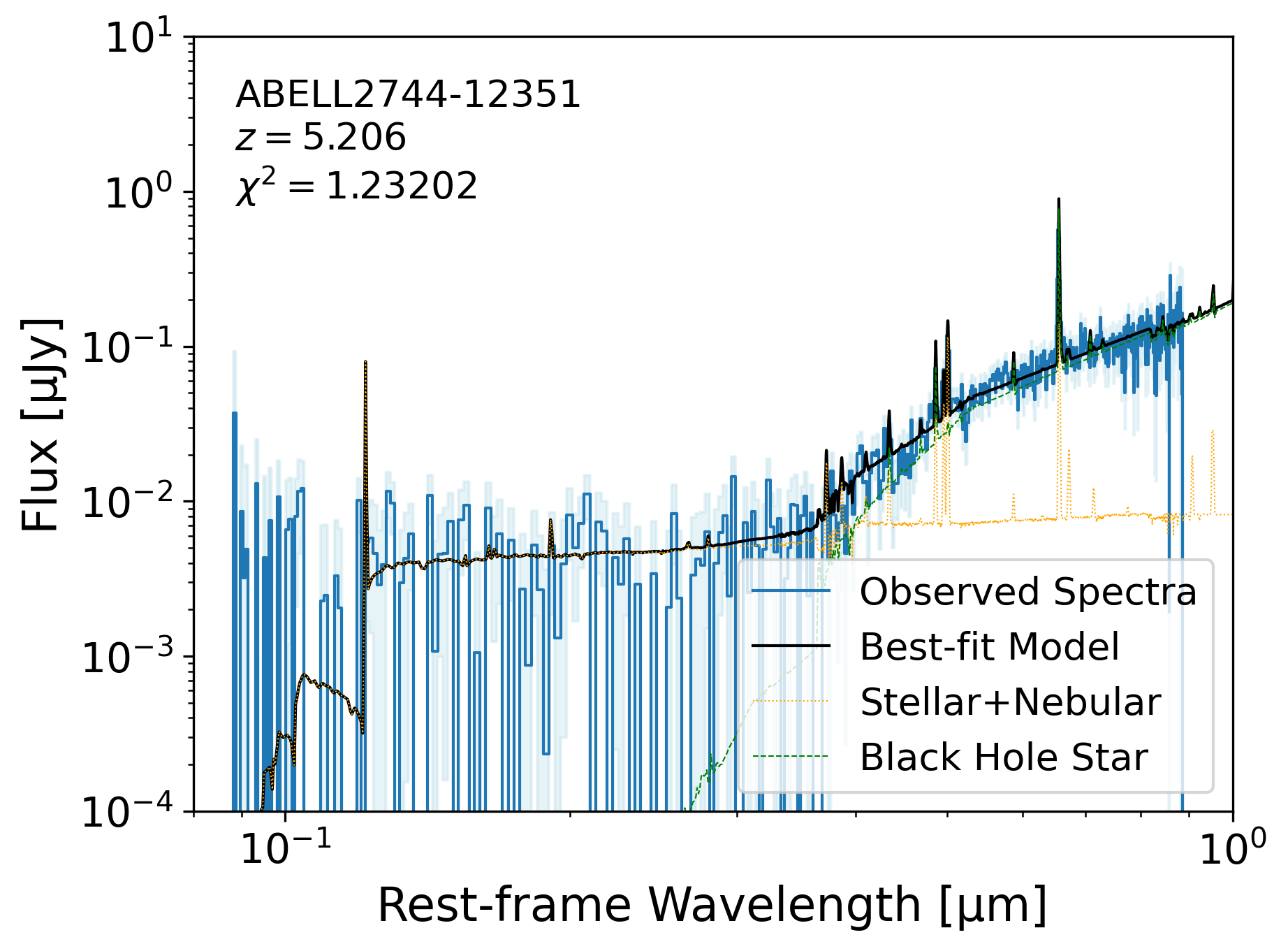}
    \end{minipage}
    \hfill
    \begin{minipage}[t]{0.32\linewidth}
        \centering
        \includegraphics[width=\linewidth]{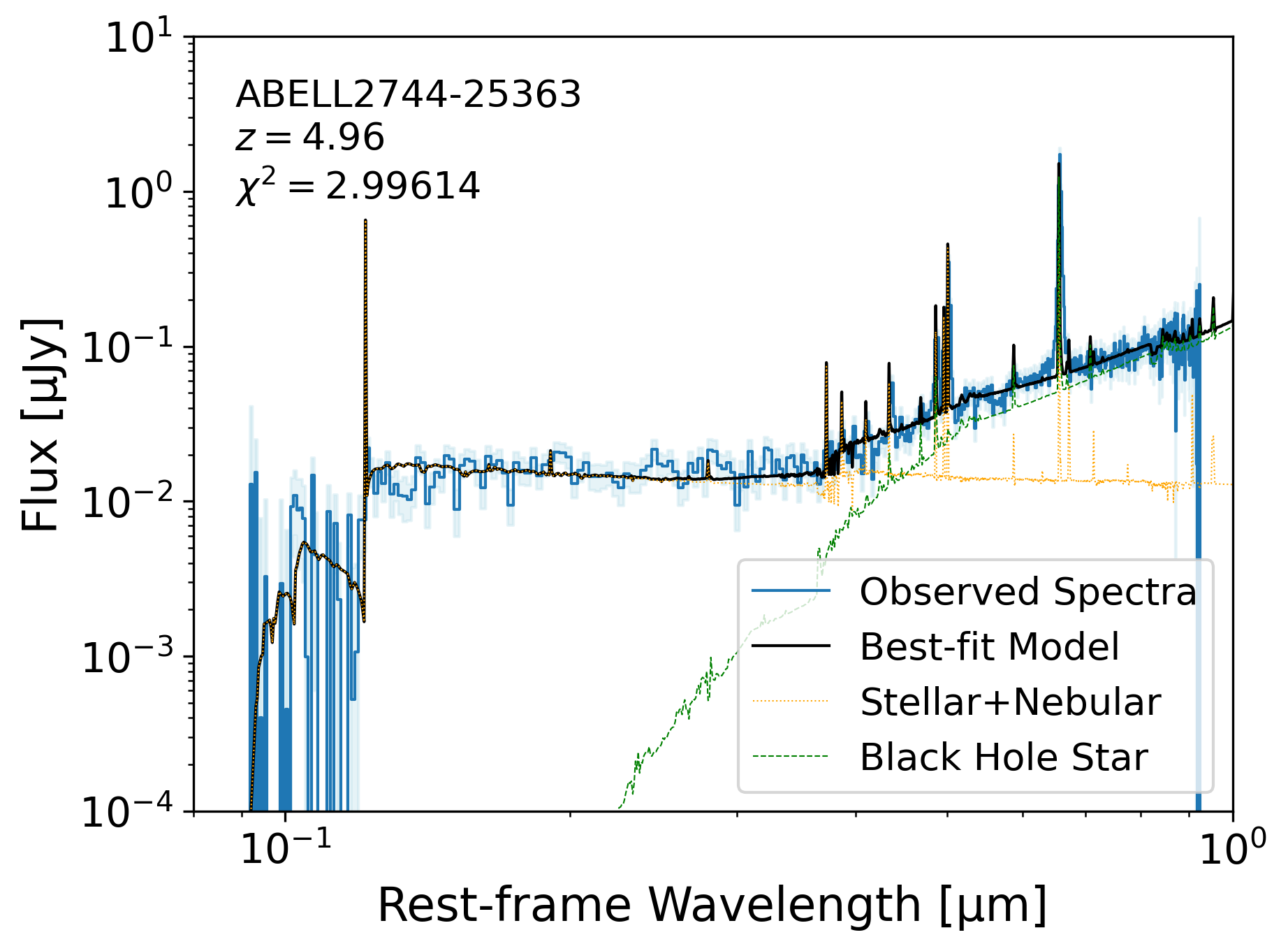}
    \end{minipage}

    \vspace{0.5cm}

    \begin{minipage}[t]{0.32\linewidth}
        \centering
        \includegraphics[width=\linewidth]{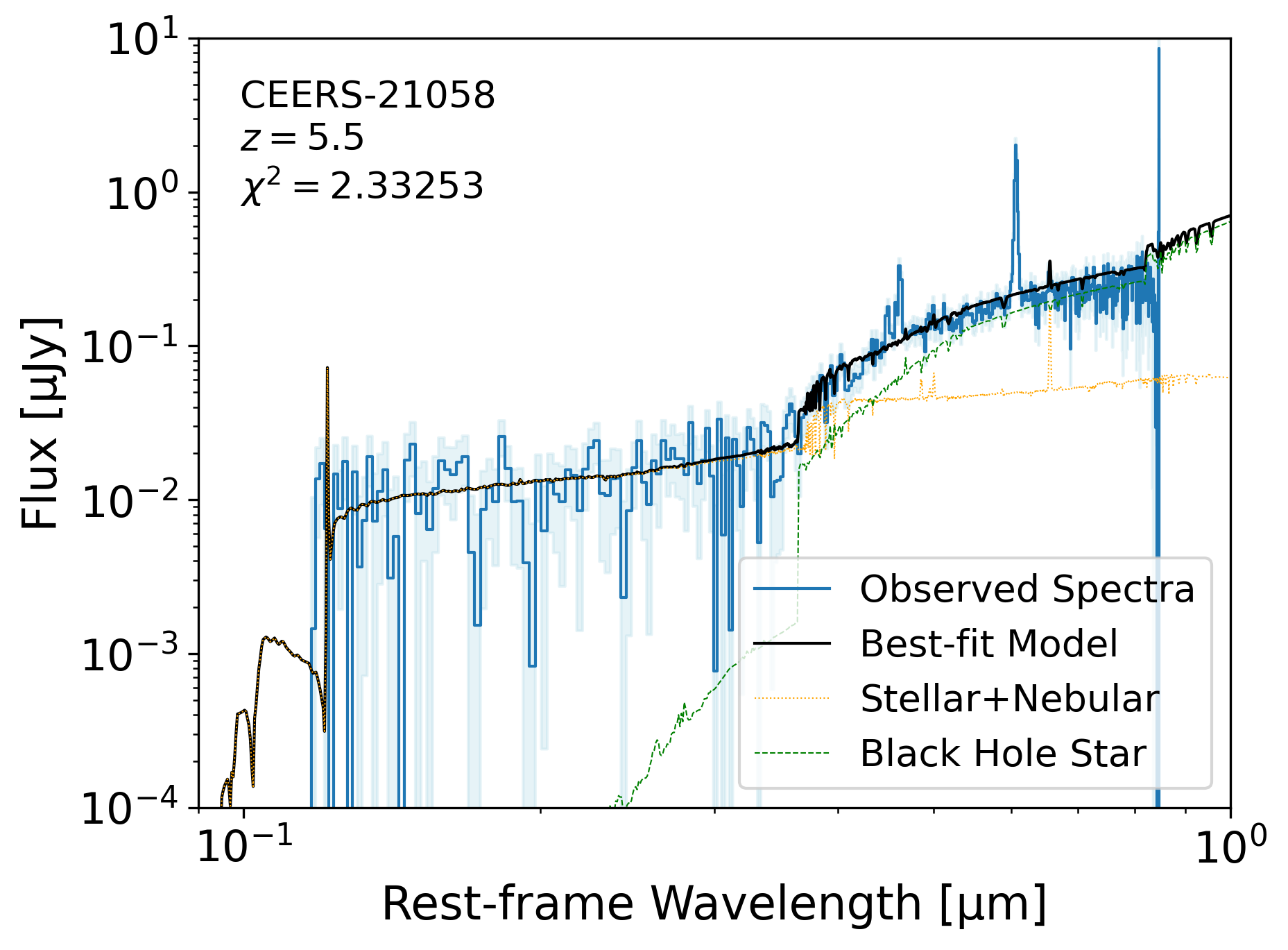}
    \end{minipage}
    \hfill
    \begin{minipage}[t]{0.32\linewidth}
        \centering
        \includegraphics[width=\linewidth]{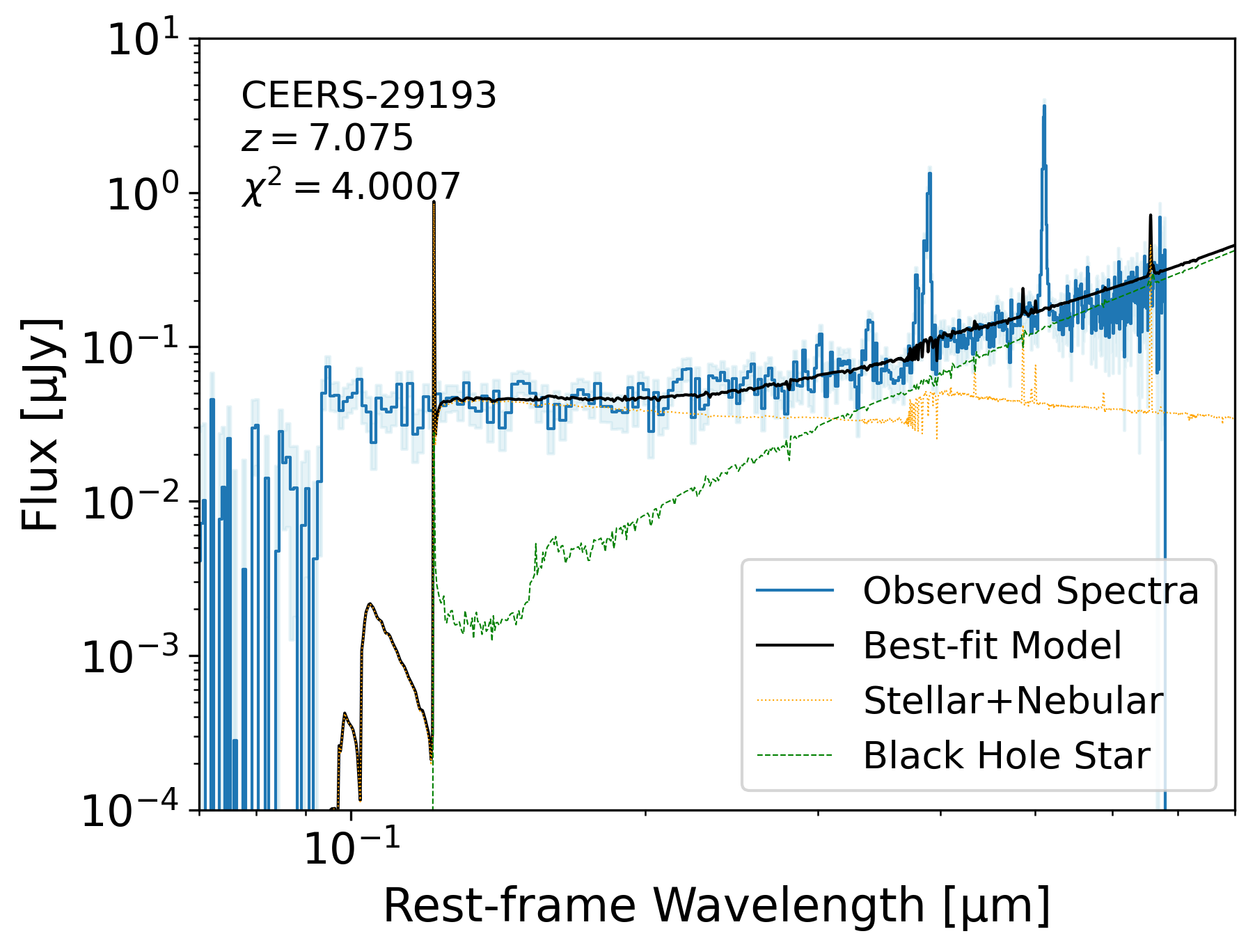}
    \end{minipage}
    \hfill
    \begin{minipage}[t]{0.32\linewidth}
        \centering
        \includegraphics[width=\linewidth]{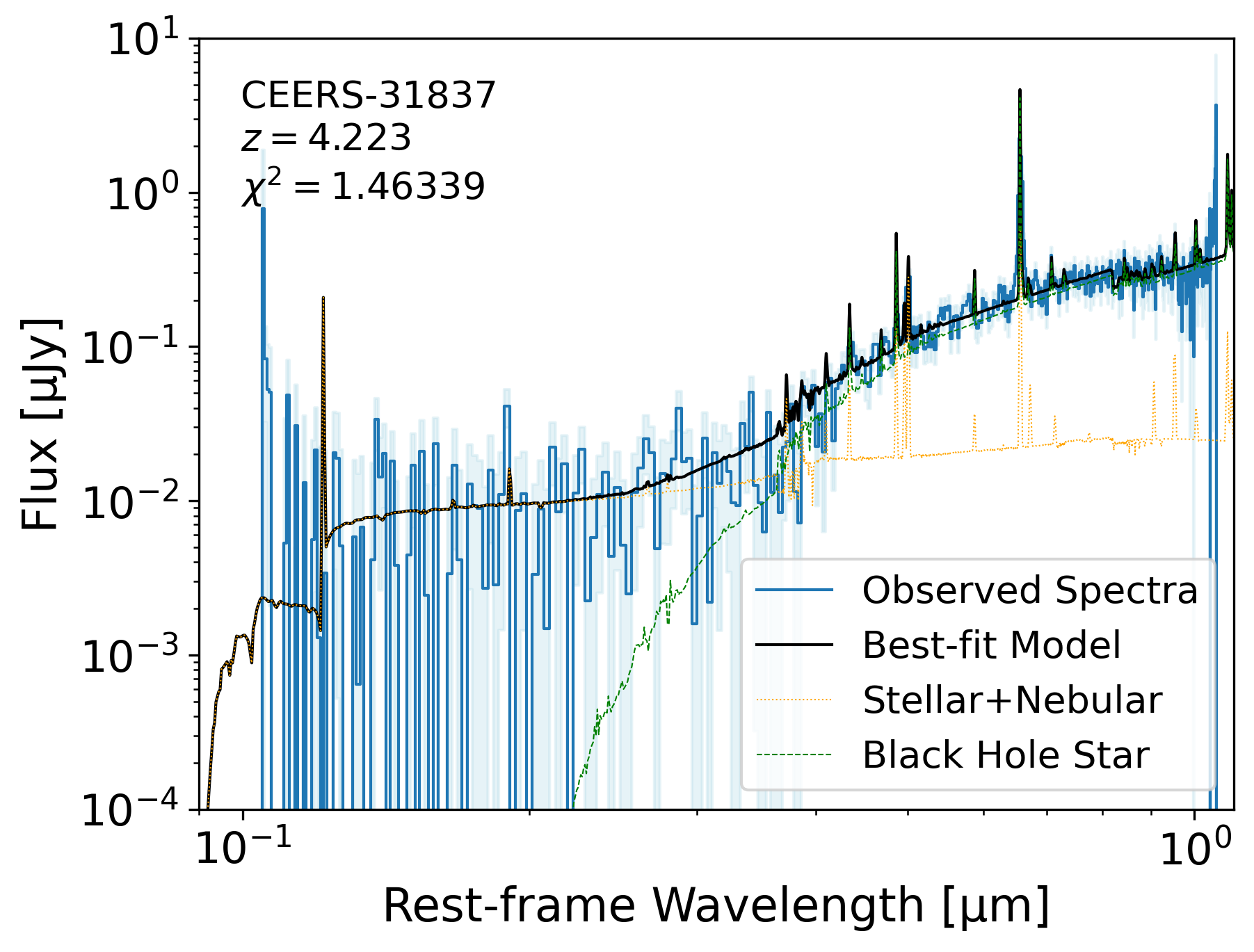}
    \end{minipage}

    \vspace{0.5cm}

    \begin{minipage}[t]{0.32\linewidth}
        \centering
        \includegraphics[width=\linewidth]{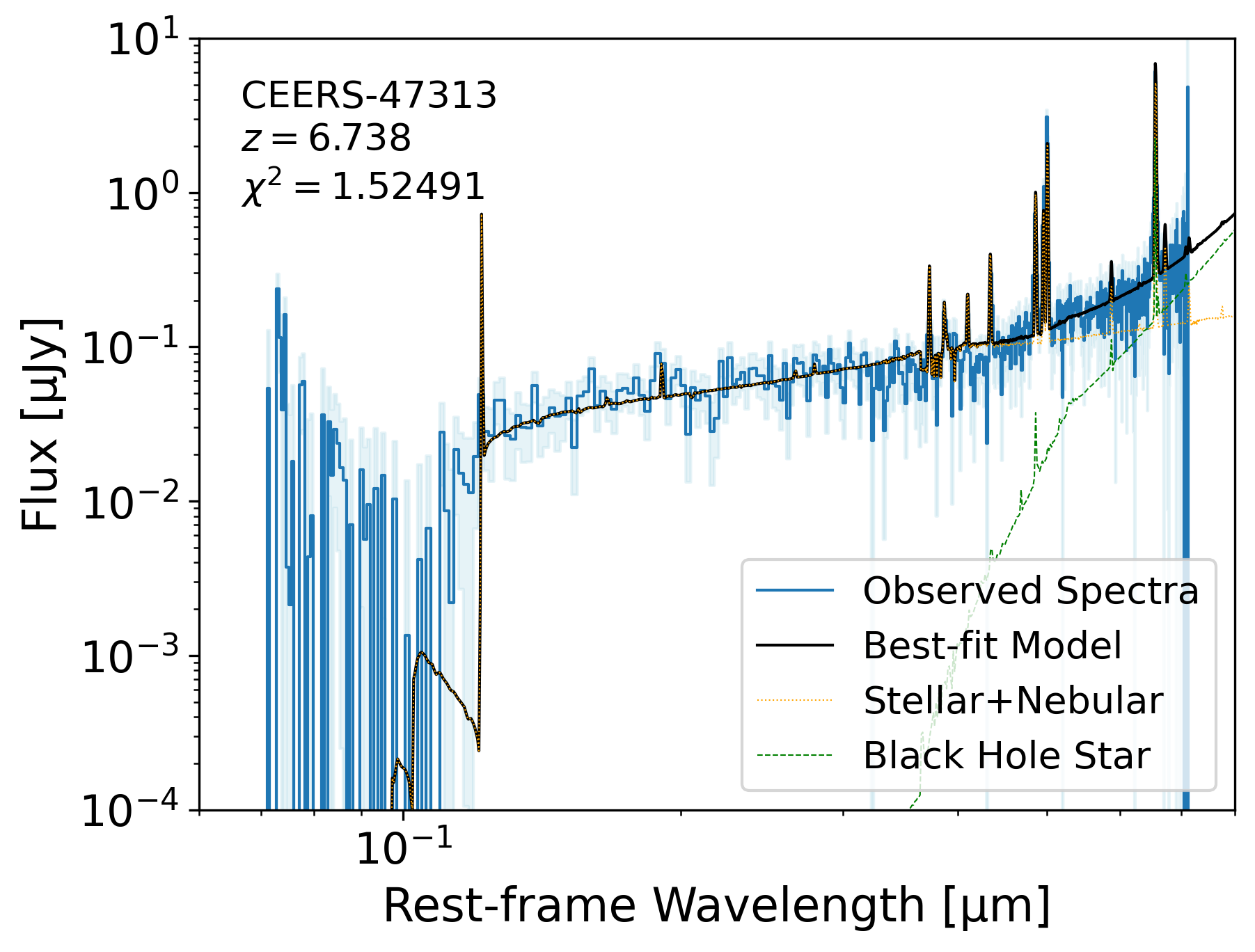}
    \end{minipage}
    \hfill
    \begin{minipage}[t]{0.32\linewidth}
        \centering
        \includegraphics[width=\linewidth]{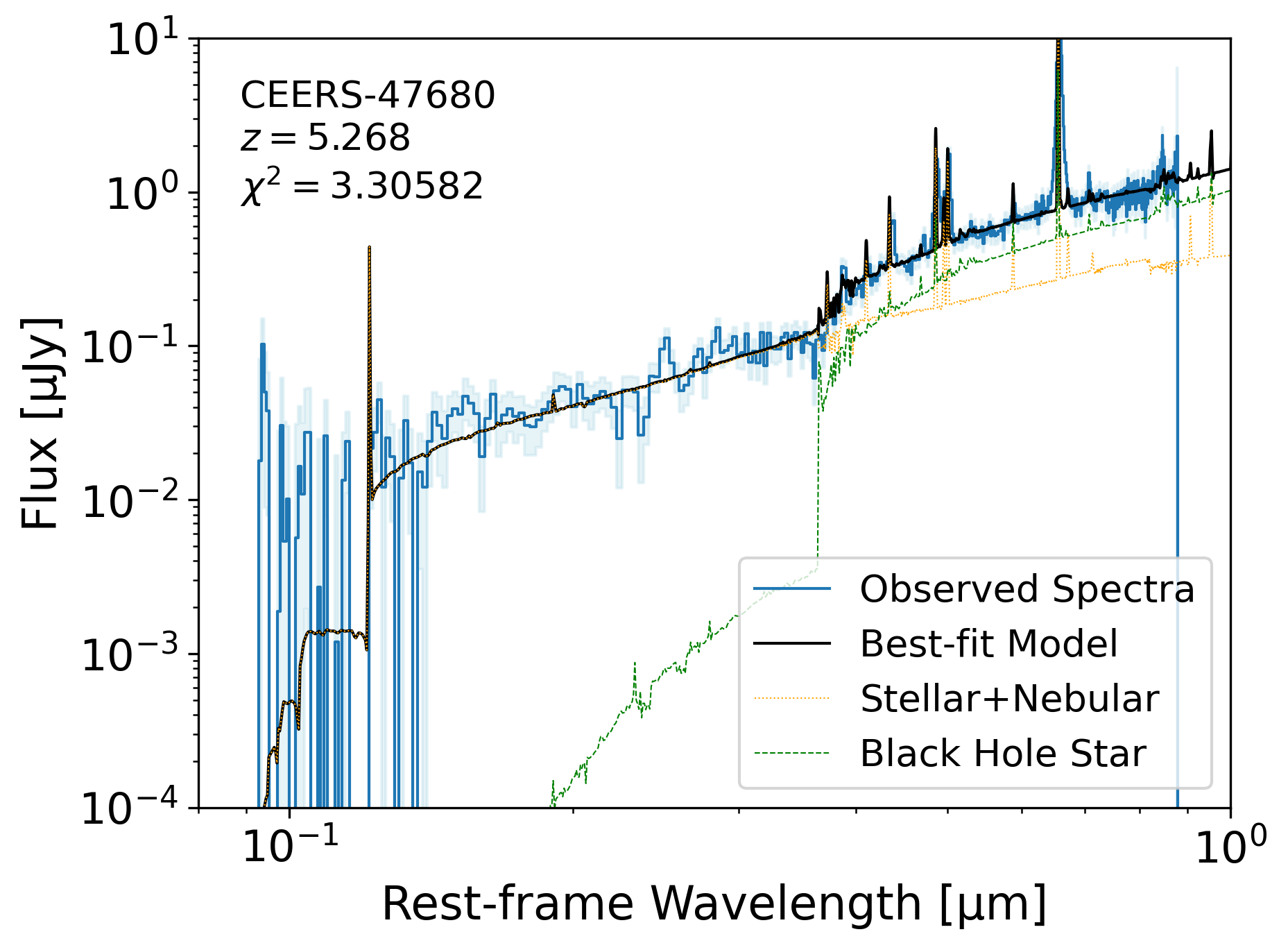}
    \end{minipage}
    \hfill
    \begin{minipage}[t]{0.32\linewidth}
        \centering
        \includegraphics[width=\linewidth]{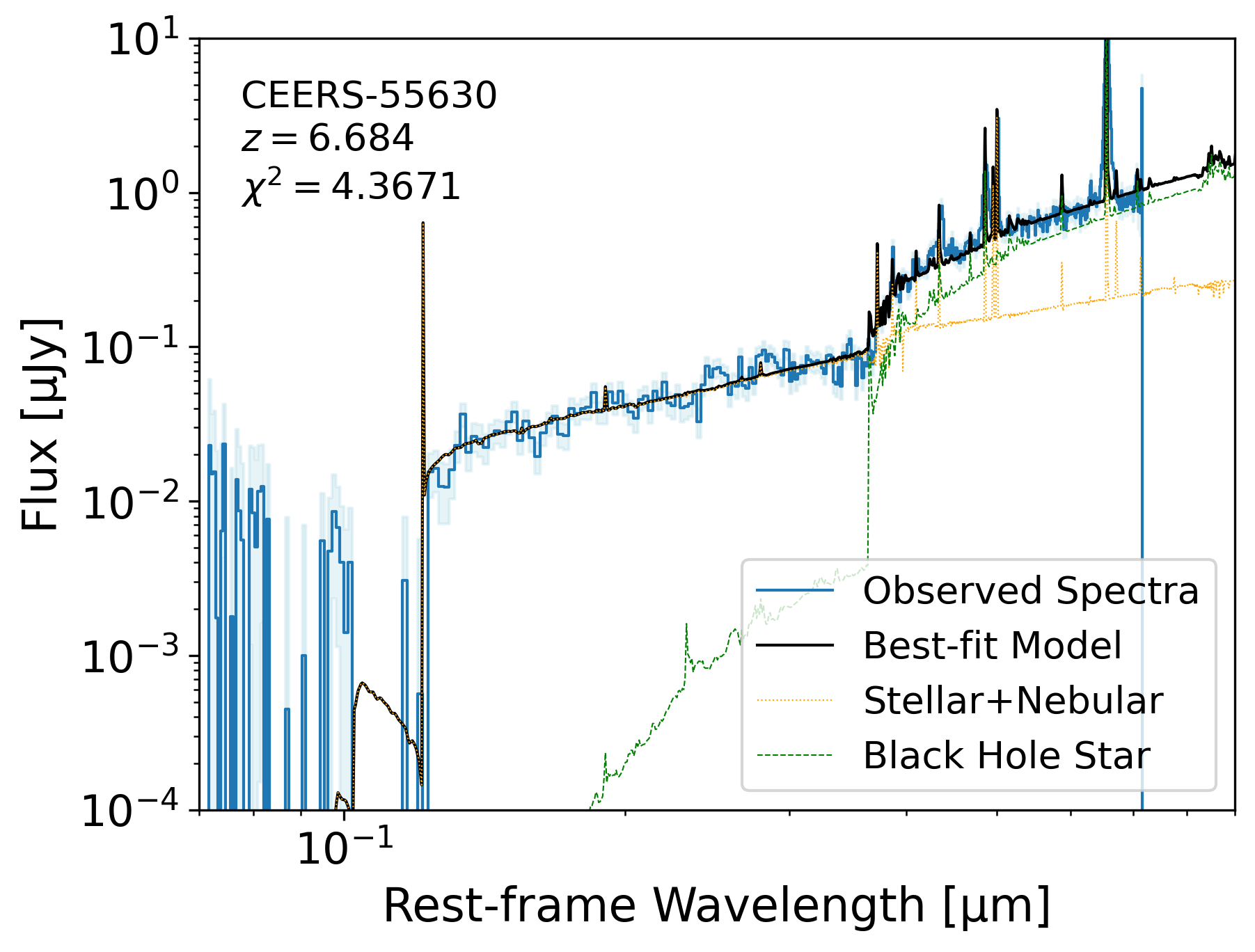}
    \end{minipage}
    \caption{Best-fit SEDs for the first half of our full LRD sample. The colored lines represent the same physical components as defined in Figure \ref{fig:example}.}
    \label{fig:12}
\end{figure}

\begin{figure}[htbp]
    \centering
    \begin{minipage}[t]{0.32\linewidth}
        \centering
        \includegraphics[width=\linewidth]{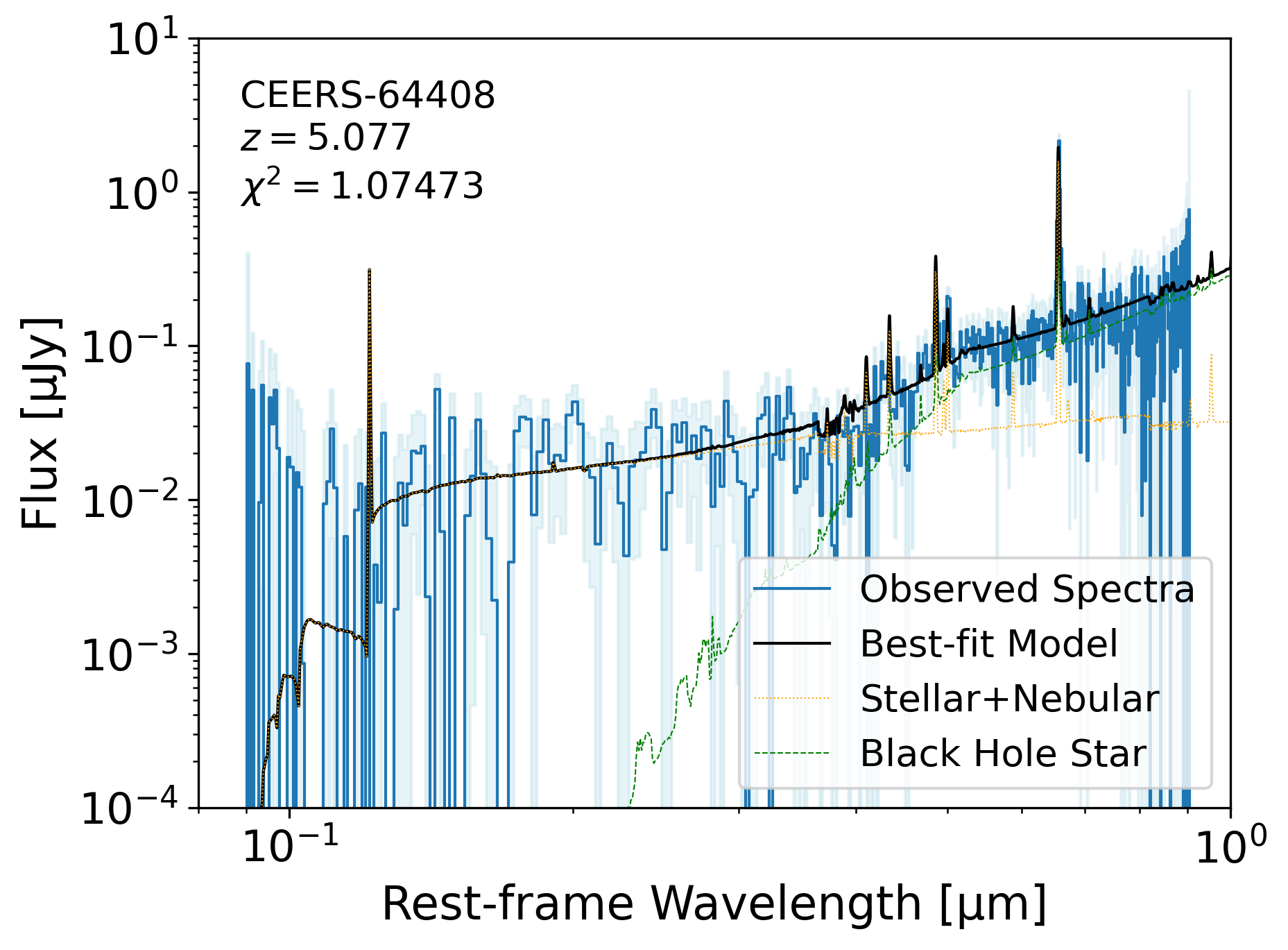}
    \end{minipage}
    \hfill
    \begin{minipage}[t]{0.32\linewidth}
        \centering
        \includegraphics[width=\linewidth]{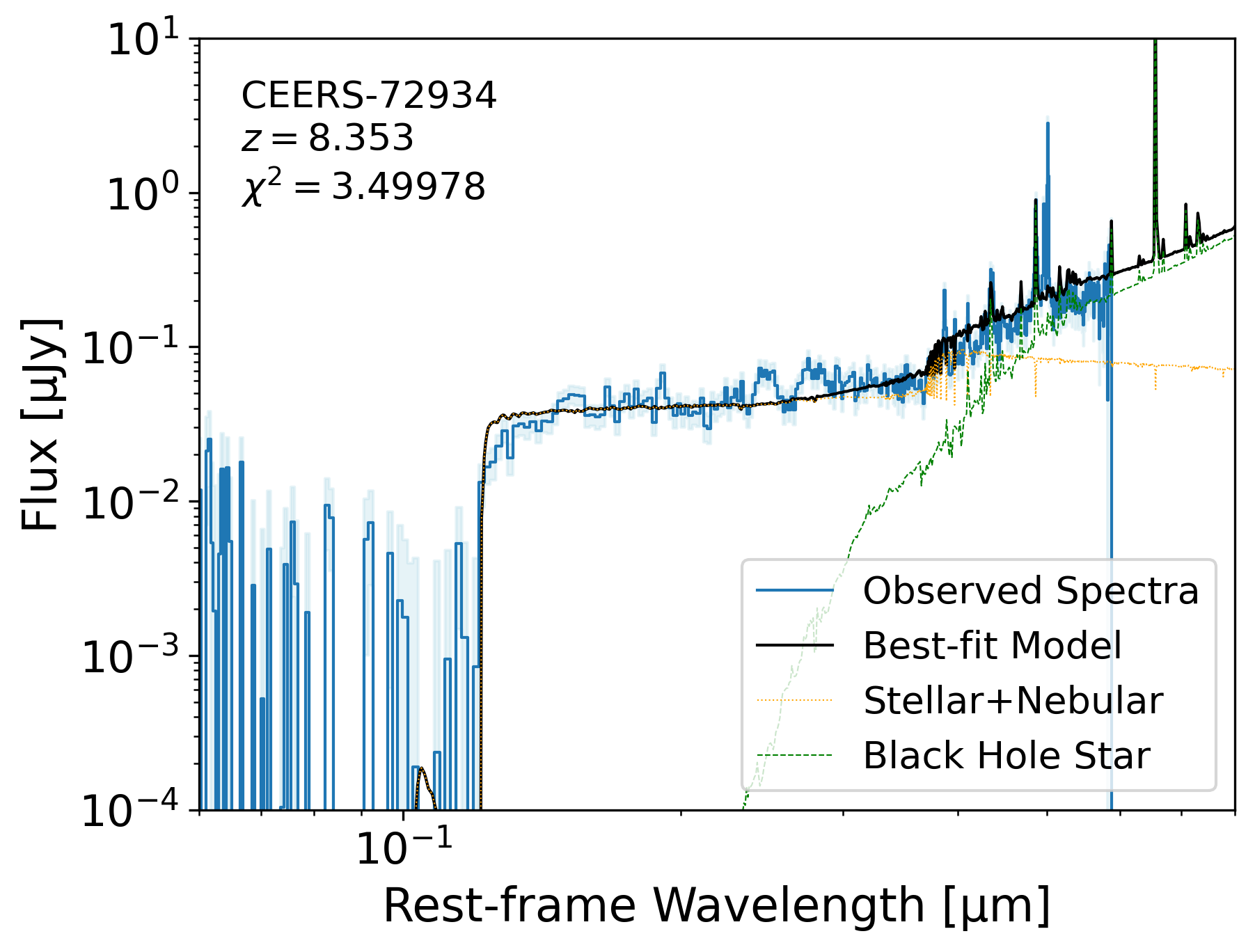}
    \end{minipage}
    \hfill
    \begin{minipage}[t]{0.32\linewidth}
        \centering
        \includegraphics[width=\linewidth]{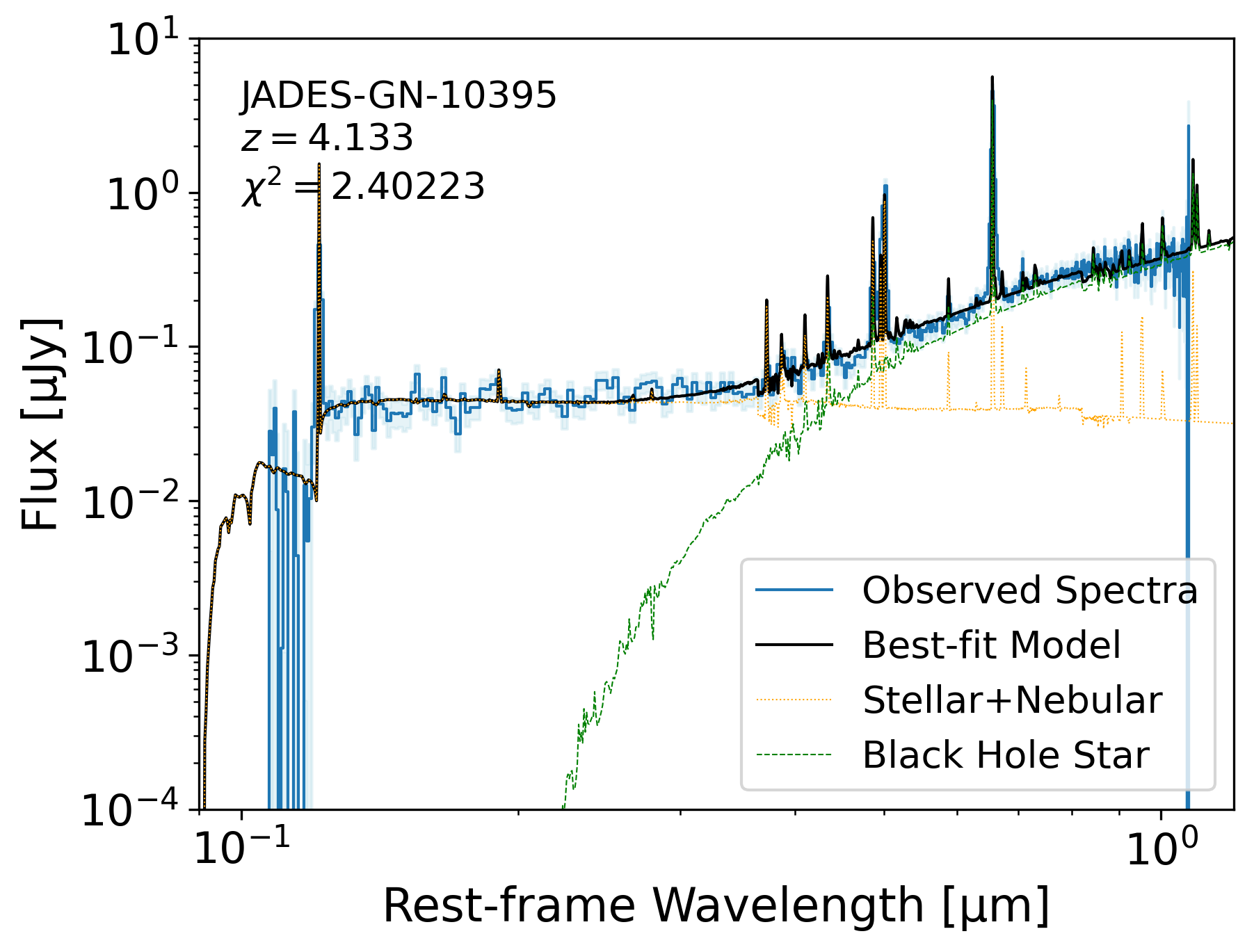}
    \end{minipage}

    \vspace{0.5cm}

    \begin{minipage}[t]{0.32\linewidth}
        \centering
        \includegraphics[width=\linewidth]{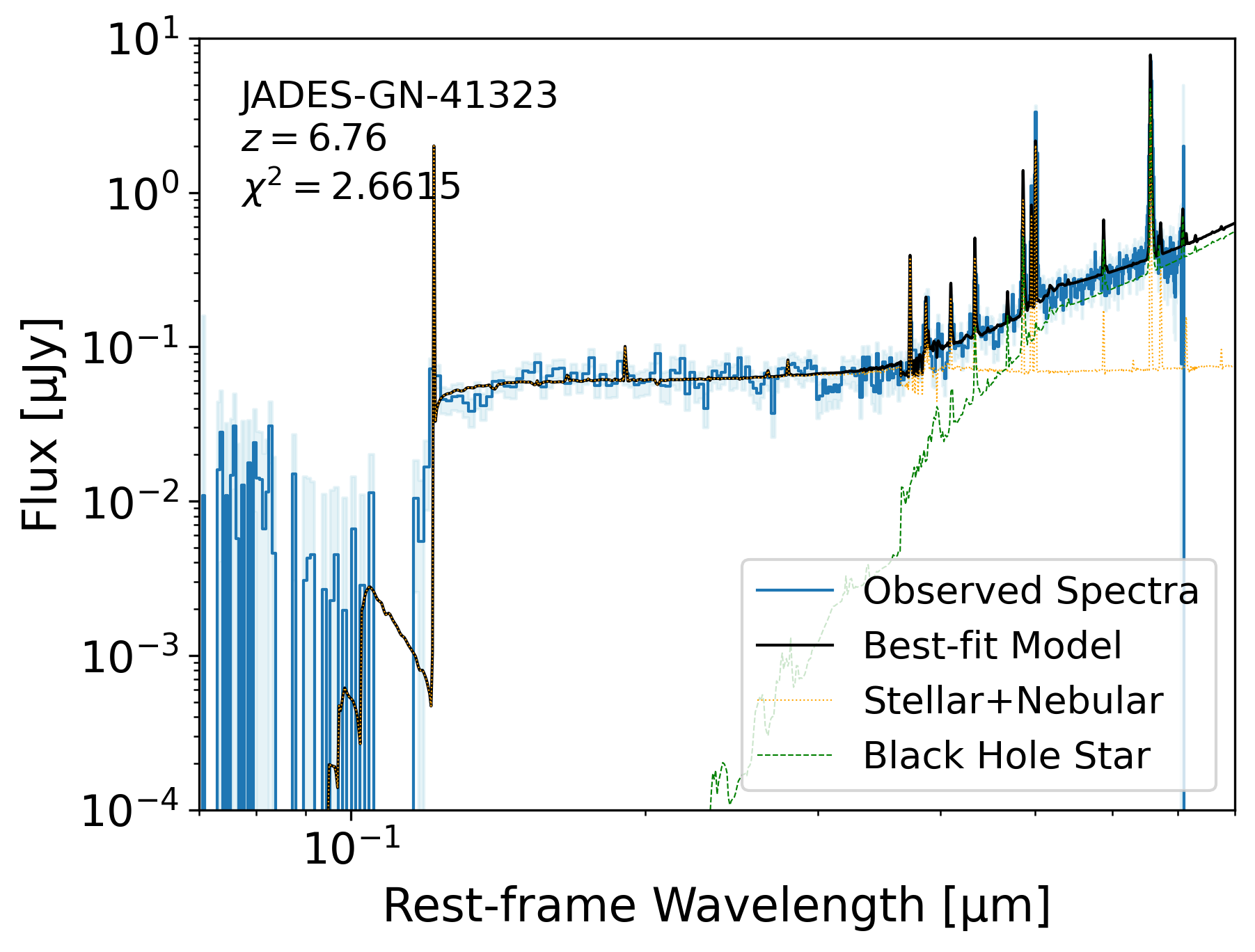}
    \end{minipage}
    \hfill
    \begin{minipage}[t]{0.32\linewidth}
        \centering
        \includegraphics[width=\linewidth]{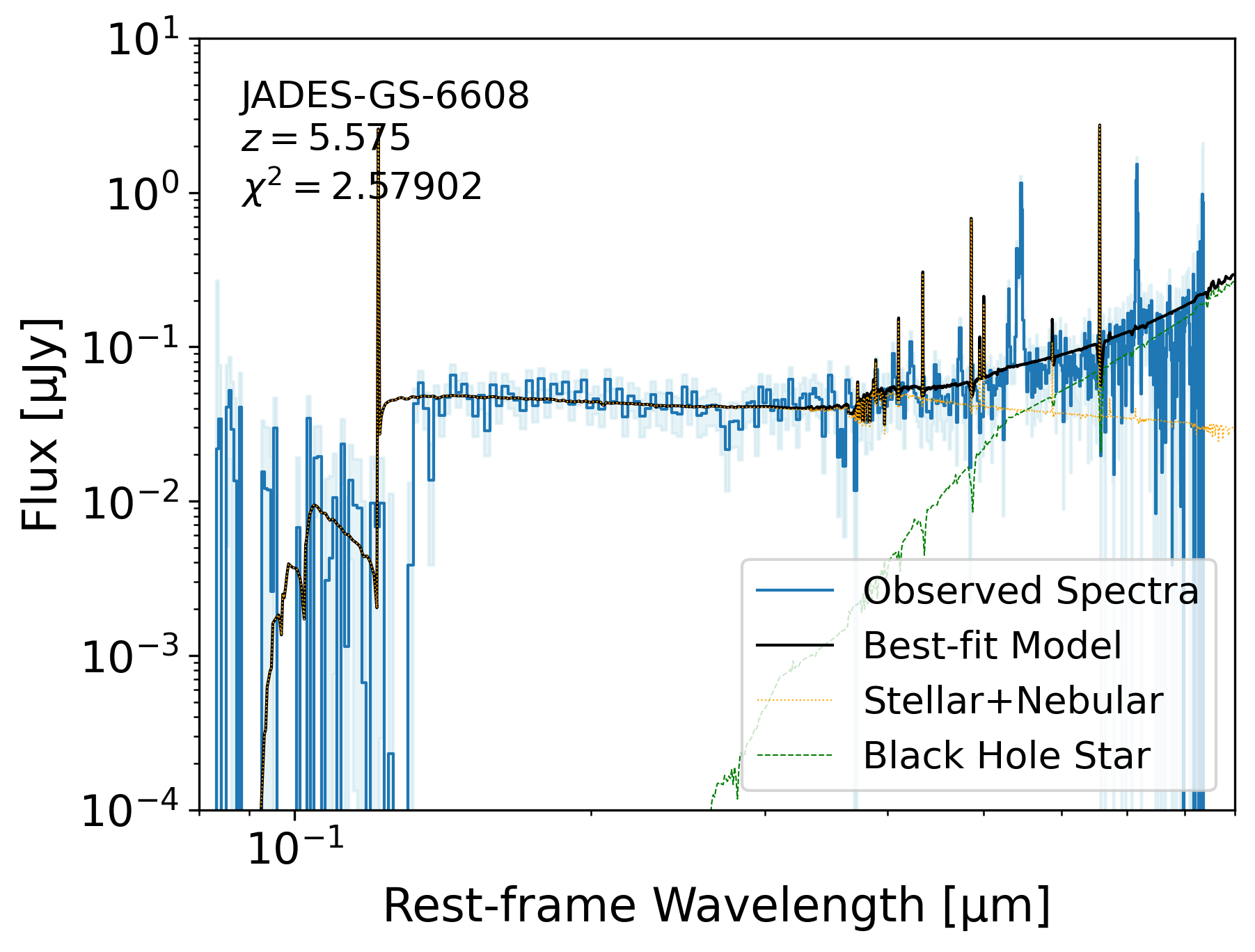}
    \end{minipage}
    \hfill
    \begin{minipage}[t]{0.32\linewidth}
        \centering
        \includegraphics[width=\linewidth]{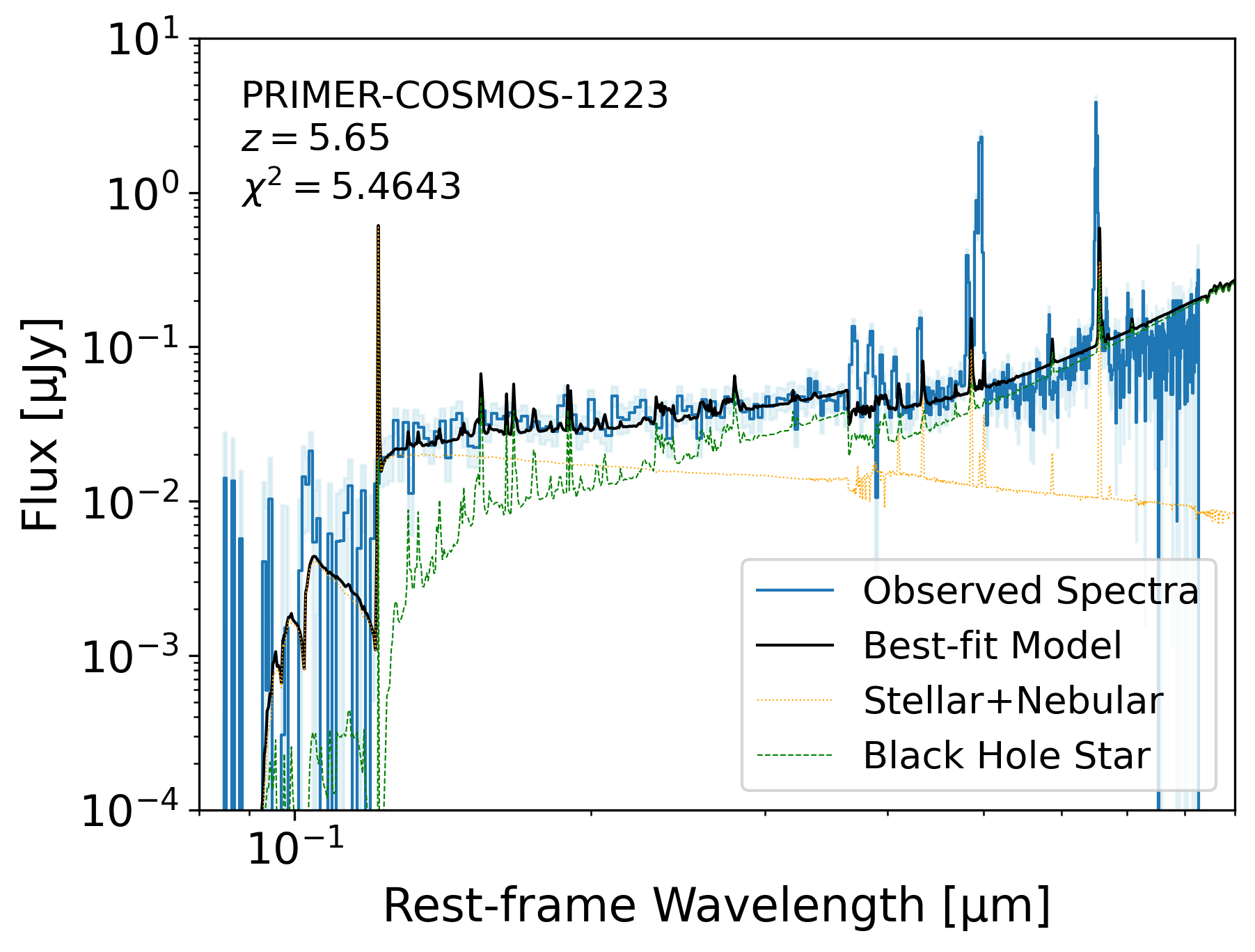}
    \end{minipage}

    \vspace{0.5cm}

    \begin{minipage}[t]{0.32\linewidth}
        \centering
        \includegraphics[width=\linewidth]{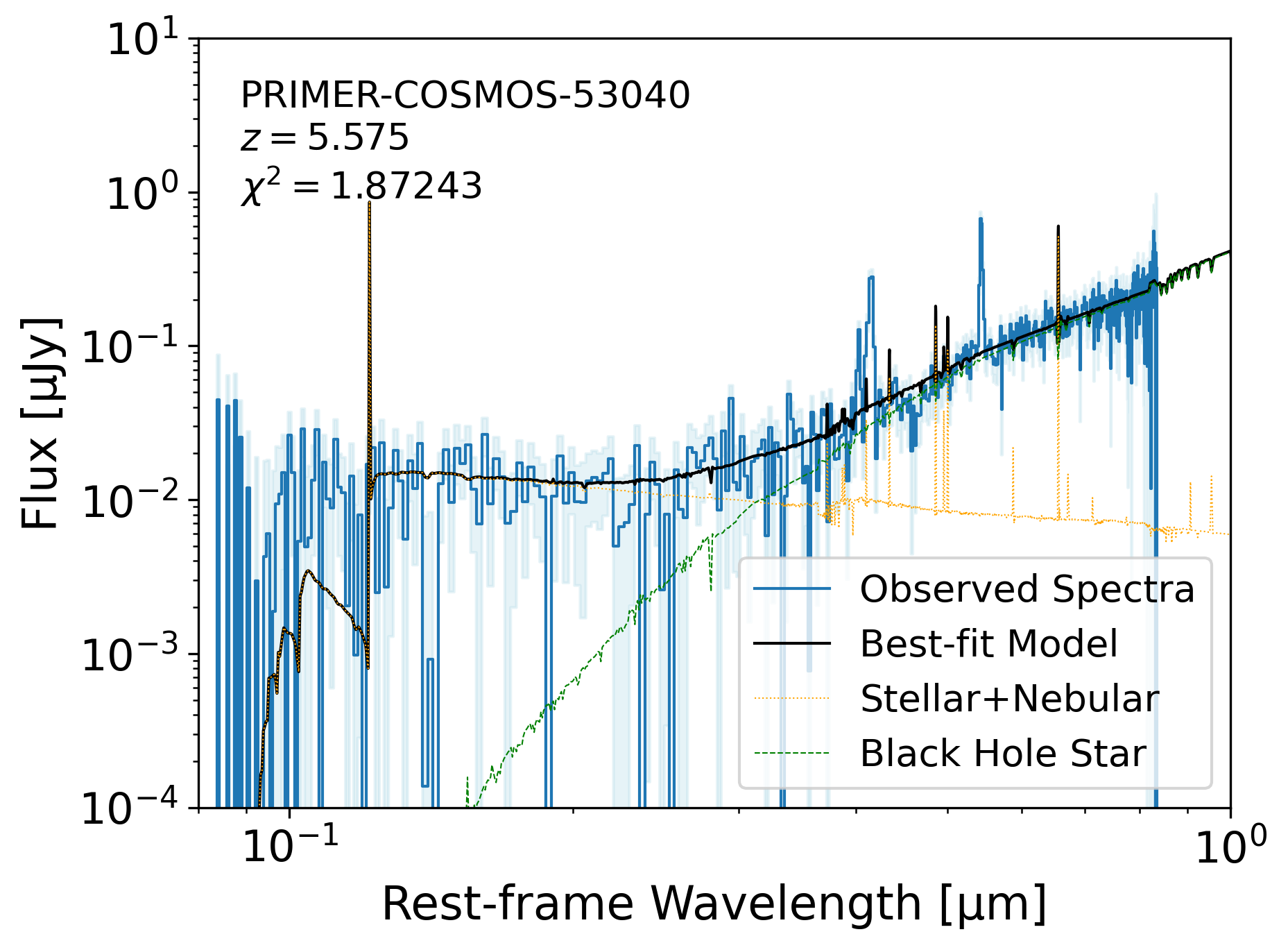}
    \end{minipage}
    \hfill
    \begin{minipage}[t]{0.32\linewidth}
        \centering
        \includegraphics[width=\linewidth]{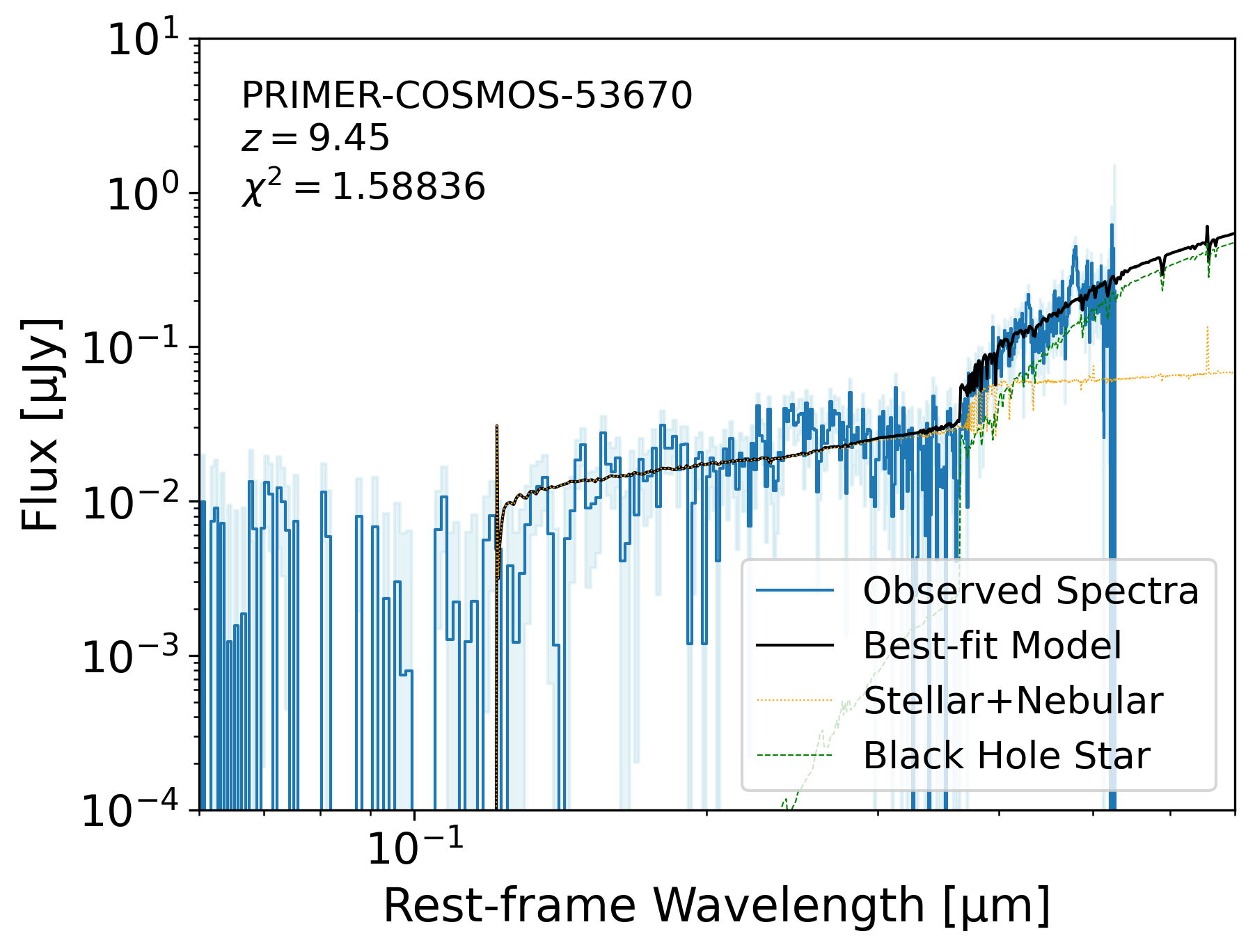}
    \end{minipage}
    \hfill
    \begin{minipage}[t]{0.32\linewidth}
        \centering
        \includegraphics[width=\linewidth]{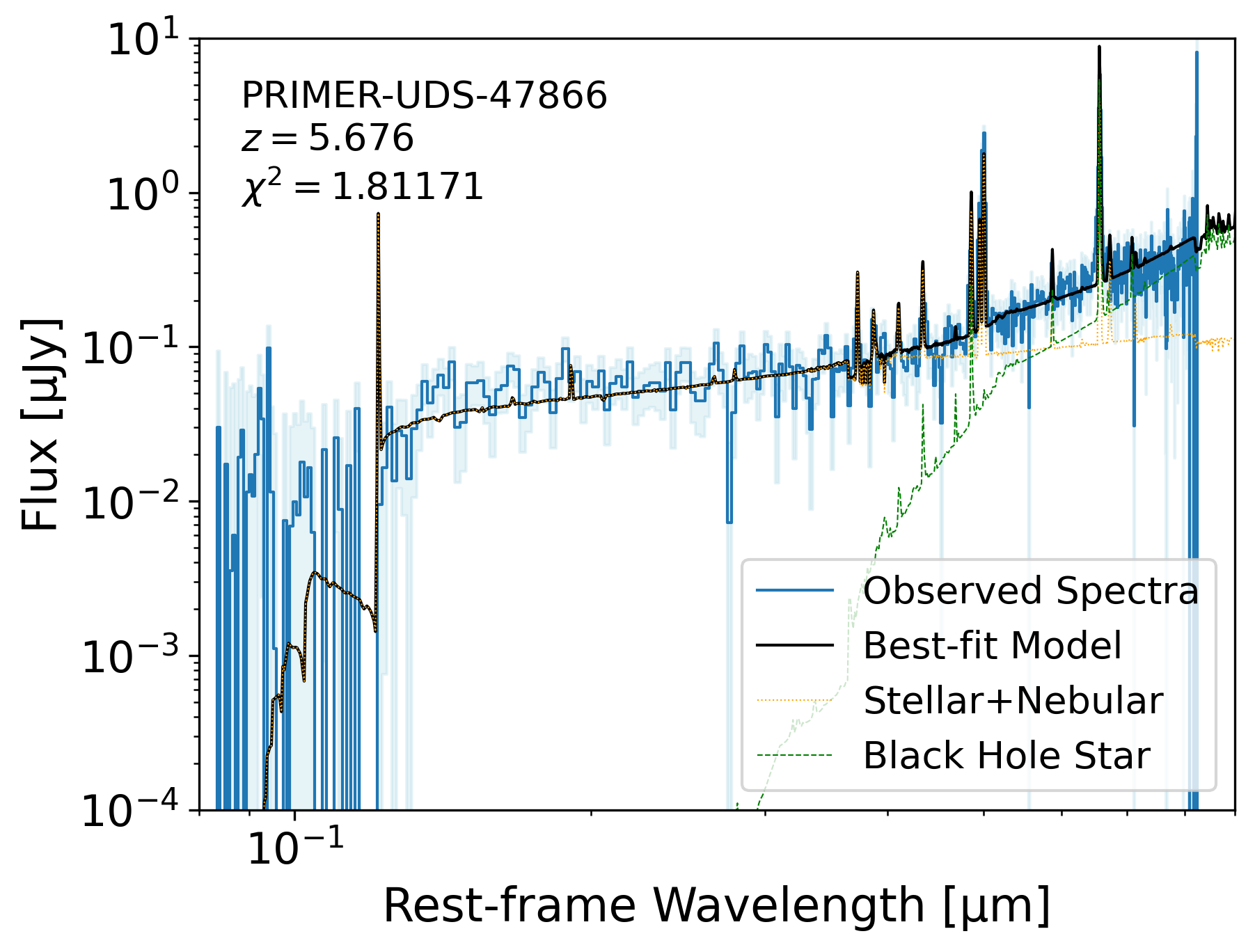}
    \end{minipage}

    \vspace{0.5cm}

    \begin{minipage}[t]{0.32\linewidth}
        \centering
        \includegraphics[width=\linewidth]{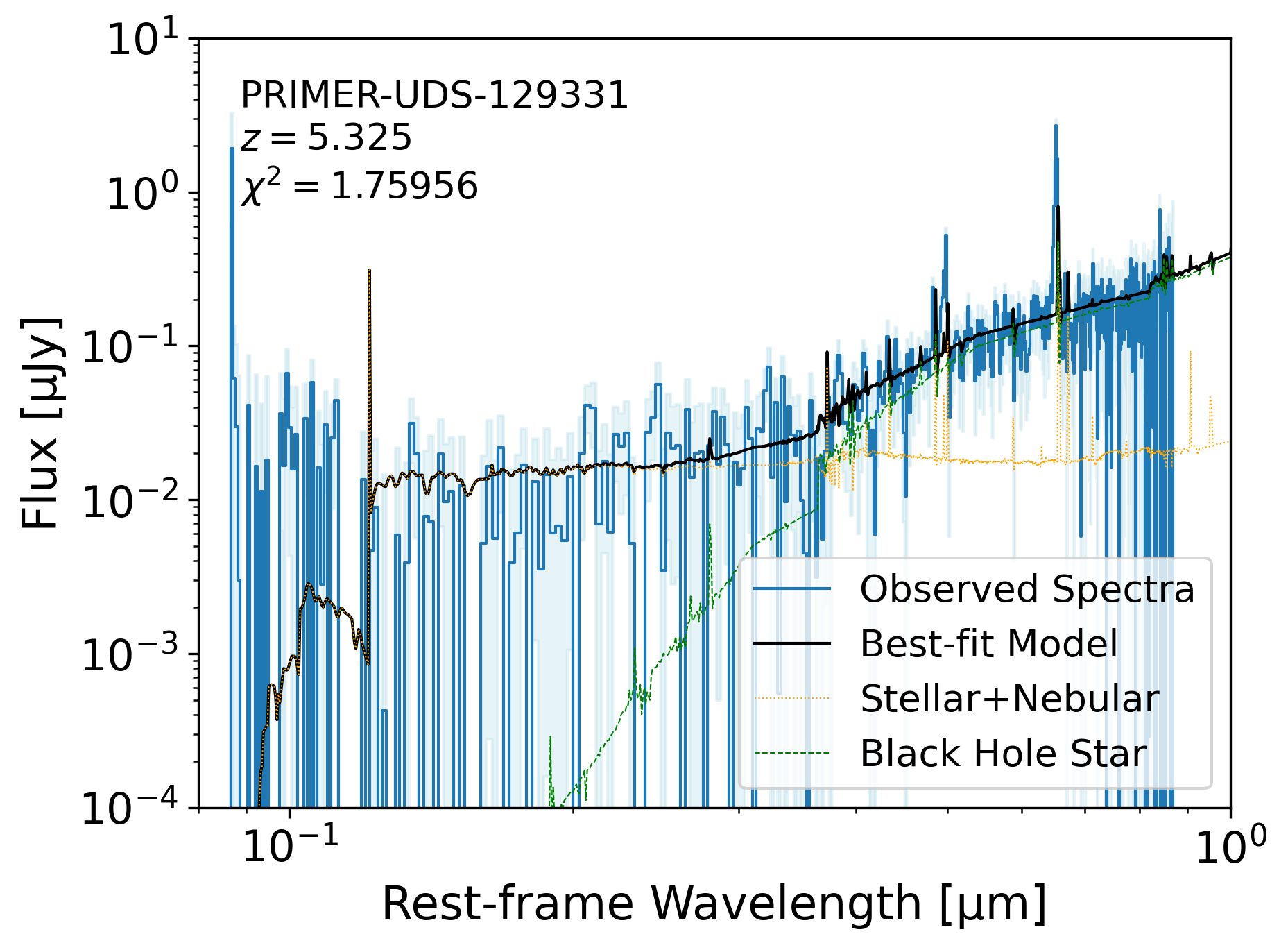}
    \end{minipage}

    \caption{Best-fit SEDs for the second half of our full LRD sample. The colored lines represent the same physical components as defined in Figure \ref{fig:example}. }
    \label{fig:24}
\end{figure}

\bibliography{sample701}{}

@ARTICLE{Merlin2024,
       author = {{Merlin}, E. and {Santini}, P. and {Paris}, D. and {Castellano}, M. and {Fontana}, A. and {Treu}, T. and {Finkelstein}, S.~L. and {Dunlop}, J.~S. and {Arrabal Haro}, P. and {Bagley}, M. and {Boyett}, K. and {Calabr{\`o}}, A. and {Correnti}, M. and {Davis}, K. and {Dickinson}, M. and {Donnan}, C.~T. and {Ferguson}, H.~C. and {Fortuni}, F. and {Giavalisco}, M. and {Glazebrook}, K. and {Grazian}, A. and {Grogin}, N.~A. and {Hathi}, N. and {Hirschmann}, M. and {Kartaltepe}, J.~S. and {Kewley}, L.~J. and {Kirkpatrick}, A. and {Kocevski}, D.~D. and {Koekemoer}, A.~M. and {Leung}, G. and {Lotz}, J.~M. and {Lucas}, R.~A. and {Magee}, D.~K. and {Marchesini}, D. and {Mascia}, S. and {McLeod}, D.~J. and {McLure}, R.~J. and {Nanayakkara}, T. and {Napolitano}, L. and {Nonino}, M. and {Papovich}, C. and {Pentericci}, L. and {P{\'e}rez-Gonz{\'a}lez}, P.~G. and {Pirzkal}, N. and {Ravindranath}, S. and {Roberts-Borsani}, G. and {Somerville}, R.~S. and {Trenti}, M. and {Trump}, J.~R. and {Vulcani}, B. and {Wang}, X. and {Watson}, P.~J. and {Wilkins}, S.~M. and {Yang}, G. and {Yung}, L.~Y.~A.},
        title = "{ASTRODEEP-JWST: NIRCam-HST multi-band photometry and redshifts for half a million sources in six extragalactic deep fields}",
      journal = {\aap},
         year = 2024,
        month = nov,
       volume = {691},
          eid = {A240},
        pages = {A240},
          doi = {10.1051/0004-6361/202451409},
archivePrefix = {arXiv},
       eprint = {2409.00169},
 primaryClass = {astro-ph.GA},
       adsurl = {https://ui.adsabs.harvard.edu/abs/2024A&A...691A.240M}
}

@ARTICLE{2022ApJ...935..110T,
       author = {{Treu}, T. and {Roberts-Borsani}, G. and {Bradac}, M. and {Brammer}, G. and {Fontana}, A. and {Henry}, A. and {Mason}, C. and {Morishita}, T. and {Pentericci}, L. and {Wang}, X. and {Acebron}, A. and {Bagley}, M. and {Bergamini}, P. and {Belfiori}, D. and {Bonchi}, A. and {Boyett}, K. and {Boutsia}, K. and {Calabr{\'o}}, A. and {Caminha}, G.~B. and {Castellano}, M. and {Dressler}, A. and {Glazebrook}, K. and {Grillo}, C. and {Jacobs}, C. and {Jones}, T. and {Kelly}, P.~L. and {Leethochawalit}, N. and {Malkan}, M.~A. and {Marchesini}, D. and {Mascia}, S. and {Mercurio}, A. and {Merlin}, E. and {Nanayakkara}, T. and {Nonino}, M. and {Paris}, D. and {Poggianti}, B. and {Rosati}, P. and {Santini}, P. and {Scarlata}, C. and {Shipley}, H.~V. and {Strait}, V. and {Trenti}, M. and {Tubthong}, C. and {Vanzella}, E. and {Vulcani}, B. and {Yang}, L.},
        title = "{The GLASS-JWST Early Release Science Program. I. Survey Design and Release Plans}",
      journal = {\apj},
         year = 2022,
        month = aug,
       volume = {935},
       number = {2},
          eid = {110},
        pages = {110},
          doi = {10.3847/1538-4357/ac8158},
archivePrefix = {arXiv},
       eprint = {2206.07978},
 primaryClass = {astro-ph.GA},
       adsurl = {https://ui.adsabs.harvard.edu/abs/2022ApJ...935..110T}
}

@ARTICLE{2024ApJ...974...92B,
       author = {{Bezanson}, Rachel and {Labbe}, Ivo and {Whitaker}, Katherine E. and {Leja}, Joel and {Price}, Sedona H. and {Franx}, Marijn and {Brammer}, Gabriel and {Marchesini}, Danilo and {Zitrin}, Adi and {Wang}, Bingjie and {Weaver}, John R. and {Furtak}, Lukas J. and {Atek}, Hakim and {Coe}, Dan and {Cutler}, Sam E. and {Dayal}, Pratika and {van Dokkum}, Pieter and {Feldmann}, Robert and {F{\"o}rster Schreiber}, Natascha M. and {Fujimoto}, Seiji and {Geha}, Marla and {Glazebrook}, Karl and {de Graaff}, Anna and {Greene}, Jenny E. and {Juneau}, St{\'e}phanie and {Kassin}, Susan and {Kriek}, Mariska and {Khullar}, Gourav and {Maseda}, Michael and {Mowla}, Lamiya A. and {Muzzin}, Adam and {Nanayakkara}, Themiya and {Nelson}, Erica J. and {Oesch}, Pascal A. and {Pacifici}, Camilla and {Pan}, Richard and {Papovich}, Casey and {Setton}, David J. and {Shapley}, Alice E. and {Smit}, Renske and {Stefanon}, Mauro and {Taylor}, Edward N. and {Williams}, Christina C.},
        title = "{The JWST UNCOVER Treasury Survey: Ultradeep NIRSpec and NIRCam Observations before the Epoch of Reionization}",
      journal = {\apj},
         year = 2024,
        month = oct,
       volume = {974},
       number = {1},
          eid = {92},
        pages = {92},
          doi = {10.3847/1538-4357/ad66cf},
archivePrefix = {arXiv},
       eprint = {2212.04026},
 primaryClass = {astro-ph.GA},
       adsurl = {https://ui.adsabs.harvard.edu/abs/2024ApJ...974...92B}
}

@ARTICLE{Morishita_2025,
       author = {{Morishita}, Takahiro and {Mason}, Charlotte A. and {Kreilgaard}, Kimi C. and {Trenti}, Michele and {Treu}, Tommaso and {Vulcani}, Benedetta and {Zhang}, Yechi and {Abdurro'uf} and {Alavi}, Anahita and {Atek}, Hakim and {Bah{\'e}}, Yannick and {Brada{\v{c}}}, Maru{\v{s}}a and {Bradley}, Larry D. and {Bunker}, Andrew J. and {Coe}, Dan and {Colbert}, James and {Gelli}, Viola and {Hayes}, Matthew J. and {Jones}, Tucker and {Kodama}, Tadayuki and {Leethochawalit}, Nicha and {Liu}, Zhaoran and {Malkan}, Matthew A. and {Mehta}, Vihang and {Metha}, Benjamin and {Newman}, Andrew B. and {Rafelski}, Marc and {Roberts-Borsani}, Guido and {Rutkowski}, Michael J. and {Scarlata}, Claudia and {Stiavelli}, Massimo and {Sutanto}, Ryo A. and {Takahashi}, Kosuke and {Teplitz}, Harry I. and {Wang}, Xin},
        title = "{BEACON: JWST NIRCam Pure-parallel Imaging Survey. I. Survey Design and Initial Results}",
      journal = {\apj},
         year = 2025,
        month = apr,
       volume = {983},
       number = {2},
          eid = {152},
        pages = {152},
          doi = {10.3847/1538-4357/adbbdc},
archivePrefix = {arXiv},
       eprint = {2412.04211},
 primaryClass = {astro-ph.GA},
       adsurl = {https://ui.adsabs.harvard.edu/abs/2025ApJ...983..152M}
}

@ARTICLE{eisenstein2023overviewjwstadvanceddeep,
       author = {{Eisenstein}, Daniel J. and {Willott}, Chris and {Alberts}, Stacey and {Arribas}, Santiago and {Bonaventura}, Nina and {Bunker}, Andrew J. and {Cameron}, Alex J. and {Carniani}, Stefano and {Charlot}, Stephane and {Curtis-Lake}, Emma and {D'Eugenio}, Francesco and {Ferruit}, Pierre and {Giardino}, Giovanna and {Hainline}, Kevin and {Hausen}, Ryan and {Jakobsen}, Peter and {Johnson}, Benjamin D. and {Maiolino}, Roberto and {Rauscher}, Bernard J. and {Rieke}, Marcia and {Rieke}, George and {Rix}, Hans-Walter and {Robertson}, Brant and {Stark}, Daniel P. and {Tacchella}, Sandro and {Williams}, Christina C. and {Willmer}, Christopher N.~A. and {Baker}, William M. and {Baum}, Stefi and {Bhatawdekar}, Rachana and {Boyett}, Kristan and {Chen}, Zuyi and {Chevallard}, Jacopo and {Circosta}, Chiara and {Curti}, Mirko and {Danhaive}, A. Lola and {DeCoursey}, Christa and {Endsley}, Ryan and {de Graaff}, Anna and {Dressler}, Alan and {Egami}, Eiichi and {Helton}, Jakob M. and {Hviding}, Raphael E. and {Ji}, Zhiyuan and {Jones}, Gareth C. and {Kumari}, Nimisha and {L{\"u}tzgendorf}, Nora and {Laseter}, Isaac and {Looser}, Tobias J. and {Lyu}, Jianwei and {Maseda}, Michael V. and {Nelson}, Erica and {Parlanti}, Eleonora and {Perna}, Michele and {Pusk{\'a}s}, D{\'a}vid and {Rawle}, Tim and {Rodr{\'\i}guez Del Pino}, Bruno and {Rujopakarn}, Wiphu and {Sandles}, Lester and {Saxena}, Aayush and {Scholtz}, Jan and {Sharpe}, Katherine and {Shivaei}, Irene and {Silcock}, Maddie S. and {Simmonds}, Charlotte and {Skarbinski}, Maya and {Smit}, Renske and {Stone}, Meredith and {Suess}, Katherine A. and {Sun}, Fengwu and {Tang}, Mengtao and {Topping}, Michael W. and {{\"U}bler}, Hannah and {Villanueva}, Natalia C. and {Wallace}, Imaan E.~B. and {Whitler}, Lily and {Witstok}, Joris and {Woodrum}, Charity},
        title = "{Overview of the JWST Advanced Deep Extragalactic Survey (JADES)}",
      journal = {\apjs},
         year = 2026,
        month = mar,
       volume = {283},
       number = {1},
          eid = {6},
        pages = {6},
          doi = {10.3847/1538-4365/ae3163},
archivePrefix = {arXiv},
       eprint = {2306.02465},
 primaryClass = {astro-ph.GA},
       adsurl = {https://ui.adsabs.harvard.edu/abs/2026ApJS..283....6E}
}

@ARTICLE{10.1093/mnras/stad2411,
       author = {{Oesch}, P.~A. and {Brammer}, G. and {Naidu}, R.~P. and {Bouwens}, R.~J. and {Chisholm}, J. and {Illingworth}, G.~D. and {Matthee}, J. and {Nelson}, E. and {Qin}, Y. and {Reddy}, N. and {Shapley}, A. and {Shivaei}, I. and {van Dokkum}, P. and {Weibel}, A. and {Whitaker}, K. and {Wuyts}, S. and {Covelo-Paz}, A. and {Endsley}, R. and {Fudamoto}, Y. and {Giovinazzo}, E. and {Herard-Demanche}, T. and {Kerutt}, J. and {Kramarenko}, I. and {Labbe}, I. and {Leonova}, E. and {Lin}, J. and {Magee}, D. and {Marchesini}, D. and {Maseda}, M. and {Mason}, C. and {Matharu}, J. and {Meyer}, R.~A. and {Neufeld}, C. and {Prieto Lyon}, G. and {Schaerer}, D. and {Sharma}, R. and {Shuntov}, M. and {Smit}, R. and {Stefanon}, M. and {Wyithe}, J.~S.~B. and {Xiao}, M.},
        title = "{The JWST FRESCO survey: legacy NIRCam/grism spectroscopy and imaging in the two GOODS fields}",
      journal = {\mnras},
         year = 2023,
        month = oct,
       volume = {525},
       number = {2},
        pages = {2864-2874},
          doi = {10.1093/mnras/stad2411},
archivePrefix = {arXiv},
       eprint = {2304.02026},
 primaryClass = {astro-ph.GA},
       adsurl = {https://ui.adsabs.harvard.edu/abs/2023MNRAS.525.2864O}
}

@ARTICLE{Finkelstein_2025,
       author = {{Finkelstein}, Steven L. and {Bagley}, Micaela B. and {Arrabal Haro}, Pablo and {Dickinson}, Mark and {Ferguson}, Henry C. and {Kartaltepe}, Jeyhan S. and {Kocevski}, Dale D. and {Koekemoer}, Anton M. and {Lotz}, Jennifer M. and {Papovich}, Casey and {P{\'e}rez-Gonz{\'a}lez}, Pablo G. and {Pirzkal}, Nor and {Somerville}, Rachel S. and {Trump}, Jonathan R. and {Yang}, Guang and {Yung}, L.~Y. Aaron and {Fontana}, Adriano and {Grazian}, Andrea and {Grogin}, Norman A. and {Kewley}, Lisa J. and {Kirkpatrick}, Allison and {Larson}, Rebecca L. and {Pentericci}, Laura and {Ravindranath}, Swara and {Wilkins}, Stephen M. and {Almaini}, Omar and {Amor{\'\i}n}, Ricardo O. and {Barro}, Guillermo and {Bhatawdekar}, Rachana and {Bisigello}, Laura and {Brooks}, Madisyn and {Buat}, V{\'e}ronique and {Buitrago}, Fernando and {Burgarella}, Denis and {Calabr{\`o}}, Antonello and {Castellano}, Marco and {Cheng}, Yingjie and {Cleri}, Nikko J. and {Cole}, Justin W. and {Cooper}, M.~C. and {Cooper}, Olivia R. and {Costantin}, Luca and {Cox}, Isa G. and {Croton}, Darren and {Daddi}, Emanuele and {Davis}, Kelcey and {Dekel}, Avishai and {Elbaz}, David and {Fern{\'a}ndez}, Vital and {Fujimoto}, Seiji and {Gandolfi}, Giovanni and {Gardner}, Jonathan P. and {Gawiser}, Eric and {Giavalisco}, Mauro and {G{\'o}mez-Guijarro}, Carlos and {Guo}, Yuchen and {Gupta}, Ansh R. and {Hathi}, Nimish P. and {Harish}, Santosh and {Henry}, Aur{\'e}lien and {Hirschmann}, Michaela and {Hu}, Weida and {Hutchison}, Taylor A. and {Iyer}, Kartheik G. and {Jaskot}, Anne E. and {Jha}, Saurabh W. and {Jung}, Intae and {Kassin}, Susan A. and {Kokorev}, Vasily and {Kurczynski}, Peter and {Leung}, Gene C.~K. and {Llerena}, Mario and {Long}, Arianna S. and {Lucas}, Ray A. and {Lu}, Shiying and {McGrath}, Elizabeth J. and {McIntosh}, Daniel H. and {Merlin}, Emiliano and {Mobasher}, Bahram and {Morales}, Alexa M. and {Napolitano}, Lorenzo and {Pacucci}, Fabio and {Pandya}, Viraj and {Rafelski}, Marc and {Rodighiero}, Giulia and {Rose}, Caitlin and {Santini}, Paola and {Seill{\'e}}, Lise-Marie and {Simons}, Raymond C. and {Shen}, Lu and {Straughn}, Amber N. and {Tacchella}, Sandro and {Taylor}, Anthony J. and {Vanderhoof}, Brittany N. and {Vega-Ferrero}, Jes{\'u}s and {Weiner}, Benjamin J. and {Willmer}, Christopher N.~A. and {Zhu}, Peixin and {Bell}, Eric F. and {Wuyts}, Stijn and {Holwerda}, Benne W. and {Wang}, Xin and {Wang}, Weichen and {Zavala}, Jorge A. and {CEERS Collaboration}},
        title = "{The Cosmic Evolution Early Release Science Survey (CEERS)}",
      journal = {\apjl},
         year = 2025,
        month = apr,
       volume = {983},
       number = {1},
          eid = {L4},
        pages = {L4},
          doi = {10.3847/2041-8213/adbbd3},
archivePrefix = {arXiv},
       eprint = {2501.04085},
 primaryClass = {astro-ph.GA},
       adsurl = {https://ui.adsabs.harvard.edu/abs/2025ApJ...983L...4F}
}

@ARTICLE{Bagley_2024,
       author = {{Bagley}, Micaela B. and {Pirzkal}, Nor and {Finkelstein}, Steven L. and {Papovich}, Casey and {Berg}, Danielle A. and {Lotz}, Jennifer M. and {Leung}, Gene C.~K. and {Ferguson}, Henry C. and {Koekemoer}, Anton M. and {Dickinson}, Mark and {Kartaltepe}, Jeyhan S. and {Kocevski}, Dale D. and {Somerville}, Rachel S. and {Yung}, L.~Y. Aaron and {Backhaus}, Bren E. and {Casey}, Caitlin M. and {Castellano}, Marco and {Ch{\'a}vez Ortiz}, {\'O}scar A. and {Chworowsky}, Katherine and {Cox}, Isabella G. and {Dav{\'e}}, Romeel and {Davis}, Kelcey and {Estrada-Carpenter}, Vicente and {Fontana}, Adriano and {Fujimoto}, Seiji and {Gardner}, Jonathan P. and {Giavalisco}, Mauro and {Grazian}, Andrea and {Grogin}, Norman A. and {Hathi}, Nimish P. and {Hutchison}, Taylor A. and {Jaskot}, Anne E. and {Jung}, Intae and {Kewley}, Lisa J. and {Kirkpatrick}, Allison and {Larson}, Rebecca L. and {Matharu}, Jasleen and {Natarajan}, Priyamvada and {Pentericci}, Laura and {P{\'e}rez-Gonz{\'a}lez}, Pablo G. and {Ravindranath}, Swara and {Rothberg}, Barry and {Ryan}, Russell and {Shen}, Lu and {Simons}, Raymond C. and {Snyder}, Gregory F. and {Trump}, Jonathan R. and {Wilkins}, Stephen M.},
        title = "{The Next Generation Deep Extragalactic Exploratory Public (NGDEEP) Survey}",
      journal = {\apjl},
         year = 2024,
        month = apr,
       volume = {965},
       number = {1},
          eid = {L6},
        pages = {L6},
          doi = {10.3847/2041-8213/ad2f31},
archivePrefix = {arXiv},
       eprint = {2302.05466},
 primaryClass = {astro-ph.GA},
       adsurl = {https://ui.adsabs.harvard.edu/abs/2024ApJ...965L...6B}
}

@ARTICLE{2011ApJS..197...35G,
       author = {{Grogin}, Norman A. and {Kocevski}, Dale D. and {Faber}, S.~M. and {Ferguson}, Henry C. and {Koekemoer}, Anton M. and {Riess}, Adam G. and {Acquaviva}, Viviana and {Alexander}, David M. and {Almaini}, Omar and {Ashby}, Matthew L.~N. and {Barden}, Marco and {Bell}, Eric F. and {Bournaud}, Fr{\'e}d{\'e}ric and {Brown}, Thomas M. and {Caputi}, Karina I. and {Casertano}, Stefano and {Cassata}, Paolo and {Castellano}, Marco and {Challis}, Peter and {Chary}, Ranga-Ram and {Cheung}, Edmond and {Cirasuolo}, Michele and {Conselice}, Christopher J. and {Roshan Cooray}, Asantha and {Croton}, Darren J. and {Daddi}, Emanuele and {Dahlen}, Tomas and {Dav{\'e}}, Romeel and {de Mello}, Du{\'\i}lia F. and {Dekel}, Avishai and {Dickinson}, Mark and {Dolch}, Timothy and {Donley}, Jennifer L. and {Dunlop}, James S. and {Dutton}, Aaron A. and {Elbaz}, David and {Fazio}, Giovanni G. and {Filippenko}, Alexei V. and {Finkelstein}, Steven L. and {Fontana}, Adriano and {Gardner}, Jonathan P. and {Garnavich}, Peter M. and {Gawiser}, Eric and {Giavalisco}, Mauro and {Grazian}, Andrea and {Guo}, Yicheng and {Hathi}, Nimish P. and {H{\"a}ussler}, Boris and {Hopkins}, Philip F. and {Huang}, Jia-Sheng and {Huang}, Kuang-Han and {Jha}, Saurabh W. and {Kartaltepe}, Jeyhan S. and {Kirshner}, Robert P. and {Koo}, David C. and {Lai}, Kamson and {Lee}, Kyoung-Soo and {Li}, Weidong and {Lotz}, Jennifer M. and {Lucas}, Ray A. and {Madau}, Piero and {McCarthy}, Patrick J. and {McGrath}, Elizabeth J. and {McIntosh}, Daniel H. and {McLure}, Ross J. and {Mobasher}, Bahram and {Moustakas}, Leonidas A. and {Mozena}, Mark and {Nandra}, Kirpal and {Newman}, Jeffrey A. and {Niemi}, Sami-Matias and {Noeske}, Kai G. and {Papovich}, Casey J. and {Pentericci}, Laura and {Pope}, Alexandra and {Primack}, Joel R. and {Rajan}, Abhijith and {Ravindranath}, Swara and {Reddy}, Naveen A. and {Renzini}, Alvio and {Rix}, Hans-Walter and {Robaina}, Aday R. and {Rodney}, Steven A. and {Rosario}, David J. and {Rosati}, Piero and {Salimbeni}, Sara and {Scarlata}, Claudia and {Siana}, Brian and {Simard}, Luc and {Smidt}, Joseph and {Somerville}, Rachel S. and {Spinrad}, Hyron and {Straughn}, Amber N. and {Strolger}, Louis-Gregory and {Telford}, Olivia and {Teplitz}, Harry I. and {Trump}, Jonathan R. and {van der Wel}, Arjen and {Villforth}, Carolin and {Wechsler}, Risa H. and {Weiner}, Benjamin J. and {Wiklind}, Tommy and {Wild}, Vivienne and {Wilson}, Grant and {Wuyts}, Stijn and {Yan}, Hao-Jing and {Yun}, Min S.},
        title = "{CANDELS: The Cosmic Assembly Near-infrared Deep Extragalactic Legacy Survey}",
      journal = {\apjs},
         year = 2011,
        month = dec,
       volume = {197},
       number = {2},
          eid = {35},
        pages = {35},
          doi = {10.1088/0067-0049/197/2/35},
archivePrefix = {arXiv},
       eprint = {1105.3753},
 primaryClass = {astro-ph.CO},
       adsurl = {https://ui.adsabs.harvard.edu/abs/2011ApJS..197...35G}
}

@ARTICLE{2011ApJS..197...36K,
       author = {{Koekemoer}, Anton M. and {Faber}, S.~M. and {Ferguson}, Henry C. and {Grogin}, Norman A. and {Kocevski}, Dale D. and {Koo}, David C. and {Lai}, Kamson and {Lotz}, Jennifer M. and {Lucas}, Ray A. and {McGrath}, Elizabeth J. and {Ogaz}, Sara and {Rajan}, Abhijith and {Riess}, Adam G. and {Rodney}, Steve A. and {Strolger}, Louis and {Casertano}, Stefano and {Castellano}, Marco and {Dahlen}, Tomas and {Dickinson}, Mark and {Dolch}, Timothy and {Fontana}, Adriano and {Giavalisco}, Mauro and {Grazian}, Andrea and {Guo}, Yicheng and {Hathi}, Nimish P. and {Huang}, Kuang-Han and {van der Wel}, Arjen and {Yan}, Hao-Jing and {Acquaviva}, Viviana and {Alexander}, David M. and {Almaini}, Omar and {Ashby}, Matthew L.~N. and {Barden}, Marco and {Bell}, Eric F. and {Bournaud}, Fr{\'e}d{\'e}ric and {Brown}, Thomas M. and {Caputi}, Karina I. and {Cassata}, Paolo and {Challis}, Peter J. and {Chary}, Ranga-Ram and {Cheung}, Edmond and {Cirasuolo}, Michele and {Conselice}, Christopher J. and {Roshan Cooray}, Asantha and {Croton}, Darren J. and {Daddi}, Emanuele and {Dav{\'e}}, Romeel and {de Mello}, Duilia F. and {de Ravel}, Loic and {Dekel}, Avishai and {Donley}, Jennifer L. and {Dunlop}, James S. and {Dutton}, Aaron A. and {Elbaz}, David and {Fazio}, Giovanni G. and {Filippenko}, Alexei V. and {Finkelstein}, Steven L. and {Frazer}, Chris and {Gardner}, Jonathan P. and {Garnavich}, Peter M. and {Gawiser}, Eric and {Gruetzbauch}, Ruth and {Hartley}, Will G. and {H{\"a}ussler}, Boris and {Herrington}, Jessica and {Hopkins}, Philip F. and {Huang}, Jia-Sheng and {Jha}, Saurabh W. and {Johnson}, Andrew and {Kartaltepe}, Jeyhan S. and {Khostovan}, Ali A. and {Kirshner}, Robert P. and {Lani}, Caterina and {Lee}, Kyoung-Soo and {Li}, Weidong and {Madau}, Piero and {McCarthy}, Patrick J. and {McIntosh}, Daniel H. and {McLure}, Ross J. and {McPartland}, Conor and {Mobasher}, Bahram and {Moreira}, Heidi and {Mortlock}, Alice and {Moustakas}, Leonidas A. and {Mozena}, Mark and {Nandra}, Kirpal and {Newman}, Jeffrey A. and {Nielsen}, Jennifer L. and {Niemi}, Sami and {Noeske}, Kai G. and {Papovich}, Casey J. and {Pentericci}, Laura and {Pope}, Alexandra and {Primack}, Joel R. and {Ravindranath}, Swara and {Reddy}, Naveen A. and {Renzini}, Alvio and {Rix}, Hans-Walter and {Robaina}, Aday R. and {Rosario}, David J. and {Rosati}, Piero and {Salimbeni}, Sara and {Scarlata}, Claudia and {Siana}, Brian and {Simard}, Luc and {Smidt}, Joseph and {Snyder}, Diana and {Somerville}, Rachel S. and {Spinrad}, Hyron and {Straughn}, Amber N. and {Telford}, Olivia and {Teplitz}, Harry I. and {Trump}, Jonathan R. and {Vargas}, Carlos and {Villforth}, Carolin and {Wagner}, Cory R. and {Wandro}, Pat and {Wechsler}, Risa H. and {Weiner}, Benjamin J. and {Wiklind}, Tommy and {Wild}, Vivienne and {Wilson}, Grant and {Wuyts}, Stijn and {Yun}, Min S.},
        title = "{CANDELS: The Cosmic Assembly Near-infrared Deep Extragalactic Legacy Survey{\textemdash}The Hubble Space Telescope Observations, Imaging Data Products, and Mosaics}",
      journal = {\apjs},
         year = 2011,
        month = dec,
       volume = {197},
       number = {2},
          eid = {36},
        pages = {36},
          doi = {10.1088/0067-0049/197/2/36},
archivePrefix = {arXiv},
       eprint = {1105.3754},
 primaryClass = {astro-ph.CO},
       adsurl = {https://ui.adsabs.harvard.edu/abs/2011ApJS..197...36K}
}

@ARTICLE{kocevski2025risefaintredagn,
       author = {{Kocevski}, Dale D. and {Finkelstein}, Steven L. and {Barro}, Guillermo and {Taylor}, Anthony J. and {Calabr{\`o}}, Antonello and {Laloux}, Brivael and {Buchner}, Johannes and {Trump}, Jonathan R. and {Leung}, Gene C.~K. and {Yang}, Guang and {Dickinson}, Mark and {P{\'e}rez-Gonz{\'a}lez}, Pablo G. and {Pacucci}, Fabio and {Inayoshi}, Kohei and {Somerville}, Rachel S. and {McGrath}, Elizabeth J. and {Akins}, Hollis B. and {Bagley}, Micaela B. and {Bowler}, Rebecca A.~A. and {Bisigello}, Laura and {Carnall}, Adam and {Casey}, Caitlin M. and {Cheng}, Yingjie and {Cleri}, Nikko J. and {Costantin}, Luca and {Cullen}, Fergus and {Davis}, Kelcey and {Donnan}, Callum T. and {Dunlop}, James S. and {Ellis}, Richard S. and {Ferguson}, Henry C. and {Fujimoto}, Seiji and {Fontana}, Adriano and {Giavalisco}, Mauro and {Grazian}, Andrea and {Grogin}, Norman A. and {Hathi}, Nimish P. and {Hirschmann}, Michaela and {Huertas-Company}, Marc and {Holwerda}, Benne W. and {Illingworth}, Garth and {Juneau}, St{\'e}phanie and {Kartaltepe}, Jeyhan S. and {Koekemoer}, Anton M. and {Li}, Wenxiu and {Lucas}, Ray A. and {Magee}, Dan and {Mason}, Charlotte and {McLeod}, Derek J. and {McLure}, Ross J. and {Napolitano}, Lorenzo and {Papovich}, Casey and {Pirzkal}, Nor and {Rodighiero}, Giulia and {Santini}, Paola and {Wilkins}, Stephen M. and {Yung}, L.~Y. Aaron},
        title = "{The Rise of Faint, Red Active Galactic Nuclei at z > 4: A Sample of Little Red Dots in the JWST Extragalactic Legacy Fields}",
      journal = {\apj},
         year = 2025,
        month = jun,
       volume = {986},
       number = {2},
          eid = {126},
        pages = {126},
          doi = {10.3847/1538-4357/adbc7d},
archivePrefix = {arXiv},
       eprint = {2404.03576},
 primaryClass = {astro-ph.GA},
       adsurl = {https://ui.adsabs.harvard.edu/abs/2025ApJ...986..126K}
}

@ARTICLE{2025A&A...699A.358V,
       author = {{Valentino}, F. and {Heintz}, K.~E. and {Brammer}, G. and {Ito}, K. and {Kokorev}, V. and {Whitaker}, K.~E. and {Gallazzi}, A. and {de Graaff}, A. and {Weibel}, A. and {Frye}, B.~L. and {Kamieneski}, P.~S. and {Jin}, S. and {Ceverino}, D. and {Faisst}, A. and {Farcy}, M. and {Fujimoto}, S. and {Gillman}, S. and {Gottumukkala}, R. and {Hamadouche}, M. and {Harrington}, K.~C. and {Hirschmann}, M. and {Jespersen}, C.~K. and {Kakimoto}, T. and {Kubo}, M. and {Lagos}, C. d. P. and {Lee}, M. and {Magdis}, G.~E. and {Man}, A.~W.~S. and {Onodera}, M. and {Rizzo}, F. and {Shimakawa}, R. and {Setton}, D.~J. and {Tanaka}, M. and {Toft}, S. and {Wu}, P.-F. and {Zhu}, P.},
        title = "{Gas outflows in two recently quenched galaxies at z = 4 and 7}",
      journal = {\aap},
         year = 2025,
        month = jul,
       volume = {699},
          eid = {A358},
        pages = {A358},
          doi = {10.1051/0004-6361/202553908},
archivePrefix = {arXiv},
       eprint = {2503.01990},
 primaryClass = {astro-ph.GA},
       adsurl = {https://ui.adsabs.harvard.edu/abs/2025A&A...699A.358V}
}

@article{ refId0,
	author = {{de Graaff, Anna} and {Brammer, Gabriel} and {Weibel, Andrea} and {Lewis, Zach} and {Maseda, Michael V.} and {Oesch, Pascal A.} and {Bezanson, Rachel} and {Boogaard, Leindert A.} and {Cleri, Nikko J.} and {Cooper, Olivia R.} and {Gottumukkala, Rashmi} and {Greene, Jenny E.} and {Hirschmann, Michaela} and {Hviding, Raphael E.} and {Katz, Harley} and {Labbé, Ivo} and {Leja, Joel} and {Matthee, Jorryt} and {McConachie, Ian} and {Miller, Tim B.} and {Naidu, Rohan P.} and {Price, Sedona H.} and {Rix, Hans-Walter} and {Setton, David J.} and {Suess, Katherine A.} and {Wang, Bingjie} and {Whitaker, Katherine E.} and {Williams, Christina C.}},
	title = {RUBIES: A complete census of the bright and red distant Universe with JWST/NIRSpec},
	DOI= "10.1051/0004-6361/202452186",
	url= "https://doi.org/10.1051/0004-6361/202452186",
	journal = {\aap},
	year = 2025,
	volume = 697,
	pages = "A189",
}

@article{Heintz,
	author = {{Heintz, K. E.} and {Brammer, G. B.} and {Watson, D.} and {Oesch, P. A.} and {Keating, L. C.} and {Hayes, M. J.} and {Abdurro’uf} and {Arellano-Córdova, K. Z.} and {Carnall, A. C.} and {Christiansen, C. R.} and {Cullen, F.} and {Davé, R.} and {Dayal, P.} and {Ferrara, A.} and {Finlator, K.} and {Fynbo, J. P. U.} and {Flury, S. R.} and {Gelli, V.} and {Gillman, S.} and {Gottumukkala, R.} and {Gould, K.} and {Greve, T. R.} and {Hardin, S. E.} and {Hsiao, T. Y.-Y} and {Hutter, A.} and {Jakobsson, P.} and {Killi, M.} and {Khosravaninezhad, N.} and {Laursen, P.} and {Lee, M. M.} and {Magdis, G. E.} and {Matthee, J.} and {Naidu, R. P.} and {Narayanan, D.} and {Pollock, C.} and {Prescott, M. K. M.} and {Rusakov, V.} and {Shuntov, M.} and {Sneppen, A.} and {Smit, R.} and {Tanvir, N. R.} and {Terp, C.} and {Toft, S.} and {Valentino, F.} and {Vijayan, A. P.} and {Weaver, J. R.} and {Wise, J. H.} and {Witstok, J.}},
	title = {The JWST-PRIMAL archival survey - A JWST/NIRSpec reference sample for the physical properties and Lyman-α absorption and emission of ∼600 galaxies at z = 5.0 − 13.4},
	DOI= "10.1051/0004-6361/202450243",
	url= "https://doi.org/10.1051/0004-6361/202450243",
	journal = {\aap},
	year = 2025,
	volume = 693,
	pages = "A60",
}

@ARTICLE{BC2003,
       author = {{Bruzual}, G. and {Charlot}, S.},
        title = "{Stellar population synthesis at the resolution of 2003}",
      journal = {\mnras},
         year = 2003,
        month = oct,
       volume = {344},
       number = {4},
        pages = {1000-1028},
          doi = {10.1046/j.1365-8711.2003.06897.x},
archivePrefix = {arXiv},
       eprint = {astro-ph/0309134},
 primaryClass = {astro-ph},
       adsurl = {https://ui.adsabs.harvard.edu/abs/2003MNRAS.344.1000B}
}

@ARTICLE{Chabrier_2003,
       author = {{Chabrier}, Gilles},
        title = "{Galactic Stellar and Substellar Initial Mass Function}",
      journal = {\pasp},
         year = 2003,
        month = jul,
       volume = {115},
       number = {809},
        pages = {763-795},
          doi = {10.1086/376392},
archivePrefix = {arXiv},
       eprint = {astro-ph/0304382},
 primaryClass = {astro-ph},
       adsurl = {https://ui.adsabs.harvard.edu/abs/2003PASP..115..763C}
}

@ARTICLE{Calzetti_2000,
       author = {{Calzetti}, Daniela and {Armus}, Lee and {Bohlin}, Ralph C. and {Kinney}, Anne L. and {Koornneef}, Jan and {Storchi-Bergmann}, Thaisa},
        title = "{The Dust Content and Opacity of Actively Star-forming Galaxies}",
      journal = {\apj},
         year = 2000,
        month = apr,
       volume = {533},
       number = {2},
        pages = {682-695},
          doi = {10.1086/308692},
archivePrefix = {arXiv},
       eprint = {astro-ph/9911459},
 primaryClass = {astro-ph},
       adsurl = {https://ui.adsabs.harvard.edu/abs/2000ApJ...533..682C}
}

@ARTICLE{perezgonzalez2024naturelittlereddots,
       author = {{P{\'e}rez-Gonz{\'a}lez}, Pablo G. and {Barro}, Guillermo and {Rieke}, George H. and {Lyu}, Jianwei and {Rieke}, Marcia and {Alberts}, Stacey and {Williams}, Christina C. and {Hainline}, Kevin and {Sun}, Fengwu and {Pusk{\'a}s}, D{\'a}vid and {Annunziatella}, Marianna and {Baker}, William M. and {Bunker}, Andrew J. and {Egami}, Eiichi and {Ji}, Zhiyuan and {Johnson}, Benjamin D. and {Robertson}, Brant and {Rodr{\'\i}guez Del Pino}, Bruno and {Rujopakarn}, Wiphu and {Shivaei}, Irene and {Tacchella}, Sandro and {Willmer}, Christopher N.~A. and {Willott}, Chris},
        title = "{What Is the Nature of Little Red Dots and what Is Not, MIRI SMILES Edition}",
      journal = {\apj},
         year = 2024,
        month = jun,
       volume = {968},
       number = {1},
          eid = {4},
        pages = {4},
          doi = {10.3847/1538-4357/ad38bb},
archivePrefix = {arXiv},
       eprint = {2401.08782},
 primaryClass = {astro-ph.GA},
       adsurl = {https://ui.adsabs.harvard.edu/abs/2024ApJ...968....4P}
}

@ARTICLE{Li_2025,
       author = {{Li}, Zhengrong and {Inayoshi}, Kohei and {Chen}, Kejian and {Ichikawa}, Kohei and {Ho}, Luis C.},
        title = "{Little Red Dots: Rapidly Growing Black Holes Reddened by Extended Dusty Flows}",
      journal = {\apj},
         year = 2025,
        month = feb,
       volume = {980},
       number = {1},
          eid = {36},
        pages = {36},
          doi = {10.3847/1538-4357/ada5fb},
archivePrefix = {arXiv},
       eprint = {2407.10760},
 primaryClass = {astro-ph.GA},
       adsurl = {https://ui.adsabs.harvard.edu/abs/2025ApJ...980...36L}
}

@ARTICLE{Greene_2024,
       author = {{Greene}, Jenny E. and {Labbe}, Ivo and {Goulding}, Andy D. and {Furtak}, Lukas J. and {Chemerynska}, Iryna and {Kokorev}, Vasily and {Dayal}, Pratika and {Volonteri}, Marta and {Williams}, Christina C. and {Wang}, Bingjie and {Setton}, David J. and {Burgasser}, Adam J. and {Bezanson}, Rachel and {Atek}, Hakim and {Brammer}, Gabriel and {Cutler}, Sam E. and {Feldmann}, Robert and {Fujimoto}, Seiji and {Glazebrook}, Karl and {de Graaff}, Anna and {Khullar}, Gourav and {Leja}, Joel and {Marchesini}, Danilo and {Maseda}, Michael V. and {Matthee}, Jorryt and {Miller}, Tim B. and {Naidu}, Rohan P. and {Nanayakkara}, Themiya and {Oesch}, Pascal A. and {Pan}, Richard and {Papovich}, Casey and {Price}, Sedona H. and {van Dokkum}, Pieter and {Weaver}, John R. and {Whitaker}, Katherine E. and {Zitrin}, Adi},
        title = "{UNCOVER Spectroscopy Confirms the Surprising Ubiquity of Active Galactic Nuclei in Red Sources at z > 5}",
      journal = {\apj},
         year = 2024,
        month = mar,
       volume = {964},
       number = {1},
          eid = {39},
        pages = {39},
          doi = {10.3847/1538-4357/ad1e5f},
archivePrefix = {arXiv},
       eprint = {2309.05714},
 primaryClass = {astro-ph.GA},
       adsurl = {https://ui.adsabs.harvard.edu/abs/2024ApJ...964...39G}
}

@ARTICLE{Hviding,
       author = {{Hviding}, Raphael E. and {de Graaff}, Anna and {Miller}, Tim B. and {Setton}, David J. and {Greene}, Jenny E. and {Labb{\'e}}, Ivo and {Brammer}, Gabriel and {Bezanson}, Rachel and {Boogaard}, Leindert A. and {Cleri}, Nikko J. and {Leja}, Joel and {Maseda}, Michael V. and {McConachie}, Ian and {Matthee}, Jorryt and {Naidu}, Rohan P. and {Oesch}, Pascal A. and {Wang}, Bingjie and {Whitaker}, Katherine E. and {Williams}, Christina C.},
        title = "{RUBIES: A spectroscopic census of little red dots: All point sources with v-shaped continua have broad lines}",
      journal = {\aap},
         year = 2025,
        month = oct,
       volume = {702},
          eid = {A57},
        pages = {A57},
          doi = {10.1051/0004-6361/202555816},
archivePrefix = {arXiv},
       eprint = {2506.05459},
 primaryClass = {astro-ph.GA},
       adsurl = {https://ui.adsabs.harvard.edu/abs/2025A&A...702A..57H}
}

@ARTICLE{Setton_2025,
       author = {{Setton}, David J. and {Greene}, Jenny E. and {Spilker}, Justin S. and {Williams}, Christina C. and {Labb{\'e}}, Ivo and {Ma}, Yilun and {Wang}, Bingjie and {Whitaker}, Katherine E. and {Leja}, Joel and {de Graaff}, Anna and {Alberts}, Stacey and {Bezanson}, Rachel and {Boogaard}, Leindert A. and {Brammer}, Gabriel and {Cutler}, Sam E. and {Cleri}, Nikko J. and {Cooper}, Olivia R. and {Dayal}, Pratika and {Fujimoto}, Seiji and {Furtak}, Lukas J. and {Goulding}, Andy D. and {Hirschmann}, Michaela and {Kokorev}, Vasily and {Maseda}, Michael V. and {McConachie}, Ian and {Matthee}, Jorryt and {Miller}, Tim B. and {Naidu}, Rohan P. and {Oesch}, Pascal A. and {Pan}, Richard and {Price}, Sedona H. and {Suess}, Katherine A. and {Weaver}, John R. and {Xiao}, Mengyuan and {Zhang}, Yunchong and {Zitrin}, Adi},
        title = "{A Confirmed Deficit of Hot and Cold Dust Emission in the Most Luminous Little Red Dots}",
      journal = {\apjl},
         year = 2025,
        month = sep,
       volume = {991},
       number = {1},
          eid = {L10},
        pages = {L10},
          doi = {10.3847/2041-8213/ade78b},
archivePrefix = {arXiv},
       eprint = {2503.02059},
 primaryClass = {astro-ph.GA},
       adsurl = {https://ui.adsabs.harvard.edu/abs/2025ApJ...991L..10S}
}

@ARTICLE{2025ApJ...991...37A,
       author = {{Akins}, Hollis B. and {Casey}, Caitlin M. and {Lambrides}, Erini and {Allen}, Natalie and {Andika}, Irham T. and {Brinch}, Malte and {Champagne}, Jaclyn B. and {Cooper}, Olivia and {Ding}, Xuheng and {Drakos}, Nicole E. and {Faisst}, Andreas and {Finkelstein}, Steven L. and {Franco}, Maximilien and {Fujimoto}, Seiji and {Gentile}, Fabrizio and {Gillman}, Steven and {Gozaliasl}, Ghassem and {Harish}, Santosh and {Hayward}, Christopher C. and {Hirschmann}, Michaela and {Ilbert}, Olivier and {Kartaltepe}, Jeyhan S. and {Kocevski}, Dale D. and {Koekemoer}, Anton M. and {Kokorev}, Vasily and {Liu}, Daizhong and {Long}, Arianna S. and {McCracken}, Henry Joy and {McKinney}, Jed and {Onoue}, Masafusa and {Paquereau}, Louise and {Renzini}, Alvio and {Rhodes}, Jason and {Robertson}, Brant E. and {Shuntov}, Marko and {Silverman}, John D. and {Tanaka}, Takumi S. and {Toft}, Sune and {Trakhtenbrot}, Benny and {Valentino}, Francesco and {Zavala}, Jorge},
        title = "{COSMOS-Web: The Overabundance and Physical Nature of ``Little Red Dots''{\textemdash}Implications for Early Galaxy and SMBH Assembly}",
      journal = {\apj},
         year = 2025,
        month = sep,
       volume = {991},
       number = {1},
          eid = {37},
        pages = {37},
          doi = {10.3847/1538-4357/ade984},
archivePrefix = {arXiv},
       eprint = {2406.10341},
 primaryClass = {astro-ph.GA},
       adsurl = {https://ui.adsabs.harvard.edu/abs/2025ApJ...991...37A}
}

@ARTICLE{Leung_2025,
       author = {{Leung}, Gene C.~K. and {Finkelstein}, Steven L. and {P{\'e}rez-Gonz{\'a}lez}, Pablo G. and {Morales}, Alexa M. and {Taylor}, Anthony J. and {Barro}, Guillermo and {Kocevski}, Dale D. and {Akins}, Hollis B. and {Carnall}, Adam C. and {Ch{\'a}vez Ortiz}, {\'O}scar A. and {Cleri}, Nikko J. and {Cullen}, Fergus and {Donnan}, Callum T. and {Dunlop}, James S. and {Ellis}, Richard S. and {Grogin}, Norman A. and {Hirschmann}, Michaela and {Koekemoer}, Anton M. and {Kokorev}, Vasily and {Lucas}, Ray A. and {McLeod}, Derek J. and {Papovich}, Casey and {Yung}, L.~Y. Aaron},
        title = "{Exploring the Nature of Little Red Dots: Constraints on Active Galactic Nucleus and Stellar Contributions from PRIMER MIRI Imaging}",
      journal = {\apj},
         year = 2025,
        month = oct,
       volume = {992},
       number = {1},
          eid = {26},
        pages = {26},
          doi = {10.3847/1538-4357/adfcce},
archivePrefix = {arXiv},
       eprint = {2411.12005},
 primaryClass = {astro-ph.GA},
       adsurl = {https://ui.adsabs.harvard.edu/abs/2025ApJ...992...26L}
}

@ARTICLE{Taylor_2025,
       author = {{Taylor}, Anthony J. and {Kokorev}, Vasily and {Kocevski}, Dale D. and {Akins}, Hollis B. and {Cullen}, Fergus and {Dickinson}, Mark and {Finkelstein}, Steven L. and {Arrabal Haro}, Pablo and {Bromm}, Volker and {Giavalisco}, Mauro and {Inayoshi}, Kohei and {Juneau}, St{\'e}phanie and {Leung}, Gene C.~K. and {P{\'e}rez-Gonz{\'a}lez}, Pablo G. and {Somerville}, Rachel S. and {Trump}, Jonathan R. and {Amor{\'\i}n}, Ricardo O. and {Barro}, Guillermo and {Burgarella}, Denis and {Brooks}, Madisyn and {Carnall}, Adam C. and {Casey}, Caitlin M. and {Cheng}, Yingjie and {Chisholm}, John and {Chworowsky}, Katherine and {Davis}, Kelcey and {Donnan}, Callum T. and {Dunlop}, James S. and {Ellis}, Richard S. and {Fern{\'a}ndez}, Vital and {Fujimoto}, Seiji and {Grogin}, Norman A. and {Gupta}, Ansh R. and {Hathi}, Nimish P. and {Jung}, Intae and {Hirschmann}, Michaela and {Kartaltepe}, Jeyhan S. and {Koekemoer}, Anton M. and {Larson}, Rebecca L. and {Leung}, Ho-Hin and {Llerena}, Mario and {Lucas}, Ray A. and {McLeod}, Derek J. and {McLure}, Ross and {Napolitano}, Lorenzo and {Papovich}, Casey and {Stanton}, Thomas M. and {Tripodi}, Roberta and {Wang}, Xin and {Wilkins}, Stephen M. and {Yung}, L.~Y. Aaron and {Zavala}, Jorge A.},
        title = "{CAPERS-LRD-z9: A Gas-enshrouded Little Red Dot Hosting a Broad-line Active Galactic Nucleus at z = 9.288}",
      journal = {\apjl},
         year = 2025,
        month = aug,
       volume = {989},
       number = {1},
          eid = {L7},
        pages = {L7},
          doi = {10.3847/2041-8213/ade789},
archivePrefix = {arXiv},
       eprint = {2505.04609},
 primaryClass = {astro-ph.GA},
       adsurl = {https://ui.adsabs.harvard.edu/abs/2025ApJ...989L...7T}
}

@ARTICLE{Kokorev_2023,
       author = {{Kokorev}, Vasily and {Fujimoto}, Seiji and {Labbe}, Ivo and {Greene}, Jenny E. and {Bezanson}, Rachel and {Dayal}, Pratika and {Nelson}, Erica J. and {Atek}, Hakim and {Brammer}, Gabriel and {Caputi}, Karina I. and {Chemerynska}, Iryna and {Cutler}, Sam E. and {Feldmann}, Robert and {Fudamoto}, Yoshinobu and {Furtak}, Lukas J. and {Goulding}, Andy D. and {de Graaff}, Anna and {Leja}, Joel and {Marchesini}, Danilo and {Miller}, Tim B. and {Nanayakkara}, Themiya and {Oesch}, Pascal A. and {Pan}, Richard and {Price}, Sedona H. and {Setton}, David J. and {Smit}, Renske and {Stefanon}, Mauro and {Wang}, Bingjie and {Weaver}, John R. and {Whitaker}, Katherine E. and {Williams}, Christina C. and {Zitrin}, Adi},
        title = "{UNCOVER: A NIRSpec Identification of a Broad-line AGN at z = 8.50}",
      journal = {\apjl},
         year = 2023,
        month = nov,
       volume = {957},
       number = {1},
          eid = {L7},
        pages = {L7},
          doi = {10.3847/2041-8213/ad037a},
archivePrefix = {arXiv},
       eprint = {2308.11610},
 primaryClass = {astro-ph.GA},
       adsurl = {https://ui.adsabs.harvard.edu/abs/2023ApJ...957L...7K}
}

@ARTICLE{naidu2025blackholestarreveals,
       author = {{Naidu}, Rohan P. and {Matthee}, Jorryt and {Katz}, Harley and {de Graaff}, Anna and {Oesch}, Pascal and {Smith}, Aaron and {Greene}, Jenny E. and {Brammer}, Gabriel and {Weibel}, Andrea and {Hviding}, Raphael and {Chisholm}, John and {Labb\textbackslash'e}, Ivo and {Simcoe}, Robert A. and {Witten}, Callum and {Atek}, Hakim and {Baggen}, Josephine F.~W. and {Belli}, Sirio and {Bezanson}, Rachel and {Boogaard}, Leindert A. and {Bose}, Sownak and {Covelo-Paz}, Alba and {Dayal}, Pratika and {Fudamoto}, Yoshinobu and {Furtak}, Lukas J. and {Giovinazzo}, Emma and {Goulding}, Andy and {Gronke}, Max and {Heintz}, Kasper E. and {Hirschmann}, Michaela and {Illingworth}, Garth and {Inoue}, Akio K. and {Johnson}, Benjamin D. and {Leja}, Joel and {Leonova}, Ecaterina and {McConachie}, Ian and {Maseda}, Michael V. and {Natarajan}, Priyamvada and {Nelson}, Erica and {Setton}, David J. and {Shivaei}, Irene and {Sobral}, David and {Stefanon}, Mauro and {Tacchella}, Sandro and {Toft}, Sune and {Torralba}, Alberto and {van Dokkum}, Pieter and {van der Wel}, Arjen and {Volonteri}, Marta and {Walter}, Fabian and {Wang}, Bingjie and {Watson}, Darach},
        title = "{A ``Black Hole Star'' Reveals the Remarkable Gas-Enshrouded Hearts of the Little Red Dots}",
      journal = {arXiv e-prints},
         year = 2025,
        month = mar,
          eid = {arXiv:2503.16596},
        pages = {arXiv:2503.16596},
          doi = {10.48550/arXiv.2503.16596},
archivePrefix = {arXiv},
       eprint = {2503.16596},
 primaryClass = {astro-ph.GA},
       adsurl = {https://ui.adsabs.harvard.edu/abs/2025arXiv250316596N}
}

@ARTICLE{Inayoshi_2025,
       author = {{Inayoshi}, Kohei and {Maiolino}, Roberto},
        title = "{Extremely Dense Gas around Little Red Dots and High-redshift Active Galactic Nuclei: A Nonstellar Origin of the Balmer Break and Absorption Features}",
      journal = {\apjl},
         year = 2025,
        month = feb,
       volume = {980},
       number = {2},
          eid = {L27},
        pages = {L27},
          doi = {10.3847/2041-8213/adaebd},
archivePrefix = {arXiv},
       eprint = {2409.07805},
 primaryClass = {astro-ph.GA},
       adsurl = {https://ui.adsabs.harvard.edu/abs/2025ApJ...980L..27I}
}

@ARTICLE{degraaff2025littlereddotshost,
       author = {{de Graaff}, Anna and {Hviding}, Raphael E. and {Naidu}, Rohan P. and {Greene}, Jenny E. and {Miller}, Tim B. and {Leja}, Joel and {Matthee}, Jorryt and {Brammer}, Gabriel and {Katz}, Harley and {Bezanson}, Rachel and {Boogaard}, Leindert A. and {Bose}, Sownak and {Chisholm}, John and {Cleri}, Nikko J. and {Dayal}, Pratika and {Feldmann}, Robert and {Fudamoto}, Yoshinobu and {Fujimoto}, Seiji and {Furtak}, Lukas J. and {Glazebrook}, Karl and {Gottumukkala}, Rashmi and {Heintz}, Kasper E. and {Kokorev}, Vasily and {Labbe}, Ivo and {Maseda}, Michael V. and {McConachie}, Ian and {Nanayakkara}, Themiya and {Nelson}, Erica and {Nowaczyk}, Przemys{\l}aw and {Oesch}, Pascal A. and {Rix}, Hans-Walter and {Setton}, David J. and {Torralba}, Alberto and {Walter}, Fabian and {Wang}, Bingjie and {Weibel}, Andrea and {van der Wel}, Arjen},
        title = "{Little Red Dots host Black Hole Stars: A unified family of gas-reddened AGN revealed by JWST/NIRSpec spectroscopy}",
      journal = {arXiv e-prints},
         year = 2025,
        month = nov,
          eid = {arXiv:2511.21820},
        pages = {arXiv:2511.21820},
          doi = {10.48550/arXiv.2511.21820},
archivePrefix = {arXiv},
       eprint = {2511.21820},
 primaryClass = {astro-ph.GA},
       adsurl = {https://ui.adsabs.harvard.edu/abs/2025arXiv251121820D}
}

@ARTICLE{lin2025discoverylittlereddots,
       author = {{Lin}, Xiaojing and {Fan}, Xiaohui and {Cai}, Zheng and {Bian}, Fuyan and {Liu}, Hanpu and {Sun}, Fengwu and {Ma}, Yilun and {Greene}, Jenny E. and {Strauss}, Michael A. and {Green}, Richard and {Lyu}, Jianwei and {Champagne}, Jaclyn B. and {Goulding}, Andy D. and {Inayoshi}, Kohei and {Jin}, Xiangyu and {Leung}, Gene C.~K. and {Li}, Mingyu and {Liu}, Weizhe and {Liu}, Yichen and {Mao}, Junjie and {Pudoka}, Maria Anne and {Tee}, Wei Leong and {Wang}, Ben and {Wang}, Feige and {Wu}, Yunjing and {Yang}, Jinyi and {Zhang}, Haowen and {Zhu}, Yongda},
        title = "{The Discovery of Little Red Dots in the Local Universe: Signatures of Cool Gas Envelopes}",
      journal = {\apj},
         year = 2026,
        month = feb,
       volume = {997},
       number = {2},
          eid = {364},
        pages = {364},
          doi = {10.3847/1538-4357/ae2bdf},
archivePrefix = {arXiv},
       eprint = {2507.10659},
 primaryClass = {astro-ph.GA},
       adsurl = {https://ui.adsabs.harvard.edu/abs/2026ApJ...997..364L}
}

@ARTICLE{kido,
       author = {{Kido}, Daisaburo and {Ioka}, Kunihito and {Hotokezaka}, Kenta and {Inayoshi}, Kohei and {Irwin}, Christopher M.},
        title = "{Black hole envelopes in Little Red Dots}",
      journal = {\mnras},
         year = 2025,
        month = dec,
       volume = {544},
       number = {4},
        pages = {3407-3416},
          doi = {10.1093/mnras/staf1898},
archivePrefix = {arXiv},
       eprint = {2505.06965},
 primaryClass = {astro-ph.HE},
       adsurl = {https://ui.adsabs.harvard.edu/abs/2025MNRAS.544.3407K}
}

@ARTICLE{asada2026originsuvcontinuumbalmer,
       author = {{Asada}, Yoshihisa and {Inayoshi}, Kohei and {Fei}, Qinyue and {Fujimoto}, Seiji and {Willott}, Chris},
        title = "{Origins of the UV continuum and Balmer emission lines in Little Red Dots: observational validation of dense gas envelope models enshrouding the AGN}",
      journal = {arXiv e-prints},
         year = 2026,
        month = jan,
          eid = {arXiv:2601.10573},
        pages = {arXiv:2601.10573},
          doi = {10.48550/arXiv.2601.10573},
archivePrefix = {arXiv},
       eprint = {2601.10573},
 primaryClass = {astro-ph.GA},
       adsurl = {https://ui.adsabs.harvard.edu/abs/2026arXiv260110573A}
}

@ARTICLE{2013ARA&A..51..511K,
       author = {{Kormendy}, John and {Ho}, Luis C.},
        title = "{Coevolution (Or Not) of Supermassive Black Holes and Host Galaxies}",
      journal = {\araa},
         year = 2013,
        month = aug,
       volume = {51},
       number = {1},
        pages = {511-653},
          doi = {10.1146/annurev-astro-082708-101811},
archivePrefix = {arXiv},
       eprint = {1304.7762},
 primaryClass = {astro-ph.CO},
       adsurl = {https://ui.adsabs.harvard.edu/abs/2013ARA&A..51..511K}
}

@ARTICLE{Izumi_2021,
       author = {{Izumi}, Takuma and {Matsuoka}, Yoshiki and {Fujimoto}, Seiji and {Onoue}, Masafusa and {Strauss}, Michael A. and {Umehata}, Hideki and {Imanishi}, Masatoshi and {Kohno}, Kotaro and {Kawaguchi}, Toshihiro and {Kawamuro}, Taiki and {Baba}, Shunsuke and {Nagao}, Tohru and {Toba}, Yoshiki and {Inayoshi}, Kohei and {Silverman}, John D. and {Inoue}, Akio K. and {Ikarashi}, Soh and {Iwasawa}, Kazushi and {Kashikawa}, Nobunari and {Hashimoto}, Takuya and {Nakanishi}, Kouichiro and {Ueda}, Yoshihiro and {Schramm}, Malte and {Lee}, Chien-Hsiu and {Suh}, Hyewon},
        title = "{Subaru High-z Exploration of Low-luminosity Quasars (SHELLQs). XIII. Large-scale Feedback and Star Formation in a Low-luminosity Quasar at z = 7.07 on the Local Black Hole to Host Mass Relation}",
      journal = {\apj},
         year = 2021,
        month = jun,
       volume = {914},
       number = {1},
          eid = {36},
        pages = {36},
          doi = {10.3847/1538-4357/abf6dc},
archivePrefix = {arXiv},
       eprint = {2104.05738},
 primaryClass = {astro-ph.GA},
       adsurl = {https://ui.adsabs.harvard.edu/abs/2021ApJ...914...36I}
}

@ARTICLE{marszewski2026littlereddotsfire,
       author = {{Marszewski}, Andrew and {Faucher-Gigu{\`e}re}, Claude-Andr{\'e} and {Sun}, Guochao and {Angl{\'e}s-Alc{\'a}zar}, Daniel and {Feldmann}, Robert and {Su}, Kung-Yi and {Miller}, Tim B. and {Roy}, Niranjan Chandra},
        title = "{Little Red Dots on FIRE: The Ability of Bursty Galaxies to Host an Abundant Population of High-redshift AGN}",
      journal = {\apjl},
         year = 2026,
        month = may,
       volume = {1002},
       number = {1},
          eid = {L13},
        pages = {L13},
          doi = {10.3847/2041-8213/ae5d45},
archivePrefix = {arXiv},
       eprint = {2601.22213},
 primaryClass = {astro-ph.GA},
       adsurl = {https://ui.adsabs.harvard.edu/abs/2026ApJ..1002L..13M}
}

@ARTICLE{Jiang_2014,
       author = {{Jiang}, Yan-Fei and {Stone}, James M. and {Davis}, Shane W.},
        title = "{A Global Three-dimensional Radiation Magneto-hydrodynamic Simulation of Super-Eddington Accretion Disks}",
      journal = {\apj},
         year = 2014,
        month = dec,
       volume = {796},
       number = {2},
          eid = {106},
        pages = {106},
          doi = {10.1088/0004-637X/796/2/106},
archivePrefix = {arXiv},
       eprint = {1410.0678},
 primaryClass = {astro-ph.HE},
       adsurl = {https://ui.adsabs.harvard.edu/abs/2014ApJ...796..106J}
}

@ARTICLE{Zhang_2025,
       author = {{Zhang}, Lizhong and {Stone}, James M. and {Mullen}, Patrick D. and {Davis}, Shane W. and {Jiang}, Yan-Fei and {White}, Christopher J.},
        title = "{Radiation GRMHD Models of Accretion onto Stellar-mass Black Holes. I. Survey of Eddington Ratios}",
      journal = {\apj},
         year = 2025,
        month = dec,
       volume = {995},
       number = {1},
          eid = {26},
        pages = {26},
          doi = {10.3847/1538-4357/ae0f91},
archivePrefix = {arXiv},
       eprint = {2506.02289},
 primaryClass = {astro-ph.HE},
       adsurl = {https://ui.adsabs.harvard.edu/abs/2025ApJ...995...26Z}
}

@ARTICLE{brazzini2026littlebluereddots,
       author = {{Brazzini}, M. and {D'Eugenio}, F. and {Maiolino}, R. and {Lyu}, J. and {DeCoursey}, C. and {{\"U}bler}, H. and {Ji}, X. and {Juod{\v{z}}balis}, I. and {Scholtz}, J. and {Jones}, G.~C. and {Hainline}, K. and {Dalla Bont{\`a}}, E. and {{\'e}rez-Gonz{\'a}lez}, P.~G. P and {Geris}, S. and {Harshan}, A. and {Feruglio}, C. and {Bischetti}, M. and {Mazzolari}, G. and {Rieke}, G. and {Alberts}, S. and {Trefoloni}, B. and {Carniani}, S. and {Parlanti}, E. and {Marconi}, A. and {Risaliti}, G. and {Ramos Almeida}, C. and {Rinaldi}, P. and {Perna}, M. and {Zamora}, S. and {Lamperti}, I. and {Venturi}, G. and {Cresci}, G. and {Bunker}, Andrew J. and {Ivey}, L.~R.},
        title = "{The Little Blue and Red Dots Rosetta Stones: Non-Gaussian broad lines, hot dust, and X-ray weakness}",
      journal = {arXiv e-prints},
         year = 2026,
        month = jan,
          eid = {arXiv:2601.22214},
        pages = {arXiv:2601.22214},
          doi = {10.48550/arXiv.2601.22214},
archivePrefix = {arXiv},
       eprint = {2601.22214},
 primaryClass = {astro-ph.GA},
       adsurl = {https://ui.adsabs.harvard.edu/abs/2026arXiv260122214B}
}

@ARTICLE{sun2026littlereddot,
       author = {{Sun}, Wendy Q. and {Naidu}, Rohan P. and {Matthee}, Jorryt and {de Graaff}, Anna and {Chisholm}, John and {Greene}, Jenny E. and {Oesch}, Pascal A. and {Torralba}, Alberto and {Hviding}, Raphael E. and {Brammer}, Gabriel and {Simcoe}, Robert A. and {Bose}, Sownak and {Bouwens}, Rychard and {Dayal}, Pratika and {Eilers}, Anna-Christina and {Fei}, Qinyue and {Furtak}, Lukas J. and {Gottumukkala}, Rashmi and {Goulding}, Andy and {Heintz}, Kasper E. and {Hirschmann}, Michaela and {Kokorev}, Vasily and {Leja}, Joel and {Liu}, Zhaoran and {Natarajan}, Priyamvada and {Santarelli}, Andrew D. and {Setton}, David J. and {Smith}, Aaron and {Tacchella}, Sandro and {Volonteri}, Marta and {Walter}, Fabian and {Weibel}, Andrea and {Williams}, Christina C.},
        title = "{Little Red Dot {\ensuremath{-}} Host Galaxy = Black Hole Star: A Gas-Enshrouded Heart at the Center of Every Little Red Dot}",
      journal = {The Open Journal of Astrophysics},
         year = 2026,
        month = may,
       volume = {9},
        pages = {62505},
          doi = {10.33232/001c.162505},
archivePrefix = {arXiv},
       eprint = {2601.20929},
 primaryClass = {astro-ph.GA},
       adsurl = {https://ui.adsabs.harvard.edu/abs/2026OJAp....962505S}
}

@ARTICLE{madau2026littlereddotsobscured,
       author = {{Madau}, Piero and {Maiolino}, Roberto},
        title = "{Little Red Dots as Obscured Little Blue Dots: A Super-Eddington Unification Model}",
      journal = {arXiv e-prints},
         year = 2026,
        month = feb,
          eid = {arXiv:2602.22386},
        pages = {arXiv:2602.22386},
          doi = {10.48550/arXiv.2602.22386},
archivePrefix = {arXiv},
       eprint = {2602.22386},
 primaryClass = {astro-ph.GA},
       adsurl = {https://ui.adsabs.harvard.edu/abs/2026arXiv260222386M}
}

@ARTICLE{fu2025discoverylittlereddots,
       author = {{Fu}, Shuqi and {Zhang}, Zijian and {Jiang}, Danyang and {Chen}, Jie and {Jiang}, Linhua and {Ho}, Luis C. and {Inayoshi}, Kohei and {Chen}, Kaiyuan and {Lyu}, Jianwei and {Sun}, Fengwu and {Wang}, Feige and {Yang}, Jinyi},
        title = "{The discovery of two little red dots in transition into quasars}",
      journal = {Nature Astronomy},
         year = 2026,
        month = jun,
          doi = {10.1038/s41550-026-02885-8},
archivePrefix = {arXiv},
       eprint = {2512.02096},
 primaryClass = {astro-ph.GA},
       adsurl = {https://ui.adsabs.harvard.edu/abs/2026NatAs.tmp..118F}
}

@ARTICLE{2023ApJ...949L..18P,
       author = {{Papovich}, Casey and {Cole}, Justin W. and {Yang}, Guang and {Finkelstein}, Steven L. and {Barro}, Guillermo and {Buat}, V{\'e}ronique and {Burgarella}, Denis and {P{\'e}rez-Gonz{\'a}lez}, Pablo G. and {Santini}, Paola and {Seill{\'e}}, Lise-Marie and {Shen}, Lu and {Arrabal Haro}, Pablo and {Bagley}, Micaela B. and {Bell}, Eric F. and {Bisigello}, Laura and {Calabr{\`o}}, Antonello and {Casey}, Caitlin M. and {Castellano}, Marco and {Chworowsky}, Katherine and {Cleri}, Nikko J. and {Costantin}, Luca and {Cooper}, M.~C. and {Dickinson}, Mark and {Ferguson}, Henry C. and {Fontana}, Adriano and {Giavalisco}, Mauro and {Grazian}, Andrea and {Grogin}, Norman A. and {Hathi}, Nimish P. and {Holwerda}, Benne W. and {Hutchison}, Taylor A. and {Kartaltepe}, Jeyhan S. and {Kewley}, Lisa J. and {Kirkpatrick}, Allison and {Kocevski}, Dale D. and {Koekemoer}, Anton M. and {Larson}, Rebecca L. and {Long}, Arianna S. and {Lucas}, Ray A. and {Pentericci}, Laura and {Pirzkal}, Nor and {Ravindranath}, Swara and {Somerville}, Rachel S. and {Trump}, Jonathan R. and {Urbano Stawinski}, Stephanie M. and {Weiner}, Benjamin J. and {Wilkins}, Stephen M. and {Yung}, L.~Y. Aaron and {Zavala}, Jorge A.},
        title = "{CEERS Key Paper. V. Galaxies at 4 < z < 9 Are Bluer than They Appear{\textendash}Characterizing Galaxy Stellar Populations from Rest-frame {\ensuremath{\sim}}1 {\ensuremath{\mu}}m Imaging}",
      journal = {\apjl},
         year = 2023,
        month = jun,
       volume = {949},
       number = {2},
          eid = {L18},
        pages = {L18},
          doi = {10.3847/2041-8213/acc948},
archivePrefix = {arXiv},
       eprint = {2301.00027},
 primaryClass = {astro-ph.GA},
       adsurl = {https://ui.adsabs.harvard.edu/abs/2023ApJ...949L..18P}
}

@ARTICLE{2024AJ....168..113C,
       author = {{Chworowsky}, Katherine and {Finkelstein}, Steven L. and {Boylan-Kolchin}, Michael and {McGrath}, Elizabeth J. and {Iyer}, Kartheik G. and {Papovich}, Casey and {Dickinson}, Mark and {Taylor}, Anthony J. and {Yung}, L.~Y. Aaron and {Arrabal Haro}, Pablo and {Bagley}, Micaela B. and {Backhaus}, Bren E. and {Bhatawdekar}, Rachana and {Cheng}, Yingjie and {Cleri}, Nikko J. and {Cole}, Justin W. and {Cooper}, M.~C. and {Costantin}, Luca and {Dekel}, Avishai and {Franco}, Maximilien and {Fujimoto}, Seiji and {Hayward}, Christopher C. and {Holwerda}, Benne W. and {Huertas-Company}, Marc and {Hirschmann}, Michaela and {Hutchison}, Taylor A. and {Koekemoer}, Anton M. and {Larson}, Rebecca L. and {Li}, Zhaozhou and {Long}, Arianna S. and {Lucas}, Ray A. and {Pirzkal}, Nor and {Rodighiero}, Giulia and {Somerville}, Rachel S. and {Vanderhoof}, Brittany N. and {de la Vega}, Alexander and {Wilkins}, Stephen M. and {Yang}, Guang and {Zavala}, Jorge A.},
        title = "{Evidence for a Shallow Evolution in the Volume Densities of Massive Galaxies at z = 4{\textendash}8 from CEERS}",
      journal = {\aj},
         year = 2024,
        month = sep,
       volume = {168},
       number = {3},
          eid = {113},
        pages = {113},
          doi = {10.3847/1538-3881/ad57c1},
archivePrefix = {arXiv},
       eprint = {2311.14804},
 primaryClass = {astro-ph.GA},
       adsurl = {https://ui.adsabs.harvard.edu/abs/2024AJ....168..113C}
}

@ARTICLE{chon2026rapidemergenceovermassiveblack,
       author = {{Chon}, Sunmyon and {Hirano}, Shingo and {Ishiyama}, Tomoaki and {Chang}, Seok-Jun and {Springel}, Volker},
        title = "{Rapid emergence of overmassive black holes in the early Universe}",
      journal = {arXiv e-prints},
         year = 2026,
        month = jan,
          eid = {arXiv:2601.04955},
        pages = {arXiv:2601.04955},
          doi = {10.48550/arXiv.2601.04955},
archivePrefix = {arXiv},
       eprint = {2601.04955},
 primaryClass = {astro-ph.GA},
       adsurl = {https://ui.adsabs.harvard.edu/abs/2026arXiv260104955C}
}

@ARTICLE{Umeda_2026,
       author = {{Umeda}, Hiroya and {Inayoshi}, Kohei and {Harikane}, Yuichi and {Murase}, Kohta},
        title = "{A Black Hole Envelope Interpretation for Cosmological Demographics of Little Red Dots}",
      journal = {\apj},
         year = 2026,
        month = mar,
       volume = {999},
       number = {2},
          eid = {183},
        pages = {183},
          doi = {10.3847/1538-4357/ae4101},
archivePrefix = {arXiv},
       eprint = {2512.04208},
 primaryClass = {astro-ph.GA},
       adsurl = {https://ui.adsabs.harvard.edu/abs/2026ApJ...999..183U}
}

@ARTICLE{Han_2012,
       author = {{Han}, Yunkun and {Han}, Zhanwen},
        title = "{Decoding Spectral Energy Distributions of Dust-obscured Starburst-Active Galactic Nucleus}",
      journal = {\apj},
         year = 2012,
        month = apr,
       volume = {749},
       number = {2},
          eid = {123},
        pages = {123},
          doi = {10.1088/0004-637X/749/2/123},
archivePrefix = {arXiv},
       eprint = {1202.6203},
 primaryClass = {astro-ph.CO},
       adsurl = {https://ui.adsabs.harvard.edu/abs/2012ApJ...749..123H}
}

@ARTICLE{Han_2014,
       author = {{Han}, Yunkun and {Han}, Zhanwen},
        title = "{BayeSED: A General Approach to Fitting the Spectral Energy Distribution of Galaxies}",
      journal = {\apjs},
         year = 2014,
        month = nov,
       volume = {215},
       number = {1},
          eid = {2},
        pages = {2},
          doi = {10.1088/0067-0049/215/1/2},
archivePrefix = {arXiv},
       eprint = {1408.6399},
 primaryClass = {astro-ph.GA},
       adsurl = {https://ui.adsabs.harvard.edu/abs/2014ApJS..215....2H}
}

@ARTICLE{Han_2019,
       author = {{Han}, Yunkun and {Han}, Zhanwen},
        title = "{A Comprehensive Bayesian Discrimination of the Simple Stellar Population Model, Star Formation History, and Dust Attenuation Law in the Spectral Energy Distribution Modeling of Galaxies}",
      journal = {\apjs},
         year = 2019,
        month = jan,
       volume = {240},
       number = {1},
          eid = {3},
        pages = {3},
          doi = {10.3847/1538-4365/aaeffa},
archivePrefix = {arXiv},
       eprint = {1811.04180},
 primaryClass = {astro-ph.GA},
       adsurl = {https://ui.adsabs.harvard.edu/abs/2019ApJS..240....3H}
}

@ARTICLE{Han_2023,
       author = {{Han}, Yunkun and {Fan}, Lulu and {Zheng}, Xian Zhong and {Bai}, Jin-Ming and {Han}, Zhanwen},
        title = "{BayeSED-GALAXIES. I. Performance Test for Simultaneous Photometric Redshift and Stellar Population Parameter Estimation of Galaxies in the CSST Wide-field Multiband Imaging Survey}",
      journal = {\apjs},
         year = 2023,
        month = dec,
       volume = {269},
       number = {2},
          eid = {39},
        pages = {39},
          doi = {10.3847/1538-4365/acfc3a},
archivePrefix = {arXiv},
       eprint = {2309.13789},
 primaryClass = {astro-ph.GA},
       adsurl = {https://ui.adsabs.harvard.edu/abs/2023ApJS..269...39H}
}

@ARTICLE{Komatsu_2011,
       author = {{Komatsu}, E. and {Smith}, K.~M. and {Dunkley}, J. and {Bennett}, C.~L. and {Gold}, B. and {Hinshaw}, G. and {Jarosik}, N. and {Larson}, D. and {Nolta}, M.~R. and {Page}, L. and {Spergel}, D.~N. and {Halpern}, M. and {Hill}, R.~S. and {Kogut}, A. and {Limon}, M. and {Meyer}, S.~S. and {Odegard}, N. and {Tucker}, G.~S. and {Weiland}, J.~L. and {Wollack}, E. and {Wright}, E.~L.},
        title = "{Seven-year Wilkinson Microwave Anisotropy Probe (WMAP) Observations: Cosmological Interpretation}",
      journal = {\apjs},
         year = 2011,
        month = feb,
       volume = {192},
       number = {2},
          eid = {18},
        pages = {18},
          doi = {10.1088/0067-0049/192/2/18},
archivePrefix = {arXiv},
       eprint = {1001.4538},
 primaryClass = {astro-ph.CO},
       adsurl = {https://ui.adsabs.harvard.edu/abs/2011ApJS..192...18K}
}

@ARTICLE{2017RMxAA..53..385F,
       author = {{Ferland}, G.~J. and {Chatzikos}, M. and {Guzm{\'a}n}, F. and {Lykins}, M.~L. and {van Hoof}, P.~A.~M. and {Williams}, R.~J.~R. and {Abel}, N.~P. and {Badnell}, N.~R. and {Keenan}, F.~P. and {Porter}, R.~L. and {Stancil}, P.~C.},
        title = "{The 2017 Release Cloudy}",
      journal = {\rmxaa},
         year = 2017,
        month = oct,
       volume = {53},
        pages = {385-438},
          doi = {10.48550/arXiv.1705.10877},
archivePrefix = {arXiv},
       eprint = {1705.10877},
 primaryClass = {astro-ph.GA},
       adsurl = {https://ui.adsabs.harvard.edu/abs/2017RMxAA..53..385F}
}

@ARTICLE{Larson_2023,
       author = {{Larson}, Rebecca L. and {Finkelstein}, Steven L. and {Kocevski}, Dale D. and {Hutchison}, Taylor A. and {Trump}, Jonathan R. and {Arrabal Haro}, Pablo and {Bromm}, Volker and {Cleri}, Nikko J. and {Dickinson}, Mark and {Fujimoto}, Seiji and {Kartaltepe}, Jeyhan S. and {Koekemoer}, Anton M. and {Papovich}, Casey and {Pirzkal}, Nor and {Tacchella}, Sandro and {Zavala}, Jorge A. and {Bagley}, Micaela and {Behroozi}, Peter and {Champagne}, Jaclyn B. and {Cole}, Justin W. and {Jung}, Intae and {Morales}, Alexa M. and {Yang}, Guang and {Zhang}, Haowen and {Zitrin}, Adi and {Amor{\'\i}n}, Ricardo O. and {Burgarella}, Denis and {Casey}, Caitlin M. and {Ch{\'a}vez Ortiz}, {\'O}scar A. and {Cox}, Isabella G. and {Chworowsky}, Katherine and {Fontana}, Adriano and {Gawiser}, Eric and {Grazian}, Andrea and {Grogin}, Norman A. and {Harish}, Santosh and {Hathi}, Nimish P. and {Hirschmann}, Michaela and {Holwerda}, Benne W. and {Juneau}, St{\'e}phanie and {Leung}, Gene C.~K. and {Lucas}, Ray A. and {McGrath}, Elizabeth J. and {P{\'e}rez-Gonz{\'a}lez}, Pablo G. and {Rigby}, Jane R. and {Seill{\'e}}, Lise-Marie and {Simons}, Raymond C. and {de La Vega}, Alexander and {Weiner}, Benjamin J. and {Wilkins}, Stephen M. and {Yung}, L.~Y. Aaron and {Ceers Team}},
        title = "{A CEERS Discovery of an Accreting Supermassive Black Hole 570 Myr after the Big Bang: Identifying a Progenitor of Massive z > 6 Quasars}",
      journal = {\apjl},
         year = 2023,
        month = aug,
       volume = {953},
       number = {2},
          eid = {L29},
        pages = {L29},
          doi = {10.3847/2041-8213/ace619},
archivePrefix = {arXiv},
       eprint = {2303.08918},
 primaryClass = {astro-ph.GA},
       adsurl = {https://ui.adsabs.harvard.edu/abs/2023ApJ...953L..29L}
}

@ARTICLE{Kocevski_2023,
       author = {{Kocevski}, Dale D. and {Onoue}, Masafusa and {Inayoshi}, Kohei and {Trump}, Jonathan R. and {Arrabal Haro}, Pablo and {Grazian}, Andrea and {Dickinson}, Mark and {Finkelstein}, Steven L. and {Kartaltepe}, Jeyhan S. and {Hirschmann}, Michaela and {Aird}, James and {Holwerda}, Benne W. and {Fujimoto}, Seiji and {Juneau}, St{\'e}phanie and {Amor{\'\i}n}, Ricardo O. and {Backhaus}, Bren E. and {Bagley}, Micaela B. and {Barro}, Guillermo and {Bell}, Eric F. and {Bisigello}, Laura and {Calabr{\`o}}, Antonello and {Cleri}, Nikko J. and {Cooper}, M.~C. and {Ding}, Xuheng and {Grogin}, Norman A. and {Ho}, Luis C. and {Hutchison}, Taylor A. and {Inoue}, Akio K. and {Jiang}, Linhua and {Jones}, Brenda and {Koekemoer}, Anton M. and {Li}, Wenxiu and {Li}, Zhengrong and {McGrath}, Elizabeth J. and {Molina}, Juan and {Papovich}, Casey and {P{\'e}rez-Gonz{\'a}lez}, Pablo G. and {Pirzkal}, Nor and {Wilkins}, Stephen M. and {Yang}, Guang and {Yung}, L.~Y. Aaron},
        title = "{Hidden Little Monsters: Spectroscopic Identification of Low-mass, Broad-line AGNs at z > 5 with CEERS}",
      journal = {\apjl},
         year = 2023,
        month = sep,
       volume = {954},
       number = {1},
          eid = {L4},
        pages = {L4},
          doi = {10.3847/2041-8213/ace5a0},
archivePrefix = {arXiv},
       eprint = {2302.00012},
 primaryClass = {astro-ph.GA},
       adsurl = {https://ui.adsabs.harvard.edu/abs/2023ApJ...954L...4K}
}

@ARTICLE{Harikane_2023,
       author = {{Harikane}, Yuichi and {Zhang}, Yechi and {Nakajima}, Kimihiko and {Ouchi}, Masami and {Isobe}, Yuki and {Ono}, Yoshiaki and {Hatano}, Shun and {Xu}, Yi and {Umeda}, Hiroya},
        title = "{A JWST/NIRSpec First Census of Broad-line AGNs at z = 4-7: Detection of 10 Faint AGNs with M $_{BH}$ {}10$^{6}$-{}10$^{8}$ M $_{☉}$ and Their Host Galaxy Properties}",
      journal = {\apj},
         year = 2023,
        month = dec,
       volume = {959},
       number = {1},
          eid = {39},
        pages = {39},
          doi = {10.3847/1538-4357/ad029e},
archivePrefix = {arXiv},
       eprint = {2303.11946},
 primaryClass = {astro-ph.GA},
       adsurl = {https://ui.adsabs.harvard.edu/abs/2023ApJ...959...39H}
}

@ARTICLE{Wang_2024,
       author = {{Wang}, Bingjie and {Leja}, Joel and {de Graaff}, Anna and {Brammer}, Gabriel B. and {Weibel}, Andrea and {van Dokkum}, Pieter and {Baggen}, Josephine F.~W. and {Suess}, Katherine A. and {Greene}, Jenny E. and {Bezanson}, Rachel and {Cleri}, Nikko J. and {Hirschmann}, Michaela and {Labb{\'e}}, Ivo and {Matthee}, Jorryt and {McConachie}, Ian and {Naidu}, Rohan P. and {Nelson}, Erica and {Oesch}, Pascal A. and {Setton}, David J. and {Williams}, Christina C.},
        title = "{RUBIES: Evolved Stellar Populations with Extended Formation Histories at z {\ensuremath{\sim}} 7─8 in Candidate Massive Galaxies Identified with JWST/NIRSpec}",
      journal = {\apjl},
         year = 2024,
        month = jul,
       volume = {969},
       number = {1},
          eid = {L13},
        pages = {L13},
          doi = {10.3847/2041-8213/ad55f7},
archivePrefix = {arXiv},
       eprint = {2405.01473},
 primaryClass = {astro-ph.GA},
       adsurl = {https://ui.adsabs.harvard.edu/abs/2024ApJ...969L..13W}
}

@ARTICLE{Kokorev_2024,
       author = {{Kokorev}, Vasily and {Chisholm}, John and {Endsley}, Ryan and {Finkelstein}, Steven L. and {Greene}, Jenny E. and {Akins}, Hollis B. and {Bromm}, Volker and {Casey}, Caitlin M. and {Fujimoto}, Seiji and {Labb{\'e}}, Ivo and {Larson}, Rebecca L.},
        title = "{Silencing the Giant: Evidence of Active Galactic Nucleus Feedback and Quenching in a Little Red Dot at z = 4.13}",
      journal = {\apj},
         year = 2024,
        month = nov,
       volume = {975},
       number = {2},
          eid = {178},
        pages = {178},
          doi = {10.3847/1538-4357/ad7d03},
archivePrefix = {arXiv},
       eprint = {2407.20320},
 primaryClass = {astro-ph.GA},
       adsurl = {https://ui.adsabs.harvard.edu/abs/2024ApJ...975..178K}
}

@article{Maiolino,
	author = {{Maiolino, Roberto} and {Scholtz, Jan} and {Curtis-Lake, Emma} and {Carniani, Stefano} and {Baker, William} and {de Graaff, Anna} and {Tacchella, Sandro} and {Übler, Hannah} and {D’Eugenio, Francesco} and {Witstok, Joris} and {Curti, Mirko} and {Arribas, Santiago} and {Bunker, Andrew J.} and {Charlot, Stéphane} and {Chevallard, Jacopo} and {Eisenstein, Daniel J.} and {Egami, Eiichi} and {Ji, Zhiyuan} and {Jones, Gareth C.} and {Lyu, Jianwei} and {Rawle, Tim} and {Robertson, Brant} and {Rujopakarn, Wiphu} and {Perna, Michele} and {Sun, Fengwu} and {Venturi, Giacomo} and {Williams, Christina C.} and {Willott, Chris}},
	title = {JADES - The diverse population of infant black holes at 4 < z < 11: Merging, tiny, poor, but mighty},
	DOI= "10.1051/0004-6361/202347640",
	url= "https://doi.org/10.1051/0004-6361/202347640",
	journal = {\aap},
	year = 2024,
	volume = 691,
	pages = "A145",
}

@ARTICLE{Li_2023,
       author = {{Li}, Jennifer I.-Hsiu and {Shen}, Yue and {Ho}, Luis C. and {Brandt}, W.~N. and {Grier}, Catherine J. and {Hall}, Patrick B. and {Homayouni}, Y. and {Koekemoer}, Anton M. and {Schneider}, Donald P. and {Trump}, Jonathan R.},
        title = "{The Sloan Digital Sky Survey Reverberation Mapping Project: The Black Hole Mass-Stellar Mass Relations at 0.2 {\ensuremath{\lesssim}} z {\ensuremath{\lesssim}} 0.8}",
      journal = {\apj},
         year = 2023,
        month = sep,
       volume = {954},
       number = {2},
          eid = {173},
        pages = {173},
          doi = {10.3847/1538-4357/acddda},
archivePrefix = {arXiv},
       eprint = {2301.04177},
 primaryClass = {astro-ph.GA},
       adsurl = {https://ui.adsabs.harvard.edu/abs/2023ApJ...954..173L}
}

@ARTICLE{Suh_2020,
       author = {{Suh}, Hyewon and {Civano}, Francesca and {Trakhtenbrot}, Benny and {Shankar}, Francesco and {Hasinger}, G{\"u}nther and {Sanders}, David B. and {Allevato}, Viola},
        title = "{No Significant Evolution of Relations between Black Hole Mass and Galaxy Total Stellar Mass Up to z {\ensuremath{\sim}} 2.5}",
      journal = {\apj},
         year = 2020,
        month = jan,
       volume = {889},
       number = {1},
          eid = {32},
        pages = {32},
          doi = {10.3847/1538-4357/ab5f5f},
archivePrefix = {arXiv},
       eprint = {1912.02824},
 primaryClass = {astro-ph.GA},
       adsurl = {https://ui.adsabs.harvard.edu/abs/2020ApJ...889...32S}
}

@ARTICLE{Sun_2025,
       author = {{Sun}, Yang and {Lyu}, Jianwei and {Rieke}, George H. and {Ji}, Zhiyuan and {Sun}, Fengwu and {Zhu}, Yongda and {Bunker}, Andrew J. and {Cargile}, Phillip A. and {Circosta}, Chiara and {D'Eugenio}, Francesco and {Egami}, Eiichi and {Hainline}, Kevin and {Helton}, Jakob M. and {Rinaldi}, Pierluigi and {Robertson}, Brant E. and {Scholtz}, Jan and {Shivaei}, Irene and {Stone}, Meredith A. and {Tacchella}, Sandro and {Williams}, Christina C. and {Willmer}, Christopher N.~A. and {Willott}, Chris},
        title = "{No Evidence for a Significant Evolution of M$_{{\ensuremath{\bullet}}}$─M. Relation in Massive Galaxies up to z {\ensuremath{\sim}} 4}",
      journal = {\apj},
         year = 2025,
        month = jan,
       volume = {978},
       number = {1},
          eid = {98},
        pages = {98},
          doi = {10.3847/1538-4357/ad973b},
archivePrefix = {arXiv},
       eprint = {2409.06796},
 primaryClass = {astro-ph.GA},
       adsurl = {https://ui.adsabs.harvard.edu/abs/2025ApJ...978...98S}
}

@ARTICLE{McGreer_2018,
       author = {{McGreer}, Ian D. and {Fan}, Xiaohui and {Jiang}, Linhua and {Cai}, Zheng},
        title = "{The Faint End of the z = 5 Quasar Luminosity Function from the CFHTLS}",
      journal = {\aj},
         year = 2018,
        month = mar,
       volume = {155},
       number = {3},
          eid = {131},
        pages = {131},
          doi = {10.3847/1538-3881/aaaab4},
archivePrefix = {arXiv},
       eprint = {1710.09390},
 primaryClass = {astro-ph.GA},
       adsurl = {https://ui.adsabs.harvard.edu/abs/2018AJ....155..131M}
}

@ARTICLE{Matsuoka_2018,
       author = {{Matsuoka}, Yoshiki and {Strauss}, Michael A. and {Kashikawa}, Nobunari and {Onoue}, Masafusa and {Iwasawa}, Kazushi and {Tang}, Ji-Jia and {Lee}, Chien-Hsiu and {Imanishi}, Masatoshi and {Nagao}, Tohru and {Akiyama}, Masayuki and {Asami}, Naoko and {Bosch}, James and {Furusawa}, Hisanori and {Goto}, Tomotsugu and {Gunn}, James E. and {Harikane}, Yuichi and {Ikeda}, Hiroyuki and {Izumi}, Takuma and {Kawaguchi}, Toshihiro and {Kato}, Nanako and {Kikuta}, Satoshi and {Kohno}, Kotaro and {Komiyama}, Yutaka and {Lupton}, Robert H. and {Minezaki}, Takeo and {Miyazaki}, Satoshi and {Murayama}, Hitoshi and {Niida}, Mana and {Nishizawa}, Atsushi J. and {Noboriguchi}, Akatoki and {Oguri}, Masamune and {Ono}, Yoshiaki and {Ouchi}, Masami and {Price}, Paul A. and {Sameshima}, Hiroaki and {Schulze}, Andreas and {Shirakata}, Hikari and {Silverman}, John D. and {Sugiyama}, Naoshi and {Tait}, Philip J. and {Takada}, Masahiro and {Takata}, Tadafumi and {Tanaka}, Masayuki and {Toba}, Yoshiki and {Utsumi}, Yousuke and {Wang}, Shiang-Yu and {Yamashita}, Takuji},
        title = "{Subaru High-z  Exploration of Low-luminosity Quasars (SHELLQs). V. Quasar Luminosity Function and Contribution to Cosmic Reionization at z = 6}",
      journal = {\apj},
         year = 2018,
        month = dec,
       volume = {869},
       number = {2},
          eid = {150},
        pages = {150},
          doi = {10.3847/1538-4357/aaee7a},
archivePrefix = {arXiv},
       eprint = {1811.01963},
 primaryClass = {astro-ph.GA},
       adsurl = {https://ui.adsabs.harvard.edu/abs/2018ApJ...869..150M}
}

@ARTICLE{Niida_2020,
       author = {{Niida}, Mana and {Nagao}, Tohru and {Ikeda}, Hiroyuki and {Akiyama}, Masayuki and {Matsuoka}, Yoshiki and {He}, Wanqiu and {Matsuoka}, Kenta and {Toba}, Yoshiki and {Onoue}, Masafusa and {Kobayashi}, Masakazu A.~R. and {Taniguchi}, Yoshiaki and {Furusawa}, Hisanori and {Harikane}, Yuichi and {Imanishi}, Masatoshi and {Kashikawa}, Nobunari and {Kawaguchi}, Toshihiro and {Komiyama}, Yutaka and {Shirakata}, Hikari and {Terashima}, Yuichi and {Ueda}, Yoshihiro},
        title = "{The Faint End of the Quasar Luminosity Function at z {\ensuremath{\sim}} 5 from the Subaru Hyper Suprime-Cam Survey}",
      journal = {\apj},
         year = 2020,
        month = dec,
       volume = {904},
       number = {2},
          eid = {89},
        pages = {89},
          doi = {10.3847/1538-4357/abbe11},
archivePrefix = {arXiv},
       eprint = {2010.00481},
 primaryClass = {astro-ph.GA},
       adsurl = {https://ui.adsabs.harvard.edu/abs/2020ApJ...904...89N}
}

@ARTICLE{Matsuoka_2023,
       author = {{Matsuoka}, Yoshiki and {Onoue}, Masafusa and {Iwasawa}, Kazushi and {Strauss}, Michael A. and {Kashikawa}, Nobunari and {Izumi}, Takuma and {Nagao}, Tohru and {Imanishi}, Masatoshi and {Akiyama}, Masayuki and {Silverman}, John D. and {Asami}, Naoko and {Bosch}, James and {Furusawa}, Hisanori and {Goto}, Tomotsugu and {Gunn}, James E. and {Harikane}, Yuichi and {Ikeda}, Hiroyuki and {Inayoshi}, Kohei and {Ishimoto}, Rikako and {Kawaguchi}, Toshihiro and {Kikuta}, Satoshi and {Kohno}, Kotaro and {Komiyama}, Yutaka and {Lee}, Chien-Hsiu and {Lupton}, Robert H. and {Minezaki}, Takeo and {Miyazaki}, Satoshi and {Murayama}, Hitoshi and {Nishizawa}, Atsushi J. and {Oguri}, Masamune and {Ono}, Yoshiaki and {Oogi}, Taira and {Ouchi}, Masami and {Price}, Paul A. and {Sameshima}, Hiroaki and {Sugiyama}, Naoshi and {Tait}, Philip J. and {Takada}, Masahiro and {Takahashi}, Ayumi and {Takata}, Tadafumi and {Tanaka}, Masayuki and {Toba}, Yoshiki and {Wang}, Shiang-Yu and {Yamashita}, Takuji},
        title = "{Quasar Luminosity Function at z = 7}",
      journal = {\apjl},
         year = 2023,
        month = jun,
       volume = {949},
       number = {2},
          eid = {L42},
        pages = {L42},
          doi = {10.3847/2041-8213/acd69f},
archivePrefix = {arXiv},
       eprint = {2305.11225},
 primaryClass = {astro-ph.GA},
       adsurl = {https://ui.adsabs.harvard.edu/abs/2023ApJ...949L..42M}
}

@ARTICLE{2024ApJ...963..129M,
       author = {{Matthee}, Jorryt and {Naidu}, Rohan P. and {Brammer}, Gabriel and {Chisholm}, John and {Eilers}, Anna-Christina and {Goulding}, Andy and {Greene}, Jenny and {Kashino}, Daichi and {Labbe}, Ivo and {Lilly}, Simon J. and {Mackenzie}, Ruari and {Oesch}, Pascal A. and {Weibel}, Andrea and {Wuyts}, Stijn and {Xiao}, Mengyuan and {Bordoloi}, Rongmon and {Bouwens}, Rychard and {van Dokkum}, Pieter and {Illingworth}, Garth and {Kramarenko}, Ivan and {Maseda}, Michael V. and {Mason}, Charlotte and {Meyer}, Romain A. and {Nelson}, Erica J. and {Reddy}, Naveen A. and {Shivaei}, Irene and {Simcoe}, Robert A. and {Yue}, Minghao},
        title = "{Little Red Dots: An Abundant Population of Faint Active Galactic Nuclei at z {\ensuremath{\sim}} 5 Revealed by the EIGER and FRESCO JWST Surveys}",
      journal = {\apj},
         year = 2024,
        month = mar,
       volume = {963},
       number = {2},
          eid = {129},
        pages = {129},
          doi = {10.3847/1538-4357/ad2345},
archivePrefix = {arXiv},
       eprint = {2306.05448},
 primaryClass = {astro-ph.GA},
       adsurl = {https://ui.adsabs.harvard.edu/abs/2024ApJ...963..129M}
}

@ARTICLE{Speagle2014,
       author = {{Speagle}, J.~S. and {Steinhardt}, C.~L. and {Capak}, P.~L. and {Silverman}, J.~D.},
        title = "{A Highly Consistent Framework for the Evolution of the Star-Forming ``Main Sequence'' from z \raisebox{-0.5ex}\textasciitilde 0-6}",
      journal = {\apjs},
         year = 2014,
        month = oct,
       volume = {214},
       number = {2},
          eid = {15},
        pages = {15},
          doi = {10.1088/0067-0049/214/2/15},
archivePrefix = {arXiv},
       eprint = {1405.2041},
 primaryClass = {astro-ph.GA},
       adsurl = {https://ui.adsabs.harvard.edu/abs/2014ApJS..214...15S}
}

@ARTICLE{2025MNRAS.538.1921M,
       author = {{Maiolino}, Roberto and {Risaliti}, Guido and {Signorini}, Matilde and {Trefoloni}, Bartolomeo and {Juod{\v{z}}balis}, Ignas and {Scholtz}, Jan and {{\"U}bler}, Hannah and {D'Eugenio}, Francesco and {Carniani}, Stefano and {Fabian}, Andy and {Ji}, Xihan and {Mazzolari}, Giovanni and {Bertola}, Elena and {Brusa}, Marcella and {Bunker}, Andrew J. and {Charlot}, Stephane and {Comastri}, Andrea and {Cresci}, Giovanni and {DeCoursey}, Christa Noel and {Egami}, Eiichi and {Fiore}, Fabrizio and {Gilli}, Roberto and {Perna}, Michele and {Tacchella}, Sandro and {Venturi}, Giacomo},
        title = "{JWST meets Chandra: a large population of Compton thick, feedback-free, and intrinsically X-ray weak AGN, with a sprinkle of SNe}",
      journal = {\mnras},
         year = 2025,
        month = apr,
       volume = {538},
       number = {3},
        pages = {1921-1943},
          doi = {10.1093/mnras/staf359},
archivePrefix = {arXiv},
       eprint = {2405.00504},
 primaryClass = {astro-ph.GA},
       adsurl = {https://ui.adsabs.harvard.edu/abs/2025MNRAS.538.1921M}
}

@ARTICLE{2024ApJ...974L..26Y,
       author = {{Yue}, Minghao and {Eilers}, Anna-Christina and {Ananna}, Tonima Tasnim and {Panagiotou}, Christos and {Kara}, Erin and {Miyaji}, Takamitsu},
        title = "{Stacking X-Ray Observations of ``Little Red Dots'': Implications for Their Active Galactic Nucleus Properties}",
      journal = {\apjl},
         year = 2024,
        month = oct,
       volume = {974},
       number = {2},
          eid = {L26},
        pages = {L26},
          doi = {10.3847/2041-8213/ad7eba},
archivePrefix = {arXiv},
       eprint = {2404.13290},
 primaryClass = {astro-ph.GA},
       adsurl = {https://ui.adsabs.harvard.edu/abs/2024ApJ...974L..26Y}
}

@ARTICLE{2024ApJ...975L...4C,
       author = {{Casey}, Caitlin M. and {Akins}, Hollis B. and {Kokorev}, Vasily and {McKinney}, Jed and {Cooper}, Olivia R. and {Long}, Arianna S. and {Franco}, Maximilien and {Manning}, Sinclaire M.},
        title = "{Dust in Little Red Dots}",
      journal = {\apjl},
         year = 2024,
        month = nov,
       volume = {975},
       number = {1},
          eid = {L4},
        pages = {L4},
          doi = {10.3847/2041-8213/ad7ba7},
archivePrefix = {arXiv},
       eprint = {2407.05094},
 primaryClass = {astro-ph.GA},
       adsurl = {https://ui.adsabs.harvard.edu/abs/2024ApJ...975L...4C}
}

@ARTICLE{2025ApJ...991L..10S,
       author = {{Setton}, David J. and {Greene}, Jenny E. and {Spilker}, Justin S. and {Williams}, Christina C. and {Labb{\'e}}, Ivo and {Ma}, Yilun and {Wang}, Bingjie and {Whitaker}, Katherine E. and {Leja}, Joel and {de Graaff}, Anna and {Alberts}, Stacey and {Bezanson}, Rachel and {Boogaard}, Leindert A. and {Brammer}, Gabriel and {Cutler}, Sam E. and {Cleri}, Nikko J. and {Cooper}, Olivia R. and {Dayal}, Pratika and {Fujimoto}, Seiji and {Furtak}, Lukas J. and {Goulding}, Andy D. and {Hirschmann}, Michaela and {Kokorev}, Vasily and {Maseda}, Michael V. and {McConachie}, Ian and {Matthee}, Jorryt and {Miller}, Tim B. and {Naidu}, Rohan P. and {Oesch}, Pascal A. and {Pan}, Richard and {Price}, Sedona H. and {Suess}, Katherine A. and {Weaver}, John R. and {Xiao}, Mengyuan and {Zhang}, Yunchong and {Zitrin}, Adi},
        title = "{A Confirmed Deficit of Hot and Cold Dust Emission in the Most Luminous Little Red Dots}",
      journal = {\apjl},
         year = 2025,
        month = sep,
       volume = {991},
       number = {1},
          eid = {L10},
        pages = {L10},
          doi = {10.3847/2041-8213/ade78b},
archivePrefix = {arXiv},
       eprint = {2503.02059},
 primaryClass = {astro-ph.GA},
       adsurl = {https://ui.adsabs.harvard.edu/abs/2025ApJ...991L..10S}
}

@ARTICLE{2025A&A...693L...2P,
       author = {{Perger}, K. and {Fogasy}, J. and {Frey}, S. and {Gab{\'a}nyi}, K. {\'E}.},
        title = "{Deep silence: Radio properties of little red dots}",
      journal = {\aap},
         year = 2025,
        month = jan,
       volume = {693},
          eid = {L2},
        pages = {L2},
          doi = {10.1051/0004-6361/202452422},
archivePrefix = {arXiv},
       eprint = {2411.19518},
 primaryClass = {astro-ph.GA},
       adsurl = {https://ui.adsabs.harvard.edu/abs/2025A&A...693L...2P}
}

@ARTICLE{2026ApJ..1000...90I,
       author = {{Inayoshi}, Kohei and {Murase}, Kohta and {Kashiyama}, Kazumi},
        title = "{Spectral Uniformity of Little Red Dots: A Natural Outcome of Coevolving Seed Black Holes and Nascent Starbursts}",
      journal = {\apj},
         year = 2026,
        month = mar,
       volume = {1000},
       number = {1},
          eid = {90},
        pages = {90},
          doi = {10.3847/1538-4357/ae42ce},
archivePrefix = {arXiv},
       eprint = {2509.19422},
 primaryClass = {astro-ph.GA},
       adsurl = {https://ui.adsabs.harvard.edu/abs/2026ApJ..1000...90I}
}

@ARTICLE{2014ApJ...782....9H,
       author = {{Hickox}, Ryan C. and {Mullaney}, James R. and {Alexander}, David M. and {Chen}, Chien-Ting J. and {Civano}, Francesca M. and {Goulding}, Andy D. and {Hainline}, Kevin N.},
        title = "{Black Hole Variability and the Star Formation-Active Galactic Nucleus Connection: Do All Star-forming Galaxies Host an Active Galactic Nucleus?}",
      journal = {\apj},
         year = 2014,
        month = feb,
       volume = {782},
       number = {1},
          eid = {9},
        pages = {9},
          doi = {10.1088/0004-637X/782/1/9},
archivePrefix = {arXiv},
       eprint = {1306.3218},
 primaryClass = {astro-ph.CO},
       adsurl = {https://ui.adsabs.harvard.edu/abs/2014ApJ...782....9H}
}

@ARTICLE{2014ARA&A..52..373L,
       author = {{Lutz}, Dieter},
        title = "{Far-Infrared Surveys of Galaxy Evolution}",
      journal = {\araa},
         year = 2014,
        month = aug,
       volume = {52},
        pages = {373-414},
          doi = {10.1146/annurev-astro-081913-035953},
archivePrefix = {arXiv},
       eprint = {1403.3334},
 primaryClass = {astro-ph.GA},
       adsurl = {https://ui.adsabs.harvard.edu/abs/2014ARA&A..52..373L}
}

@ARTICLE{1996ApJ...457..645W,
       author = {{Wang}, Boqi and {Heckman}, Timothy M.},
        title = "{Internal Absorption and the Luminosity of Disk Galaxies}",
      journal = {\apj},
         year = 1996,
        month = feb,
       volume = {457},
        pages = {645},
          doi = {10.1086/176760},
       adsurl = {https://ui.adsabs.harvard.edu/abs/1996ApJ...457..645W}
}

@ARTICLE{2001ApJ...558...72S,
       author = {{Sullivan}, Mark and {Mobasher}, Bahram and {Chan}, Ben and {Cram}, Lawrence and {Ellis}, Richard and {Treyer}, Marie and {Hopkins}, Andrew},
        title = "{A Comparison of Independent Star Formation Diagnostics for an Ultraviolet-selected Sample of Nearby Galaxies}",
      journal = {\apj},
         year = 2001,
        month = sep,
       volume = {558},
       number = {1},
        pages = {72-80},
          doi = {10.1086/322451},
archivePrefix = {arXiv},
       eprint = {astro-ph/0104425},
 primaryClass = {astro-ph},
       adsurl = {https://ui.adsabs.harvard.edu/abs/2001ApJ...558...72S}
}

@ARTICLE{2002A&A...393...33B,
       author = {{Buat}, V. and {Burgarella}, D. and {Deharveng}, J.~M. and {Kunth}, D.},
        title = "{Spectral Energy Distributions of starburst galaxies in the 900-1200 {\r{A}} range}",
      journal = {\aap},
         year = 2002,
        month = oct,
       volume = {393},
        pages = {33-42},
          doi = {10.1051/0004-6361:20020925},
archivePrefix = {arXiv},
       eprint = {astro-ph/0206136},
 primaryClass = {astro-ph},
       adsurl = {https://ui.adsabs.harvard.edu/abs/2002A&A...393...33B}
}

@ARTICLE{2009ApJ...698L.116P,
       author = {{Pannella}, M. and {Carilli}, C.~L. and {Daddi}, E. and {McCracken}, H.~J. and {Owen}, F.~N. and {Renzini}, A. and {Strazzullo}, V. and {Civano}, F. and {Koekemoer}, A.~M. and {Schinnerer}, E. and {Scoville}, N. and {Smol{\v{c}}i{\'c}}, V. and {Taniguchi}, Y. and {Aussel}, H. and {Kneib}, J.~P. and {Ilbert}, O. and {Mellier}, Y. and {Salvato}, M. and {Thompson}, D. and {Willott}, C.~J.},
        title = "{Star Formation and Dust Obscuration at z {\ensuremath{\approx}} 2: Galaxies at the Dawn of Downsizing}",
      journal = {\apjl},
         year = 2009,
        month = jun,
       volume = {698},
       number = {2},
        pages = {L116-L120},
          doi = {10.1088/0004-637X/698/2/L116},
archivePrefix = {arXiv},
       eprint = {0905.1674},
 primaryClass = {astro-ph.CO},
       adsurl = {https://ui.adsabs.harvard.edu/abs/2009ApJ...698L.116P}
}

@ARTICLE{2013MNRAS.429.1113H,
       author = {{Heinis}, S. and {Buat}, V. and {B{\'e}thermin}, M. and {Aussel}, H. and {Bock}, J. and {Boselli}, A. and {Burgarella}, D. and {Conley}, A. and {Cooray}, A. and {Farrah}, D. and {Ibar}, E. and {Ilbert}, O. and {Ivison}, R.~J. and {Magdis}, G. and {Marsden}, G. and {Oliver}, S.~J. and {Page}, M.~J. and {Rodighiero}, G. and {Roehlly}, Y. and {Schulz}, B. and {Scott}, Douglas and {Smith}, A.~J. and {Viero}, M. and {Wang}, L. and {Zemcov}, M.},
        title = "{HERMES: unveiling obscured star formation - the far-infrared luminosity function of ultraviolet-selected galaxies at z {\ensuremath{\sim}} 1.5}",
      journal = {\mnras},
         year = 2013,
        month = feb,
       volume = {429},
       number = {2},
        pages = {1113-1132},
          doi = {10.1093/mnras/sts397},
archivePrefix = {arXiv},
       eprint = {1211.4336},
 primaryClass = {astro-ph.CO},
       adsurl = {https://ui.adsabs.harvard.edu/abs/2013MNRAS.429.1113H}
}

@ARTICLE{2014MNRAS.437.1268H,
       author = {{Heinis}, S. and {Buat}, V. and {B{\'e}thermin}, M. and {Bock}, J. and {Burgarella}, D. and {Conley}, A. and {Cooray}, A. and {Farrah}, D. and {Ilbert}, O. and {Magdis}, G. and {Marsden}, G. and {Oliver}, S.~J. and {Rigopoulou}, D. and {Roehlly}, Y. and {Schulz}, B. and {Symeonidis}, M. and {Viero}, M. and {Xu}, C.~K. and {Zemcov}, M.},
        title = "{HerMES: dust attenuation and star formation activity in ultraviolet-selected samples from z{\ensuremath{\sim}} 4 to {\ensuremath{\sim}} 1.5}",
      journal = {\mnras},
         year = 2014,
        month = jan,
       volume = {437},
       number = {2},
        pages = {1268-1283},
          doi = {10.1093/mnras/stt1960},
archivePrefix = {arXiv},
       eprint = {1310.3227},
 primaryClass = {astro-ph.CO},
       adsurl = {https://ui.adsabs.harvard.edu/abs/2014MNRAS.437.1268H}
}

@ARTICLE{2024A&A...690A.290C,
       author = {{Calabr{\`o}}, A. and {Pentericci}, L. and {Santini}, P. and {Ferrara}, A. and {Llerena}, M. and {Mascia}, S. and {Napolitano}, L. and {Yung}, L.~Y.~A. and {Bisigello}, L. and {Castellano}, M. and {Cleri}, N.~J. and {Dekel}, A. and {Dickinson}, M. and {Franco}, M. and {Giavalisco}, M. and {Hirschmann}, M. and {Holwerda}, B.~W. and {Koekemoer}, A.~M. and {Lucas}, R.~A. and {Pacucci}, F. and {Pirzkal}, N. and {Roberts-Borsani}, G. and {Seill{\'e}}, L.~M. and {Tacchella}, S. and {Wilkins}, S. and {Amor{\'\i}n}, R. and {Arrabal Haro}, P. and {Bagley}, M.~B. and {Finkelstein}, S.~L. and {Kartaltepe}, J.~S. and {Papovich}, C.},
        title = "{Evolution of the star formation rate and {\ensuremath{\Sigma}}$_{SFR}$ of galaxies at cosmic morning (4 < z < 10)}",
      journal = {\aap},
         year = 2024,
        month = oct,
       volume = {690},
          eid = {A290},
        pages = {A290},
          doi = {10.1051/0004-6361/202449768},
archivePrefix = {arXiv},
       eprint = {2402.17829},
 primaryClass = {astro-ph.GA},
       adsurl = {https://ui.adsabs.harvard.edu/abs/2024A&A...690A.290C}
}

@ARTICLE{2015ApJ...806..259R,
       author = {{Reddy}, Naveen A. and {Kriek}, Mariska and {Shapley}, Alice E. and {Freeman}, William R. and {Siana}, Brian and {Coil}, Alison L. and {Mobasher}, Bahram and {Price}, Sedona H. and {Sanders}, Ryan L. and {Shivaei}, Irene},
        title = "{The MOSDEF Survey: Measurements of Balmer Decrements and the Dust Attenuation Curve at Redshifts z \raisebox{-0.5ex}\textasciitilde 1.4-2.6}",
      journal = {\apj},
         year = 2015,
        month = jun,
       volume = {806},
       number = {2},
          eid = {259},
        pages = {259},
          doi = {10.1088/0004-637X/806/2/259},
archivePrefix = {arXiv},
       eprint = {1504.02782},
 primaryClass = {astro-ph.GA},
       adsurl = {https://ui.adsabs.harvard.edu/abs/2015ApJ...806..259R}
}

@ARTICLE{2023ApJ...951...29L,
       author = {{Lorenz}, Brian and {Kriek}, Mariska and {Shapley}, Alice E. and {Reddy}, Naveen A. and {Sanders}, Ryan L. and {Barro}, Guillermo and {Coil}, Alison L. and {Mobasher}, Bahram and {Price}, Sedona H. and {Runco}, Jordan N. and {Shivaei}, Irene and {Siana}, Brian and {Weisz}, Daniel R.},
        title = "{An Updated Dust-to-Star Geometry: Dust Attenuation Does Not Depend on Inclination in 1.3 {\ensuremath{\leq}}z {\ensuremath{\leq}}2.6 Star-forming Galaxies from MOSDEF}",
      journal = {\apj},
         year = 2023,
        month = jul,
       volume = {951},
       number = {1},
          eid = {29},
        pages = {29},
          doi = {10.3847/1538-4357/accdd1},
archivePrefix = {arXiv},
       eprint = {2304.08521},
 primaryClass = {astro-ph.GA},
       adsurl = {https://ui.adsabs.harvard.edu/abs/2023ApJ...951...29L}
}

@ARTICLE{2023MNRAS.523.4568F,
       author = {{Furtak}, Lukas J. and {Zitrin}, Adi and {Weaver}, John R. and {Atek}, Hakim and {Bezanson}, Rachel and {Labb{\'e}}, Ivo and {Whitaker}, Katherine E. and {Leja}, Joel and {Price}, Sedona H. and {Brammer}, Gabriel B. and {Wang}, Bingjie and {Marchesini}, Danilo and {Pan}, Richard and {Dayal}, Pratika and {van Dokkum}, Pieter and {Feldmann}, Robert and {Fujimoto}, Seiji and {Franx}, Marijn and {Khullar}, Gourav and {Nelson}, Erica J. and {Mowla}, Lamiya A.},
        title = "{UNCOVERing the extended strong lensing structures of Abell 2744 with the deepest JWST imaging}",
      journal = {\mnras},
         year = 2023,
        month = aug,
       volume = {523},
       number = {3},
        pages = {4568-4582},
          doi = {10.1093/mnras/stad1627},
archivePrefix = {arXiv},
       eprint = {2212.04381},
 primaryClass = {astro-ph.GA},
       adsurl = {https://ui.adsabs.harvard.edu/abs/2023MNRAS.523.4568F}
}
\bibliographystyle{aasjournalv7}

\end{CJK*}
\end{document}